%% file: main.tex
\documentclass[lettersize,journal]{IEEEtran}

\usepackage[a-1b]{pdfx}
\usepackage{xcolor}
\usepackage{soul}
\definecolor{NavyBlue}{RGB}{0, 0, 128}
\usepackage{url}
\usepackage[utf8]{inputenc}

\usepackage{amsthm}
\usepackage{amsmath}
\usepackage{amsthm}
\usepackage{amssymb}
\usepackage{graphicx,color}
\usepackage{epstopdf}
\usepackage{wrapfig}
\usepackage{algorithmic}
\usepackage{dirtytalk}
\usepackage{algorithm}
\usepackage{mathrsfs}
\usepackage{float}
\usepackage{mathtools}
\usepackage{tabulary}
\usepackage{booktabs}
\usepackage{caption}
\usepackage{subfig}
\usepackage{xprintlen}
\usepackage{multirow}
\hypersetup{hidelinks=true}
\graphicspath{{./}{./files/figures_eps/}}

\PassOptionsToPackage{bookmarks={false}}{hyperref}

\newcommand{\proposed}{{\tt{HEEDFUL}}} 

\newenvironment{myitemize}
  {\begin{list}{$\bullet$}{%
    \setlength{\leftmargin}{0.4cm}%
    \setlength{\parsep}{0.0cm}%
    \setlength{\itemsep}{0.05cm}%
    \setlength{\topsep}{0.0cm}%
  }}
  {\end{list}}

\newcommand{\myitemizebegin}{\begin{list}{$\bullet$}
{
 \setlength{\leftmargin}{0.09cm}
 \setlength{\parsep}{0.0cm}
 \setlength{\itemsep}{0.05cm}
 \setlength{\topsep}{0.0cm}
}}
\newcommand{\myitemizeend}{\end{list}}

\newcommand{\comment}[1]{ }

\usepackage{titlesec}
\titlespacing*{\section}
{0pt}{1.5ex}{0ex} 
\titlespacing*{\subsection}
{0pt}{0.8ex}{0ex}  
\titlespacing*{\subsubsection}
{0pt}{0.5ex}{0ex} 

\begin{document}

\title{HEEDFUL: Leveraging Sequential Transfer Learning for Robust WiFi Device Fingerprinting Amid Hardware Warm-Up Effects}

\author{Abdurrahman Elmaghbub,~\IEEEmembership{Student Member,~IEEE}, and Bechir Hamdaoui,~\IEEEmembership{Senior Member,~IEEE}%
    \thanks{A. Elmaghbub and B. Hamdaoui are with Oregon State University, Corvallis, OR 97330 USA (e-mails: \{elmaghba,hamdaoui\}@oregonstate.edu).}%
    \thanks{Corresponding author: A. Elmaghbub (email: elmaghba@oregonstate.edu).}%
    \thanks{This work is supported in part by NSF Awards No. 2350214 and 2003273.}%
}







\maketitle


\begin{abstract}
\input{1-abstract}
\end{abstract}


 \maketitle

\section{INTRODUCTION}
\label{sec:into}
\input{2-introduction}


\section{Related Works}
\label{sec:related-work}
\input{related}

\section{EXPOSING the EFFECT of HARDWARE WARM-UP During Stabilization}
\label{sec:exposing}
\input{3-exposing}



\section{The ANATOMY of RF FINGERPRINTS}
\label{sec:anatomy}
\input{4-anatomy}

\subsection{Hardware Impairments Analysis}
\label{subsec:Imp_behavior}
\input{5-behavior}

\section{The PROPOSED FRAMEWORK: \proposed}
\label{sec:proposed}
\input{8-framework}

\section{PERFORMANCE EVALUATION and ANALYSIS}
\label{sec:evaluation}
\input{9-evaluation}

\section{CONCLUSIONS and FUTURE DIRECTIONS}
\label{sec:conc}
\input{10-conclusion}

\bibliographystyle{IEEEtran}
\bibliography{IEEEexample, refs-wisec24-journal}

\begin{IEEEbiography}{Abdurrahman Elmaghbub}~received the B.S. degree with summa cum laude, and MS in Electrical and Computer Engineering from Oregon State University in 2019, 2021, respectively, and is currently pursuing his Ph.D. degree in the School of Electrical Engineering and Computer Science at Oregon State University. His research interests are in the area of wireless communication and networking with a current focus on applying deep learning to wireless device classification.
\end{IEEEbiography}

\begin{IEEEbiography}{Bechir Hamdaoui}~is a Professor in the School of Electrical Engineering and Computer Science at Oregon State University. He received M.S. degrees in both ECE (2002) and CS (2004), and the Ph.D. degree in ECE (2005) all from the University of Wisconsin-Madison. His general interests are on theoretical and experimental research that enhances the cybersecurity \& resiliency of future intelligent networked systems, including connected \& autonomous vehicles, 5G/6G wireless, networked drones, smart cities, and cloud data centers. He is the Founding Director of the NetSTAR Laboratory at Oregon State University.
Dr. Hamdaoui and his team have won several awards, including the ISSIP 2020 Distinguished Recognition Award, the 2009 NSF CAREER Award, the ICC 2017 Best Paper Award, and the 2016 EECS Outstanding Research Award. He serves/served as an Associate Editor for several IEEE journals and magazines and chaired \& organized many IEEE/ACM conference symposia \& workshop programs. He served as a Distinguished Lecturer for the IEEE Communication Society in 2016 and 2017 and served as the Chair \& Co-chair of the IEEE Communications Society's Wireless Technical Committee (WTC) from January 2019 until December 2022.
\end{IEEEbiography}
\end{document}

%% file: 1-abstract.tex
Deep Learning-based RF fingerprinting approaches struggle to perform well in cross-domain scenarios, particularly during hardware warm-up. This often-overlooked vulnerability has been jeopardizing their reliability and their  adoption in practical settings. To address this critical gap, in this work, we first dive deep into the anatomy of RF fingerprints, revealing insights into the temporal fingerprinting variations during and post hardware stabilization. Introducing \proposed, a novel framework harnessing sequential transfer learning and targeted impairment estimation, we then address these challenges with remarkable consistency, eliminating blind spots even during challenging warm-up phases. Our evaluation showcases \proposed's efficacy, achieving remarkable classification accuracies of up to 96\% during the initial device operation intervals---far surpassing traditional models. Furthermore, cross-day and cross-protocol assessments confirm \proposed’s superiority, achieving and maintaining high accuracy during both the stable and initial warm-up phases when tested on WiFi signals. Additionally, we release WiFi type B and N RF fingerprint datasets that, for the first time, incorporate both the time-domain representation and real hardware impairments of the frames. This underscores the importance of leveraging hardware impairment data, enabling a deeper understanding of fingerprints and facilitating the development of more robust RF fingerprinting solutions.

%% file: 2-introduction.tex

\IEEEPARstart{D}{eep} learning (DL)-based RF fingerprinting holds great potential for enabling a wide range of network security services, including safeguarding against unauthorized network access, detecting malicious network activities, and fortifying the overall security of interconnected cyber-physical infrastructures \cite{hamdaoui2020deep, sankhe2019oracle, jian2020deep, jagannath2022comprehensive}. It enables automated device authentication through the extraction of hardware fingerprints from received RF signals that exist due to inherent hardware imperfections introduced during the manufacturing of the RF components~\cite{sankhe2019no, elmaghbub2020widescan, rajendran2022rf, smaini2012rf}. 
Because such hardware imperfections are random and differ from device to device, they create per-device-unique signatures that can be leveraged to identify and distinguish devices from one another.


\subsection{Temporal Sensitivity Challenges}
Despite their promising capabilities, DL-based RF fingerprinting approaches have shown a notable sensitivity to temporal variations \cite{elmaghbub2021, adl, li2022radionet, hamdaoui2022deep, al2020exposing, hanna_wisig_2022}. For instance, when trained on data captured one day and tested on data from another day, a significant performance decline has been observed \cite{elmaghbub2021,al2020exposing}. While this degradation is commonly attributed to changes in the wireless channel~\cite{al2020exposing, yan2022rrf}, further examination is warranted since similar degradation trends were also observed in wired scenarios \cite{elmaghbub2021, al2020exposing}. This decline in performance persists even when the time gap between training and testing data collection is as short as 5 minutes in a static environment \cite{adl}, challenging the prevailing belief about the wireless channel's role in the performance drop. 
An additional insight comes from the results reported in \cite{adl}, indicating that the performance of a deep learning model degrades sharply when trained on data captured during the first two minutes of device operation, with accuracy declining as the time gap between training and testing increases---until the gap no longer matters. This suggests the issue stems from hardware changes during early operation rather than channel effects, highlighting the need to study hardware stability over time to fully understand and mitigate these performance issues.

\subsection{Fingerprinting During Hardware Warm-Up}
\begin{figure}
\centering
    \begin{minipage}{\linewidth} 
        \centering 
        \includegraphics[width=\linewidth,height=.55\linewidth]{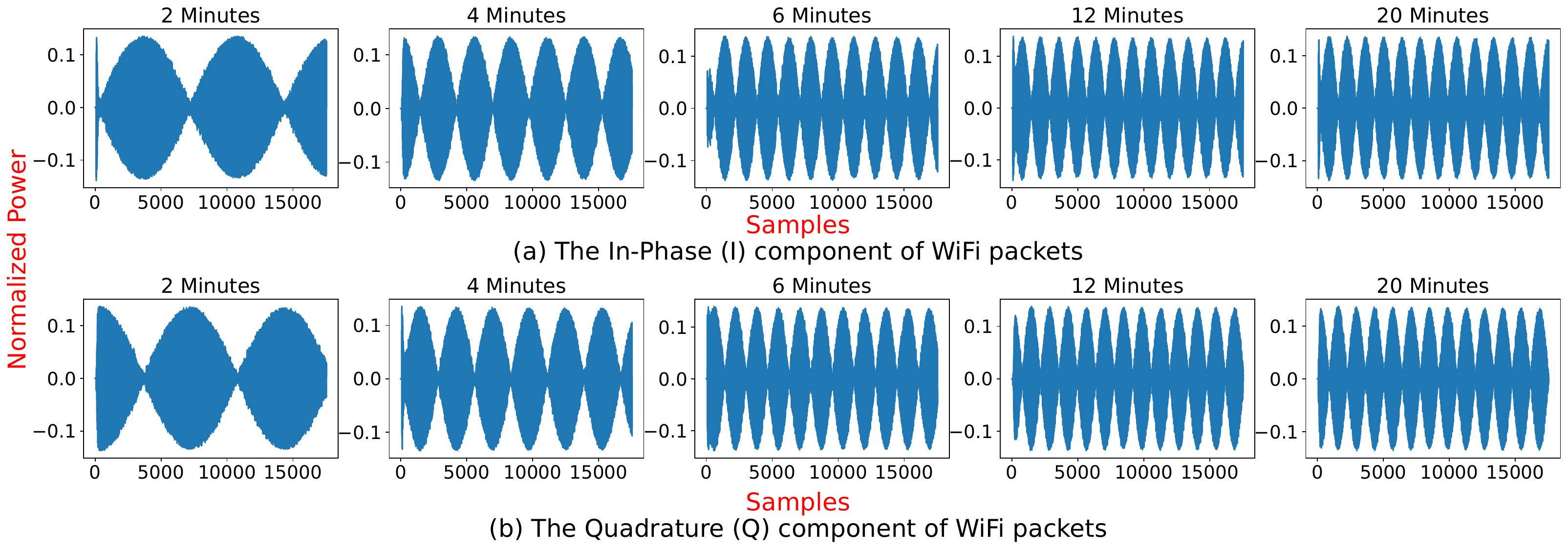}
       \caption{I/Q signal behavior observed at different times during the device's initial operation period.} 
       \label{fig:warm_up_interval}
    \end{minipage}
    
\end{figure}

In Fig.~\ref{fig:warm_up_interval}, we depict time-domain I/Q representations of WiFi frames captured at different intervals during devices' initial operation: the far-left figures represent signals recorded two minutes after device activation, while the far-right figures correspond to signals recorded at 20 minutes post-activation, with intermediate figures corresponding to recordings at subsequent time intervals. 
Observe that as the time-of-capture increases, both the I and Q shapes undergo changes, characterized by a frequency-increasing envelope in the signal over time. Notably, the I and Q signal components become more stable around the 12-minute mark from device activation, which we will refer to as the end of `warm-up' period. The alterations observed during the warm-up period are primarily attributed to the complicated interplay of thermal effects on electronic RF components \cite{vanier1992aging}. As these devices power on and initiate transmission, the increase in temperature triggers changes in the electrical and mechanical properties of critical components. For instance, as a result of the thermal expansion and contraction of materials, oscillators' resonance frequencies may drift, amplifiers' gain may fluctuate, and power supplies' output voltages may exhibit transient deviations. These thermal-induced variations ripple through the RF signal path, introducing fluctuations in signal amplitude, phase, and frequency. Consequently, the transient alterations in RF emissions become particularly pronounced during warm-up, as the electronic components strive to stabilize and reach their steady-state operating conditions. These thermal effects underscore the necessity of understanding and mitigating the warm-up period's impact on RF fingerprinting.


Existing literature offers only partial solutions to hardware stabilization during the warm-up period, leaving key gaps unaddressed \cite{tyler2023considerations}.
Some may inadvertently overlook the warm-up period and train their systems on data collected within that period, thereby exposing the system to unstable behaviors, undermining its reliability and accuracy. Conversely, \cite{elmaghbub2023eps, tyler2023considerations} recommend delaying data collection until after the warm-up period to ensure stable and consistent measurements.
While substantially boosting cross-day classification accuracy on stable captures \cite{elmaghbub2023eps}, it proves ineffective when tested on warm-up captures. Not accounting for such initial stage variations can result in overfitting to stable behavior, creating substantial blind spots for frames transmitted before devices stabilize. 
An alternative approach, as proposed by \cite{shen2021radio}, attempts to compensate for hardware impairments that undergo changes during the warm-up period, aiming to discourage the system from relying on them during training. However, it's crucial to note that this approach comes at the cost of reducing the feature space, potentially omitting primary features that differentiate devices during stable phases. 

\subsection{\proposed: Overcoming the Impact of Warm-Up Time}
Despite the extensive research on improving RF fingerprinting \cite{jagannath2022comprehensive}, the effect of hardware stabilization has not received adequate attention \cite{tyler2023considerations}. This work aims then to fill this gap. It begins by unfolding the impact of hardware stabilization on RF fingerprinting performance and then proposes \proposed, a novel framework that delivers robust performance throughout the various stages of a device's life cycle, effectively managing the challenges faced during warm-up. 
 
Two crucial observations drive our design for \proposed: (1) hardware impairments exhibit stability post-stabilization, suggesting that training on stable captures can mitigate cross-day performance issues, and (2) training on stable captures falls short in recognizing devices during the warm-up phase, hinting at potential causes such as overfitting, focus on specific impairments, or unrelated feature learning. To overcome these challenges, \proposed~incorporates targeted impairment estimation and sequential transfer learning. Targeted impairment estimation enables the system to discern impairments within the entire radio signal stack, essentially creating a mapping between the signal and impairments. On the other hand, sequential transfer learning, which involves training models separately on different related tasks to transfer knowledge from the source task to enhance performance on the target task \cite{ruder2019neural}, leverages this knowledge to guide the learning of fingerprint features for the RF fingerprinting task. In doing so, the system is anchored in an immutable hardware identity, ensuring robust recognition across all phases. This dual-component approach captures and integrates hardware impairment-related information, forming the foundation for \proposed's ability to deliver consistent and reliable performance throughout a device's lifecycle. \proposed~leverages deep single-input-multi-output CNN models trained on time-domain I/Q representations of WiFi frames and eight hardware impairments, providing a strong basis for accurate impairment estimation. 
It then seamlessly integrates the pretrained impairment estimator with a device classification head, all trained on stable data. 

The evaluation demonstrates \proposed's warm-up resiliency, achieving remarkable classification accuracy of 96\% when testing within the initial 6 minutes of device operation, compared to 64\% and 40\% achieved by Residual Neural Networks (ResNet) and CNN models, respectively, for WiFi type B transmissions. Even in the most unstable intervals, \proposed~attains testing accuracies of 95\% and 90\% in 4-minute and 2-minute intervals. Cross-day assessment confirms \proposed's superiority, achieving a steady 87\% classification accuracy across warm-up intervals on the Day 2 dataset, while ResNet and CNN models reach only 20\% and 43\% accuracy in the initial 6-minute interval. For WiFi type N frames, \proposed~achieves 97\% during the initial 6-minute interval and 91\% during the initial 2-minute interval. 
\subsection{Main Contributions and Results}

  \textbf{(1)~Unfolding of the hardware warm-up impact on DL-based RF fingerprinting:}  
Our study reveals dynamic signal characteristics during warm-up, causing a notable decline in fingerprinting accuracy in cross-day evaluations. Although training and testing on stable data captured after warm-up yields high performance,  stable-phase-trained systems struggle when tested with data captured at early warm-up time.
    
\textbf{(2)~An in-depth analysis of RF fingerprint anatomy:} 
We study and analyze the behavior of 8 key hardware impairments through captured WiFi 802.11 frames, crucial for RF fingerprinting. Spanning 15 devices over their initial 30-minute operation period, our analysis provides valuable insights into DL-based RF fingerprinting's time sensitivity, laying a foundation for developing resilient methods to address warm-up challenges and cross-time adaptation.
    
\textbf{(3)~Proposal of \proposed, a novel RF fingerprinting framework, to address the hardware warm-up issue:}
\proposed~showcases a transformative approach in RF fingerprinting, achieving 96\% accuracy in recognizing devices within their initial 6 minutes of operation, significantly outperforming traditional CNN and ResNet models. Even amid the challenging early operational phases, \proposed~remains robust, with accuracies of 95\% and 90\% at the initial 4 and 2 minutes, respectively. Moreover, it consistently achieves an 87\% cross-day classification accuracy at different warm-up intervals. For WiFi type N frames in the wireless scenario, \proposed~maintains high accuracy, achieving 97\% and 91\% during the initial 6-minute and 2-minute intervals, respectively.

\textbf{(4)~Release, for the first time, of WiFi RF fingerprints datasets with the corresponding hardware impairments:} 
The datasets encompass both wired and wireless data of WiFi 802.11b and 802.11n collected over a month from 15 Pycom devices, capturing over 250,000 frames with time-domain I/Q samples and their corresponding hardware impairments. These datasets enable the assessment and understanding of device behavior in RF fingerprinting during warm-up and stable phases. They also hold potential for broader applications such as impairment estimation and compensation. 
The datasets can be downloaded from NetSTAR Lab website at {\color{blue} {\it https://research.engr.oregonstate.edu/hamdaoui/datasets}}. 
In addition, the code used in this paper is publicly available for researchers to use and can be downloaded at 
{\color{blue} {\it https://github.com/NetSTAR-Lab}}. 

For convenience, we summarize the acronyms in Table~\ref{t:notations}.


In this manuscript, we extend our previous work \cite{elmaghbub2024no} by adding several key components. First, we include a comprehensive related work section where we review literature on various aspects of domain adaptation in RF fingerprinting, such as adaptation to network environment and deployment changes, system configuration and setup changes, and hardware warm-up and stabilization (Sec.~\ref{sec:related-work}). Second, we consider a new list of impairments studied in the RF anatomy section to include I/Q gain imbalance, Quadrature Error, Pilot EVM, and I/Q Timing Skew (Sec.~\ref{sec:anatomy}). Third, we provide an in-depth analysis of the impairment estimator capabilities, featuring the t-SNE outputs for different scenarios (Sec.~\ref{sec:evaluation}). Fourth, we evaluate the impairment estimation and device classification efficacy of our proposed technique, \proposed, on both the wired scenarios of the WiFi type B dataset and the wired and wireless scenarios of the WiFi type N datasets (Sec.~\ref{sec:evaluation}.D). Fifth, we demonstrate the importance of the protocol-agnostic property of RF fingerprinting frameworks and  showcase the efficacy of \proposed~in cross-protocol scenarios (Sec.~\ref{sec:evaluation}.E). Finally, we release the wired and wireless WiFi type N datasets (Sec.~\ref{sec:evaluation}.A). These changes are reflected in the paper's structure as follows. We begin with the related work in Sec.~\ref{sec:related-work}. This is followed by an examination of the hardware warm-up effects on DL-based RF fingerprinting in Sec.~\ref{sec:exposing}. An in-depth anatomical analysis of RF fingerprints is presented in Sec.~\ref{sec:anatomy}. Our proposed framework, \proposed, is detailed in Sec.~\ref{sec:proposed}. Sec.~\ref{sec:evaluation} provides a description of our testbed and datasets, along with a comprehensive evaluation. The paper concludes with final remarks in Sec.~\ref{sec:conc}.


 \begin{table}\normalsize
\vspace{-7pt}
\caption{Acronyms}
\centering  
\resizebox{0.4\textwidth}{!}{
\label{t:notations}
\begin{tabular}{l l}
\hline
\noalign{\medskip }
\proposed & The proposed approach \\
CNN & Convolution Neural Network \\
DL & Deep Learning\\
RF & Radio Frequency \\
RFFI & Radio Frequency Fingerprint Identification \\
EPS & Envelope Power Spectrum \\
IoT & Internet of Things \\
SDR & Software Defined Radio \\
USRP & Universal Software Radio Peripheral \\
OCXO & Oven-Controlled Crystal Oscillator \\
HR/DSSS & High-Rate Direct-Sequence Spread Spectrum \\
I/Q & In-phase/Quadrature \\
PA & Power Amplifier \\
CFO & Carrier Frequency Offset \\
SCE & Symbol Clock Error \\
EVM & Error Vector Magnitude \\
OFDM & Orthogonal Frequency Division Multiplexing \\
CSE & Carrier Suppression Error \\
SCPI & Standard Commands for Programmable Instruments \\
MSE & Mean Square Error \\
MAE & Mean Absolute Error \\
t-SNE & t-Distribution Stochastic Neighbor Embedding \\
\noalign{\smallskip} \hline \noalign{\smallskip}
\end{tabular}
}
\end{table}

%% file: related.tex



\begin{table*}
  \centering
  \large
  \caption{Summary of RF Fingerprinting Approaches for Environment and Deployment Adaptation}
  \label{tab:env_deploy_adapt}
  \resizebox{\textwidth}{!}{
  \begin{tabular}{|l|l|l|l|l|}
    \hline
    \textbf{Reference} & \textbf{Dataset Type} & \textbf{Adaptation Strategy} & \textbf{Key Features} & \textbf{Limitations} \\
    \hline
    Brown et al.~\cite{brown2105charrnets} & Synthetic \& real WiFi & Complex-valued CNNs with group-theoretic frameworks & Handles multipath propagation channels & Average cross-day accuracy of 55\% on real datasets \\
    \hline
    Peng et al.~\cite{peng2024channel} & Cellular LTE Uplink & Leveraging subcarrier correlation & Channel-agnostic, distinct fingerprints & Performance varies with bandwidth configuration \\
    \hline
    Yang et al.~\cite{yang2023use} & Various Environments & PA nonlinearity quotient with transfer learning & High detection/classification accuracy & Requires alternating high and low power levels \\
    \hline
    Elmaghbub et al.~\cite{elmaghbub2023eps} & 15 802.11b Devices & Double-Sided Envelope Power Spectrum (EPS) & Over 93\% cross-day/loc accuracy & Performance drops during early warm-up phase \\
    \hline
    Li et al.~\cite{li2022radionet} & 10 HackRF Devices & Adversarial domain adaptation with k-NN classifier & 64\% accuracy over two days & Challenges in scaling; reliance on target domain data \\
    \hline
    Kong et al.~\cite{kong2024towards} & WiFi Signals & DeepCRF framework extracting micro-CSI & Robust to channel variability & Difficulty in extracting micro-CSI from noisy data \\
    \hline
    Elmaghbub et al.~\cite{adl} & 50 WiFi Devices & Disentanglement representation with adversarial learning & Short-term success in adaptation & Accuracy declines to 15\% over several days \\
    \hline
    Al-Shawabka et al.~\cite{al2023signcrf} & 5 \& 20 WiFi Devices & Cycle-consistent GAN for environment translation & 83\% accuracy with 5 devices & Accuracy drops to 34\% with 20 devices \\
    \hline
    Jiang et al.~\cite{jiang2024channel} & Satellite Signals & Channel-resilient fingerprinting with data augmentation & Incorporates channel prediction module & High latency in satellite links; specific to satellite \\
    \hline
  \end{tabular}
  }
\end{table*}

Our literature review focuses on RF fingerprinting approaches that address domain adaptation problems. We categorize them based on their adaptation to (1) network environment and deployment changes, (2) system configuration and setup changes, and (3) hardware warm-up and stabilization.



\subsection{Adaptation to Environment \& Deployment Changes}
 
Brown et al. \cite{brown2105charrnets} employ complex-valued CNNs with group-theoretic and manifold-based methods to address multipath propagation, validated on synthetic and real WiFi datasets. While performance improves, the average cross-day accuracy on real data remains 55\%, limiting practical security use.
Peng et al. \cite{peng2024channel} leverage channel correlation across adjacent subcarriers to extract channel-agnostic RF fingerprints from uplink LTE signals, achieving high classification accuracy. They also find that LTE devices yield more distinct fingerprints when using more resource blocks.
Yang et al. \cite{yang2023use} use PA nonlinearity and transfer learning to improve RF fingerprinting robustness, achieving notable gains in detection and classification. However, the method requires alternating high/low-power packets for computing the nonlinearity quotient and depends on removing preambles affected by fast-moving objects, which limits real-world applicability.
Elmaghbub et al. \cite{elmaghbub2023eps} propose the Double-Sided Envelope Power Spectrum (EPS) to improve domain adaptation in RF fingerprinting, achieving over 93\% cross-day and 95\% cross-location accuracy on 15 WiFi devices. However, they found that the performance drops when packets are transmitted during the early warm-up phase.
Li et al. \cite{li2022radionet} use adversarial domain adaptation with a k-NN classifier, achieving 64\% accuracy on 10 HackRF WiFi devices over two days, but face challenges in scaling and reliance on target domain data. 
Kong et al. \cite{kong2024towards} introduce DeepCRF, which extracts micro-CSI from WiFi CSI curves as RF fingerprints, addressing noise and variability with a robust framework.
Elmaghbub et al. \cite{adl} combine disentanglement representation and adversarial learning on 50 WiFi devices, showing short-term success but declining to 15\% accuracy over several days, indicating issues with long-term effectiveness. Al-Shawabka et al. \cite{al2023signcrf} employ a cycle-consistent generative adversarial network for environment translation, achieving 83\% accuracy with 5 WiFi devices but dropping to 34\% with 20 devices, highlighting scalability challenges.
Jiang et al. \cite{jiang2024channel} propose a channel-resilient, low-power fingerprinting method for satellite signals that leverages data augmentation. To address the high latency inherent in satellite links, they integrate a channel prediction module within the augmentation process, enabling the model to effectively capture and adapt to current channel conditions.

Table~\ref{tab:env_deploy_adapt} provides a summary that categorizes the reviewed works based on their key distinguishing characteristics.

\subsection{Adaptation to System Configuration \& Setup Changes}

\begin{table*}[!ht]
  \centering
  \large
  \caption{Summary of Receiver-Agnostic and Cross-Domain RF Fingerprinting Approaches}
  \label{tab:receiver_agnostic_rff}
    \resizebox{\textwidth}{!}{
  \begin{tabular}{|l|l|l|l|l|}
    \hline
    \textbf{Reference} & \textbf{Dataset Type} & \textbf{Adaptation Strategy} & \textbf{Key Features} & \textbf{Limitations} \\
    \hline
    Chen et al.~\cite{chen2024cross} & 10 nRF24 Txers/3 Rxers & Unsupervised pre-training with contrastive learning & Over 90\% accuracy across devices & Limited to specific transceivers \\
    \hline
    Cai et al.~\cite{cai2024toward} & Spectrograms & Semantic features using discrete wavelet transforms & Enhanced cross-domain accuracy & Requires decomposition into sub-bands \\
    \hline
    Yang et al.~\cite{yang2024mitigating} & WiSig Dataset & Domain alignment with adaptive pseudo-labeling & Significant improvements in RFFI accuracy & Scalability and generalization issues \\
    \hline
    Shen et al.~\cite{shen2023towards} & LoRaWAN (20 SDRs) & Adversarial training with collaborative inference & Up to 40\% accuracy improvement & Scalability issue in IoT networks \\
    \hline
    Zhang et al.~\cite{zhang2024domain} & Various Receivers & Federated learning with Txer/Rxer disentanglement & Preserves privacy, supports Rxer-independence & Complexity in federated learning setup \\
    \hline
    Gaskin et al.~\cite{gaskin2023deep} & LoRa \& USRP Testbed & Metric learning with calibration (Tweak) & Receiver-independent feature space & Requires calibration to generalize \\
    \hline
    Zhang et al.~\cite{zhang2022data} & ADS-B \& FIT/CorteXlab & 3S method with ResNeXt & 96\% closed-set accuracy; 6\% gain from 3S & Limited applicability to other protocols \\
    \hline
    Agadakos et al.~\cite{agadakos2019deep} & WiFi \& ADS-B & Deep Complex-valued Neural Networks & High accuracy across protocols & Scalability remains a challenge \\
    \hline
    Wu et al.~\cite{wu2024receiver} & Various Transmitters & Demodulation, reconstruction, adversarial training & 38.4\% accuracy gain at 0 dB SNR & Needs noise-level robustness validation \\
    \hline
    Zhang et al.~\cite{zhang2022variable} & Simulation Datasets & Modulation SEI with DANN + Gaussian Encoder & Tackles modulation adaptation problem & Real-world scalability remains uncertain \\
    \hline
  \end{tabular}}
\end{table*}

\begin{table*}
  \centering
  \large
  \caption{Summary of RF Fingerprinting Approaches Considering Hardware Warm-up and Temperature Effects}
  \label{tab:warmup_temperature_rff}
      \resizebox{\textwidth}{!}{
  \begin{tabular}{|l|l|l|l|l|}
    \hline
    \textbf{Reference} & \textbf{Dataset Type} & \textbf{Adaptation Strategy} & \textbf{Key Features} & \textbf{Limitations} \\
    \hline
    Elmaghbub et al.~\cite{elmaghbub2023eps} & 15 802.11b Devices & Double-Sided Envelope Power Spectrum & Robustness across time, channel, and location & Struggles with inference packets during early warm-up phase \\
    \hline
    Shen et al.~\cite{shen2021radio} & LoRa Devices & Two-step CFO compensation process & Improves accuracy from 59 to 83\% on raw IQ data & May reduce distinguishing power; risk of artificial impairments \\
    \hline
    Zhang et al.~\cite{zhang2021radio} & Simulations & CFO compensation post stability phase & High classification accuracy in simulations & Lacks real measurements; CFO-only compensation insufficient \\
    \hline
    Xiaolin et al.~\cite{gu2022terff} & Smartphones & TeRFF: Temperature-aware RF fingerprints & Mitigates CFO instability caused by temp. fluctuations & Requires extensive data collection and accurate temperature data \\
    \hline
    Yilmaz and Yazici~\cite{yilmaz2022effect} & Various Devices & Blended training across temperatures & Improves performance at edge temperatures & Low accuracy at extremes; impractical to cover all thermal states \\
    \hline
    Peggs et al.~\cite{peggs2023preamble} & 4 Devices & Training with temperature-extreme data & Enhances accuracy across -40°C to 80°C & Complex training; limited generalizability \\
    \hline
  \end{tabular}}
\end{table*}

Chen et al. \cite{chen2024cross} address the cross-receiver RF fingerprinting problem using unsupervised pre-training with contrastive learning and subdomain adaptation. They validate their approach with signals from 10 nRF24 transceivers and three types of receivers, achieving a 7\%-15\% increase in classification accuracy and maintaining over 90\% accuracy across devices. 
Cai et al. \cite{cai2024toward} improve cross-domain RF fingerprinting by linearly interpolating semantic features, using wavelet-based spectrogram decomposition to extract key features.
Yang et al. \cite{yang2024mitigating} propose a domain adaptation method combining domain alignment and adaptive pseudo-labeling to handle cross-receiver variability. Using the WiSig dataset, they achieve significant improvements in RFFI accuracy but note challenges in scalability and generalization to different environments. 
Shen et al. \cite{shen2023towards} design a receiver-agnostic RFFI system with adversarial training and collaborative inference, showing up to 40\% accuracy gain in a 10-device, 20-receiver LoRaWAN study, though scalability to diverse IoT networks remains uncertain.
%
%
Zhang et al. \cite{zhang2024domain} propose a receiver-independent emitter identification model that separates emitter-specific from receiver-specific features, and extend it with FedRIEI, a federated approach enabling distributed training without sharing raw data.
Gaskin et al. \cite{gaskin2023deep} present Tweak, combining metric learning and calibration to create a receiver-independent feature space. Validated on a LoRa and USRP testbed, their technique faces limitations in the need for calibration data and generalization to real-world environments. 
Zhang et al. \cite{zhang2022data} propose a data-enhancement-aided, protocol-agnostic transmitter recognition system using the 3S method and a ResNeXt-aided network. Validated on ADS-B signals and the FIT/CorteXlab dataset, they achieve 96\% accuracy in closed-set recognition and over 6\% improvement with the 3S method alone. 
Agadakos et al. \cite{agadakos2019deep} introduce a protocol-agnostic RFFI system using Deep Complex-valued Neural Networks. Their approach achieves high accuracy on WiFi and ADS-B datasets, demonstrating capability across multiple protocols, but faces real-world generalization and scalability challenges. 
Wu et al. \cite{wu2024receiver} propose a receiver-agnostic RF fingerprinting method that enhances transmitter feature extraction under noise through demodulation, reconstruction, and adversarial training. Their results show a 38.4\% accuracy improvement over baseline methods at 0 dB SNR.
Zhang et al. \cite{zhang2022variable} tackle modulation adaptation using Domain Adversarial Networks and a Gaussian Encoder, though real-world scalability remains unverified.

Table~\ref{tab:receiver_agnostic_rff} summaries the reviewed works based on their key distinguishing characteristics.

\subsection{Adaptation to Hardware Warm-up \& Stabilization}
As noted earlier, the effects of hardware warm-up and stabilization on RF fingerprinting have been largely overlooked, despite their critical role in ensuring the accuracy and reliability of RF fingerprinting systems.
For instance, in \cite{elmaghbub2023eps}, Elmaghbub et al. investigated the instability of hardware during the warm-up period and demonstrated its significant impact on the accuracy of deep learning models for device classification. 
They suggested allowing devices time to reach a stable state before data capture. While their EPS-based framework showed robustness across time, channel, and location, it struggled with inference packets from the first minutes after powering on the device. Moreover, restricting devices to transmit only during their stable phase is impractical for many IoT-related applications.  
Shen et al. \cite{shen2021radio} propose a two-step CFO compensation process that improves LoRa device classification accuracy from 59\% to 83\% using raw I/Q samples. Despite these gains, compensating for CFO can result in losing its distinguishing power, and inaccurate compensation may introduce artificial impairments. 
Zhang et al. \cite{zhang2021radio} recommend compensating for CFO variation by relying on IQ amplitude, phase imbalances, and PA nonlinearity for stable impairments. While achieving high accuracy in simulations, their technique was not tested through real measurements.
%
Xiaolin et al. \cite{gu2022terff} introduce TeRFF, a temperature-aware RF fingerprinting approach that mitigates CFO instability in smartphones caused by temperature fluctuations. Despite achieving 79\% accuracy at registered temperatures, their method requires extensive data collection, accurate temperature data, and introduces complexity and potential security risks. 
Yilmaz and Yazici \cite{yilmaz2022effect} examine ambient temperature effects on RF fingerprinting, noting accuracy drops when training and test data differ; blended training helps but remains ineffective at extreme temperatures. 
Peggs et al. \cite{peggs2023preamble} investigate RF fingerprinting across temperatures from -40°C to 80°C, and propose training classifiers with data from temperature extremes to improve accuracy. However, this adds complexity and limits generalizability, being tested on only four devices. 
Table~\ref{tab:warmup_temperature_rff} summaries the reviewed works based on their key features.

This paper fills this gap by looking at the bigger picture and utilizing the behavior of all key hardware impairments. Before presenting our comprehensive solution, we examine how these impairments behave during the warm-up time, laying the groundwork for more robust RF fingerprinting.

%% file: 3-exposing.tex
In this section, we run experiments to demonstrate how transient behaviors during warm-up periods can lead to significant performance drops and explore whether stabilizing the dataset post-warm-up addresses these challenges comprehensively.

\subsection{Testbed Description and Dataset Collection}
\label{sec:setup}
\input{6-setup}


\subsection{Device Fingerprinting Evaluation}
\label{sec:results}
\input{7-results}

%% file: 6-setup.tex

 \begin{figure}
 \setlength{\abovecaptionskip}{3pt}
     \subfloat[15 Pycom Transmitters.\label{subfig-1:tx}]{%
       \includegraphics[width=0.23\textwidth, height = 0.18\textwidth]{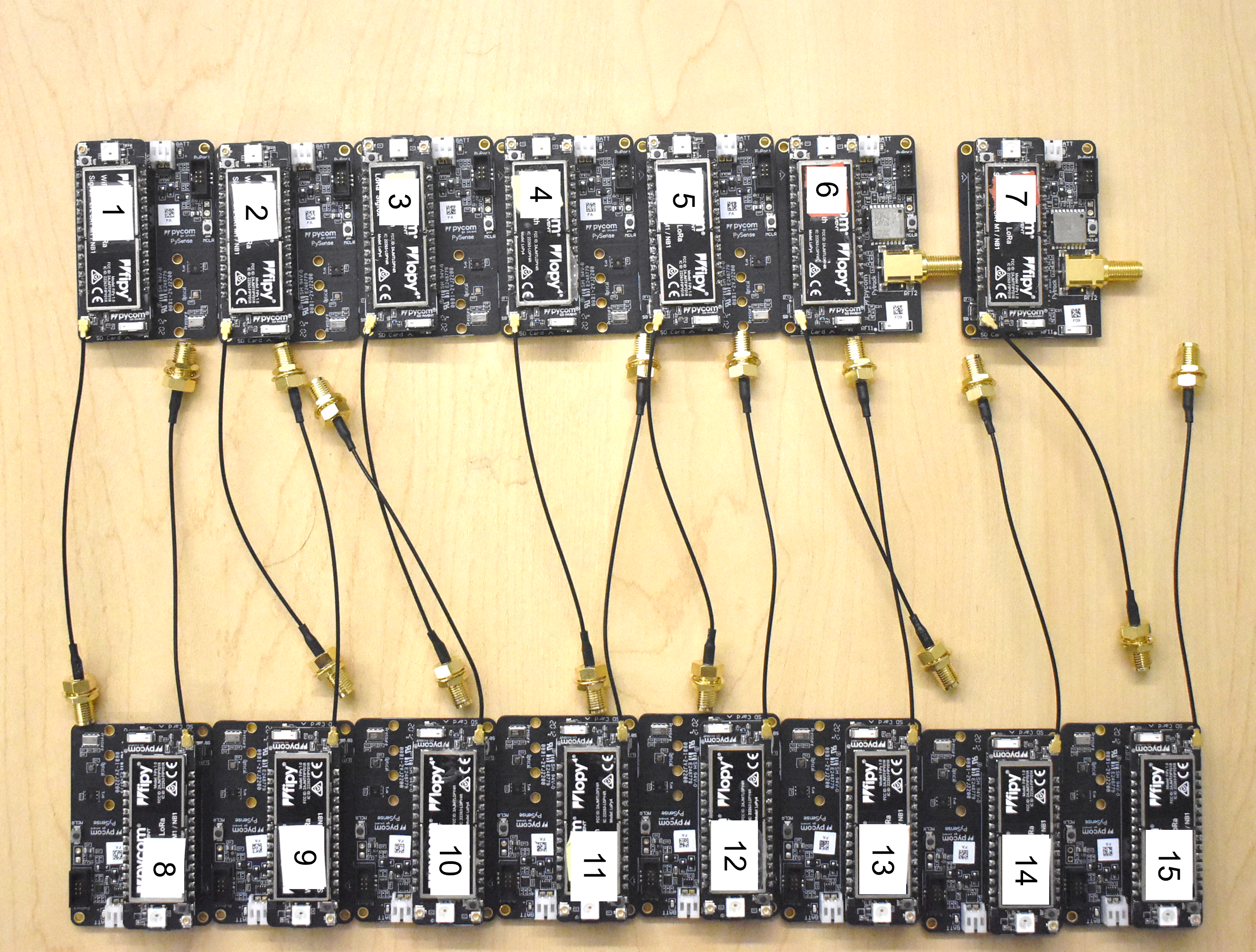}
     }
    \hspace{0.00001cm}
     \subfloat[Wired-WiFi and Wireless-WiFi.\label{subfig-2:recv}]{%
       \includegraphics[width=0.23\textwidth, height = 0.18\textwidth]{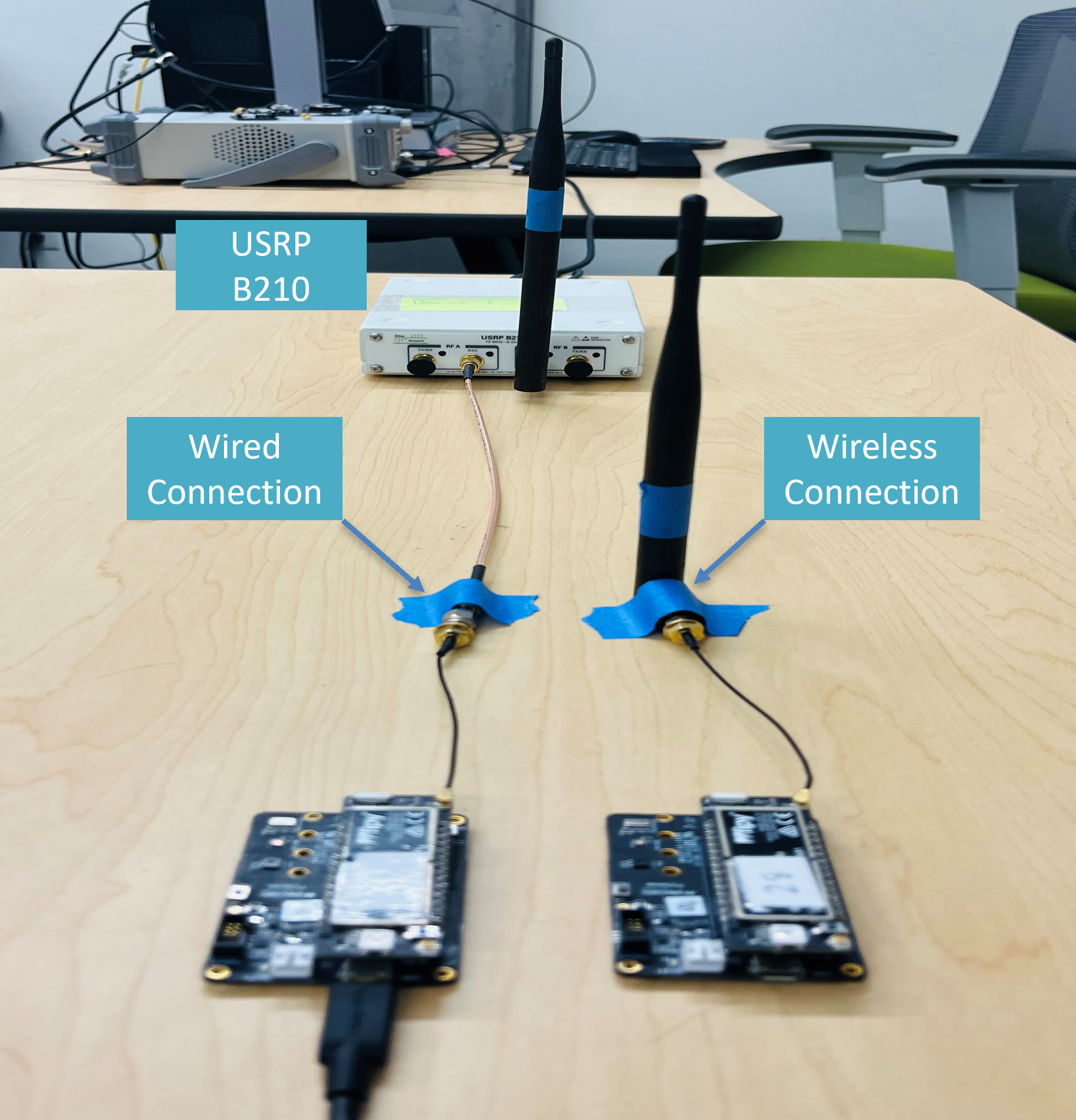}
     }
     \caption{IoT Testbed consisting of 15 Pycom transmitting devices and a USRP B210 receiving device}
     \label{fig:testbed}
     \vspace{-10pt}
\end{figure}

\subsubsection{Testbed and Experimental Setup} \label{testbed_setup_section}
Our experimental testbed, depicted in Fig.~\ref{fig:testbed}, consists of $15$ commercial off-the-shelf devices, comprising $10$ FiPy and $5$ LoPy IoT modules. Each device integrates embedded RF front-ends capable of supporting various wireless protocols, making them ideal for studying physical-layer hardware fingerprinting. Data acquisition was performed using an Ettus USRP B210 receiver, which was externally synchronized to a high-stability oven-controlled crystal oscillator (OCXO) to ensure accurate sampling and minimize drift during prolonged capture sessions. 


The experiments were conducted within a controlled indoor laboratory environment to ensure a stable testing setup with minimal external wireless interference and limited uncontrolled multipath effects. Normal WiFi traffic from nearby access points was present during the experiments; however, the environment remained static with no significant physical movement or dynamic interference throughout the data collection periods. To ensure fairness and minimize temporal biases, devices were captured in a round-robin fashion, where each Pycom transmitter was activated approximately two minutes after the previous device completed its data capture session.

Two experimental configurations were employed to capture the datasets. In the \textbf{wireless setup}, each Pycom transmitter was placed precisely $1$ meter away from the USRP B210 receiver, with both the transmitters and the receiver equipped with VERT$900$ omnidirectional antennas. Each device continuously transmitted packets over a $10$-minute period per session. This wireless experiment was repeated across three consecutive days under consistent environmental conditions, resulting in the capture of over $5000$ WiFi frames per device per day. To complement this setup and isolate the impact of the wireless channel, a \textbf{wired setup} was also employed, where each Pycom transmitter was directly connected to the USRP receiver using high-quality SMA coaxial cables. This configuration removed the influence of multipath, fading, and path loss, allowing a focused analysis of hardware-originated signal distortions. The wired data collection followed the same schedule as the wireless collection, spanning three consecutive days and yielding over $3000$ WiFi frames per device per day. Together, these two complementary setups provided a comprehensive dataset suitable for analyzing the effects of hardware impairments under both controlled and realistic transmission environments.

\subsubsection{I/Q Data Collection}

This section outlines the configuration and process used for data collection to evaluate the impact of the warm-up phase on RF fingerprinting. Additional details related to the data collected for evaluating our proposed framework are provided in Section~\ref{subsec:wifi-collection}.
For this experiment, we begin by powering on the Pycom devices one at a time via USB connections from a laptop, with each being configured to transmit identical 802.11b WiFi frames with a duration of 559us back-to-back, separated by a small gap. All transmissions took place over WiFi Channel 1, utilizing the High-Rate Direct-Sequence Spread Spectrum (HR/DSSS) physical-layer mode, and conducted at a fixed data rate of $1$ Mbps, operating at a carrier frequency of $2.412$ GHz with a $20$ MHz bandwidth.
We captured the first two minutes of transmissions using the USRP B210 at a sample rate of 45MSps. The captured signals were then digitally down-converted to the baseband and stored as I/Q samples on our computer. We refer to these initial captures as ``warm-up captures'' since the circuit components on the Pycom devices had not yet stabilized during this time. After allowing the devices sufficient time to stabilize, we continued with the data collection process. Specifically, we waited an additional 10 minutes before initiating the capture process again. This ensured that the crystal oscillators and other RF components had stabilized and were operating consistently. We refer to these subsequent captures as ``stable captures''. To ensure that the device identification relies exclusively on physical-layer hardware impairments rather than packet-level identifiers, all devices were configured to broadcast identical WiFi packets. Specifically, the transmitted frames included a spoofed MAC address that was hardcoded to be the same across all devices, along with a zero-byte payload. This design prevents the DL model from exploiting metadata, such as MAC addresses, for device recognition, and forces it instead to learn true physical-layer features embedded in the I/Q signals. Furthermore, by hardcoding the MAC address rather than removing it during preprocessing, we emulate a realistic adversarial scenario where malicious devices may spoof MAC addresses to evade detection, thus enhancing the robustness and generalizability of the fingerprinting framework.

Finally, we extracted the WiFi packets from the raw I/Q sample files and stored them in HDF5-formatted files, maintaining the order of arrival to preserve the temporal structure of the captured signals.
Finally, we extracted the WiFi packets from the raw I/Q sample files and stored them in HDF5-formatted files in the same order they were received. This method allowed us to maintain the integrity of the captured signals and ensured that they were accurately represented in the final dataset. Detailed information and the link to the datasets can be found in~\cite{elmaghbub2024no}.


%% file: 7-results.tex
We design and carry out the following three experiments: 
\newline\textbf{(1)~Experiment 1: Train and Test on Warm-up Captures.} We train the system using frames captured within the first 2 minutes following device activation, corresponding to the early phase of the hardware warm-up period. Based on our experimental observations, the full warm-up duration for our devices extends to approximately 12 minutes, during which key hardware impairments, such as Carrier Frequency Offset (CFO) and Symbol Clock Error (SCE), gradually stabilize. After training on frames from the initial 2 minutes, we test the system using frames captured at various subsequent minutes within the warm-up period. Additionally, we perform cross-day testing using frames collected during the same early interval but across three different days. This experimental setup allows us to uncover the impact of training on transitional hardware states, both in short-term and cross-day scenarios.
\newline\textbf{(2)~Experiment 2: Train and Test on Stable captures.} 
For this experiment, we train the system on stable data (i.e., collected after hardware stabilization, which is about 12 minutes from device activation), and then test it using the stable frames captured across three different days in both wired and wireless settings. By replicating the previous experiment with the only difference being that we wait until the devices stabilize, it allows us to confirm whether or not the varying signal characteristics during warm-up are the main contributor to the performance degradation in the cross-day scenario. Additionally, the inclusion of both wired and wireless setups allowed us to quantify the impact of the wireless channel. 
\newline\textbf{(3)~Experiment 3: Train on Stable Captures and Test on Warm-up Captures.} 
In this experiment, we tested the system trained on stable frames using packets captured during warm-up, specifically within the first two minutes after device activation. This enabled us to assess the system's ability to generalize effectively with devices operating in the warm-up phase and identify potential blind spots in its performance.

For all these evaluations, we used the CNN classification model employed in~\cite{elmaghbub2021}, adopting a 5-fold cross-validation method. For each experiment, we used 3200 packets per device for training and 800 packets for testing. To represent each packet, we employ a 2x8192 tensor, encompassing 8192 samples of each of the I and Q data. This window size is selected experimentally based on performance superiority.

\subsubsection{\bf{Exp. 1 Results: The Impact of the Warm-up Phase}}
\begin{figure}[t!]
\setlength{\abovecaptionskip}{3pt}
\subfloat[Short-term Evaluation]{
   \includegraphics[width=.47\columnwidth, height=.4\columnwidth]{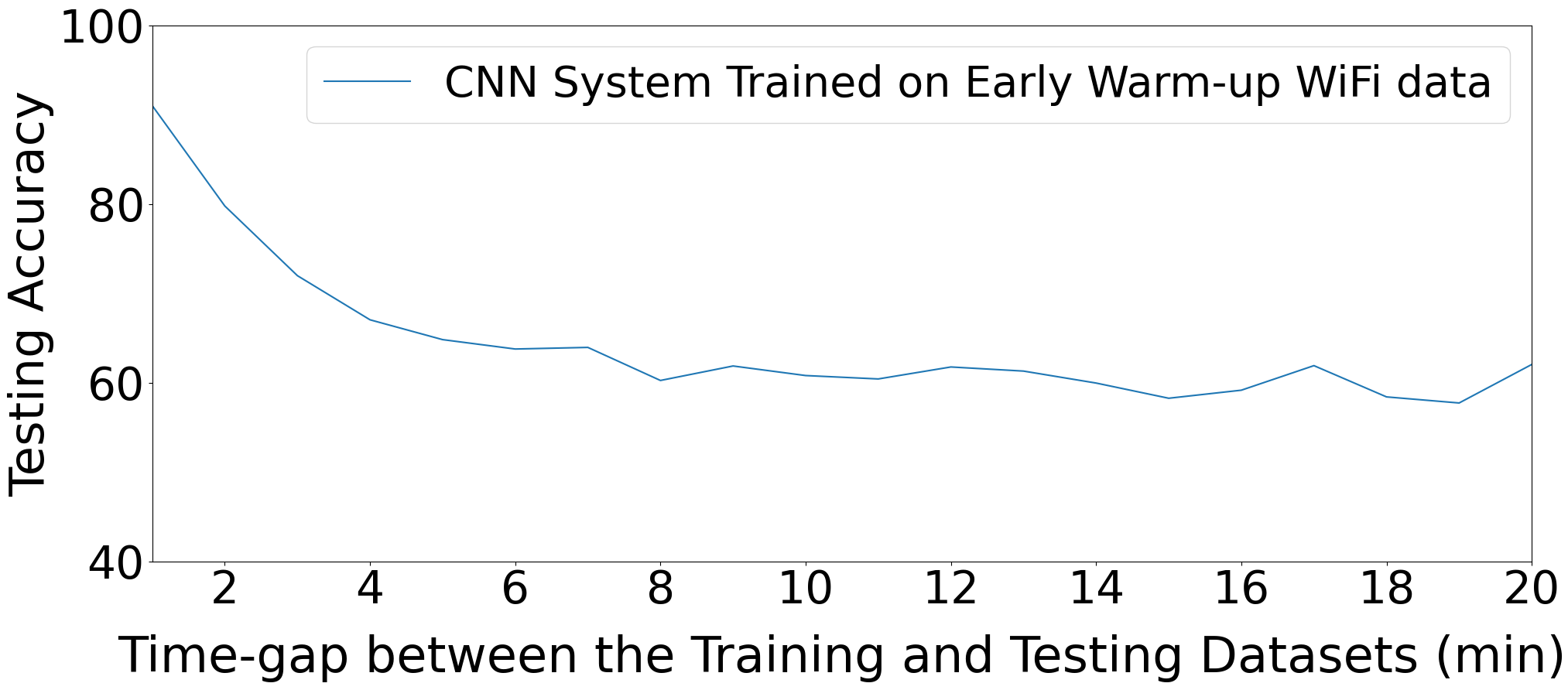}
   \label{TT_graph}}
   \hfill
\subfloat[Cross-Day Evaluation]{
   \includegraphics[width=.47\columnwidth, height=.4\columnwidth]{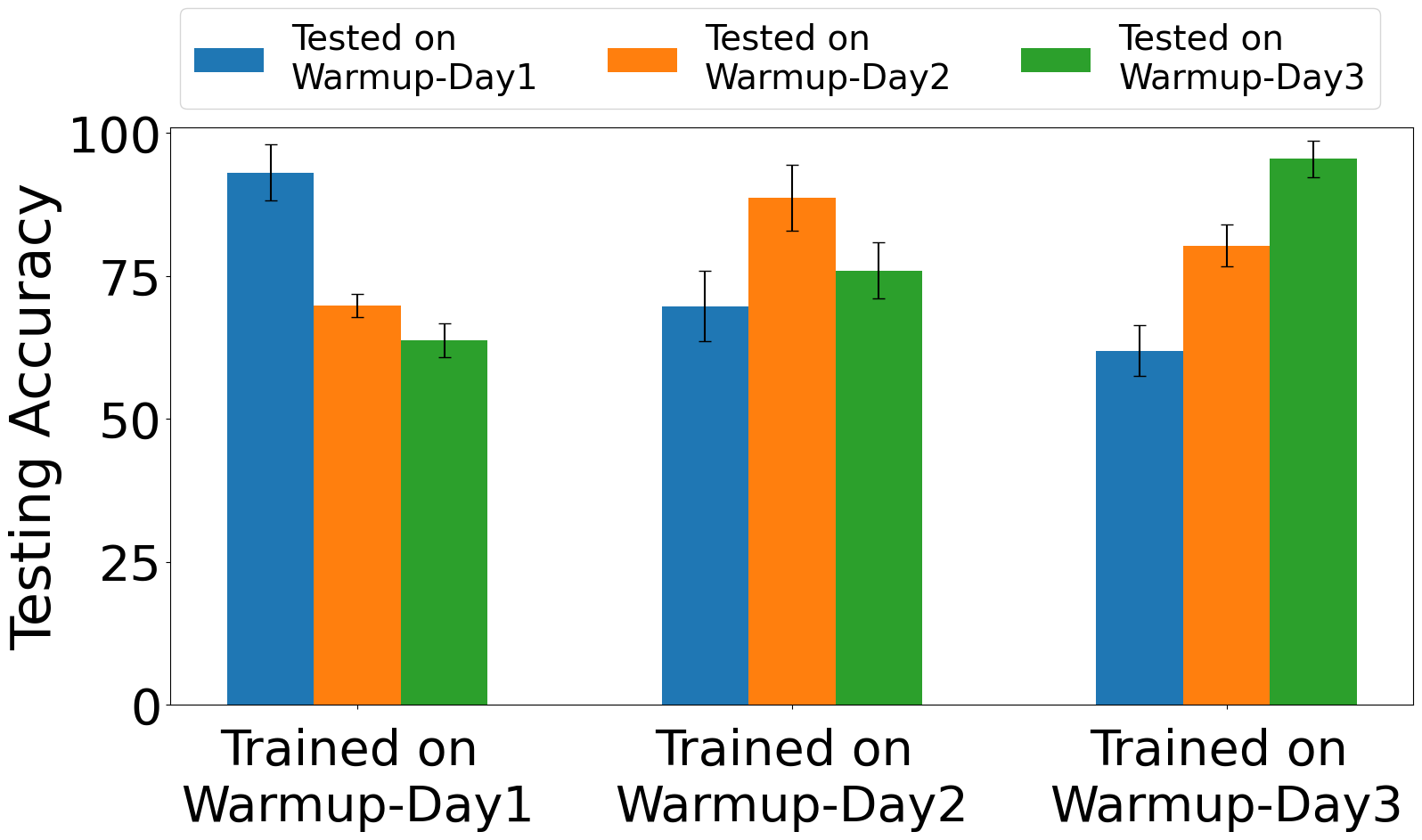}
   \label{warm_TTW}}\\

   \caption{Exp. 1: training \& testing on Warm-up captures}
\label{exp1}
\vspace{-0.2in}
\end{figure}

%
The results from Experiment 1, as depicted in Fig.~\ref{TT_graph}, provide a compelling visualization of the system's declining accuracy when trained on early warm-up data and tested across successive 20-minute intervals. This marked decrease, particularly pronounced in the initial stages post-training, underscores the dynamic nature of signal characteristics during the warm-up phase, a phenomenon further confirmed by observations in Fig.~\ref{fig:warm_up_interval}. The diminishing rate of accuracy decline, stabilizing towards the end of the warm-up phase, suggests a gradual approach to signal consistency. 
Moreover, cross-day testing reveals a substantial average accuracy drop of 30\% when the model is tested with warm-up capture (the first two minutes) on other days, as shown in Fig.~\ref{warm_TTW}, reinforcing the hypothesis that the transient signal behavior during the warm-up phase significantly contributes to cross-day performance degradation. 
This analysis underlines the critical need for careful consideration when using warm-up period data for training, to mitigate the inherent instability and ensure more reliable RF fingerprinting performance.

\vspace{-0.1pt}



\subsubsection{\bf{Exp. 2 Results:} Fingerprinting After Device Stabilization}
\begin{figure}[t!]
\setlength{\abovecaptionskip}{3pt}
  
\subfloat[Wired Setup]{
   \includegraphics[width=.47\columnwidth, height=.4\columnwidth]{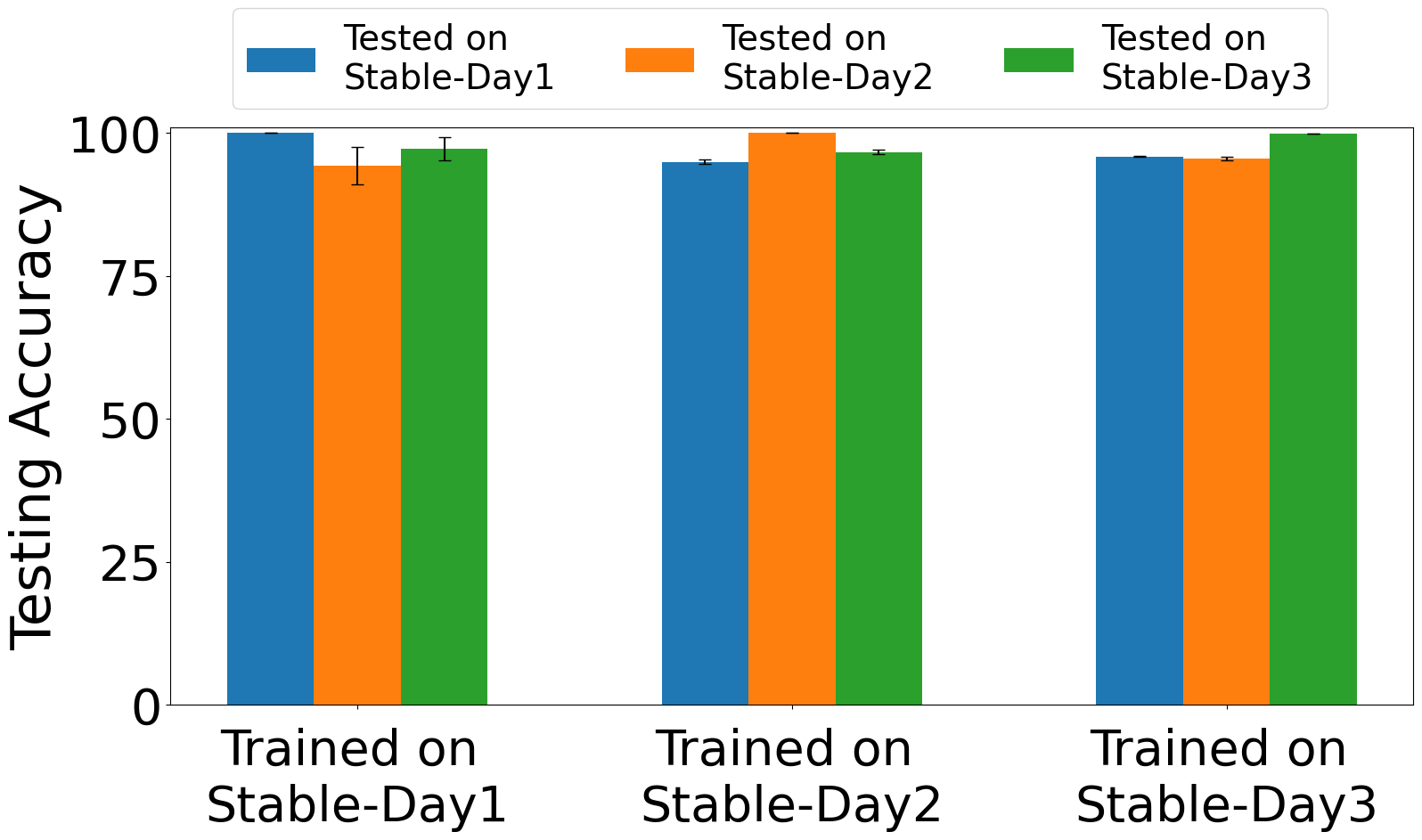}
   \label{TTS_wired}}
    \hfill
   \subfloat[Wireless Setup]{
   \includegraphics[width=.47\columnwidth,height=.4\columnwidth]{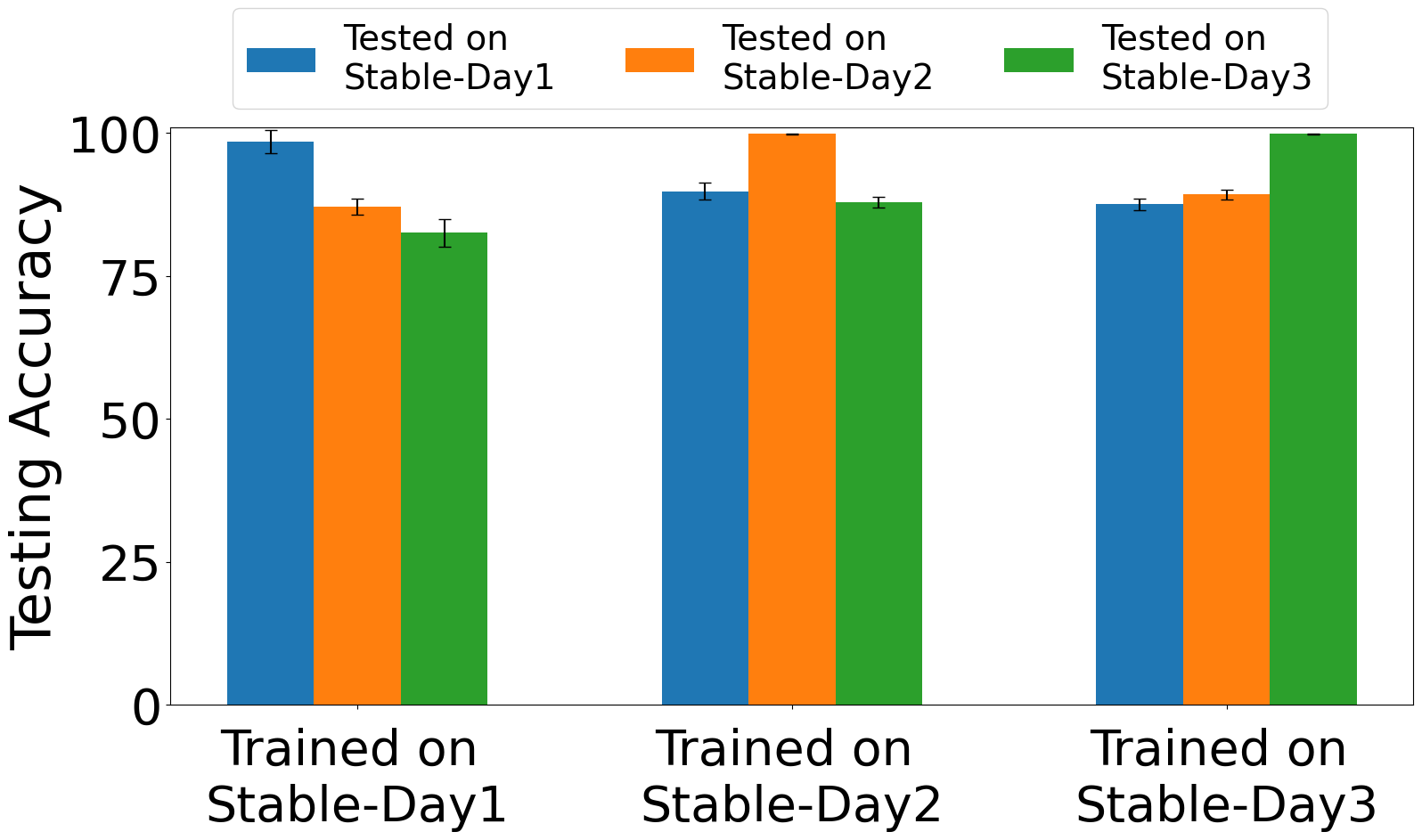}
   \label{TTS_wireless}}
    \\
   \caption{Exp. 2: training \& testing on Stable captures}
\label{exp2}
\vspace{-10pt}
\end{figure}

Now, as shown in Figs. \ref{TTS_wired} (wired) and \ref{TTS_wireless} (wireless), when both training and testing are done using Stable captures (data collected after warm up period), the testing accuracy does not suffer from the degradation observed in the previous experiment when using warm up data. 
Note that even the cross-day testing accuracy remains impressively high, with an average drop of only 4.4\% when compared to same-day testing in the wired scenario. This result indicates that the model can consistently classify devices in the wired environment, even when facing temporal variations across days. The minimal performance drop in the wired setting reinforces the hypothesis that the main driver of cross-day sensitivity observed in Experiment 1 was indeed the transient behavior of signals during the warm-up period.
In the wireless setting, depicted in Fig. \ref{TTS_wireless}, the model continues to exhibit strong performance when trained and tested on stable captures, showcasing an average cross-day testing accuracy of approximately 87.5\%. Although this accuracy is slightly lower than that achieved in the wired setting, it remains impressive, considering the additional challenges posed by the wireless channel. The difference in performance between the wired and wireless settings, with a slightly higher performance drop in the wireless environment (12\%), suggests that the wireless channel's impact contributes to the observed variance in cross-day accuracy to some extent. However, it is important to note that the impact of hardware stabilization on performance is still profound, as evidenced by the significant disparity in performance when training on warm-up data versus stable data.

\subsubsection{\bf{Exp. 3 Results: Confirming the Impact of Warm-up}}
\begin{figure}[t!]
\setlength{\abovecaptionskip}{3pt}

   \subfloat[Wired Setup]{
   \includegraphics[width=.47\columnwidth, height=.4\columnwidth]{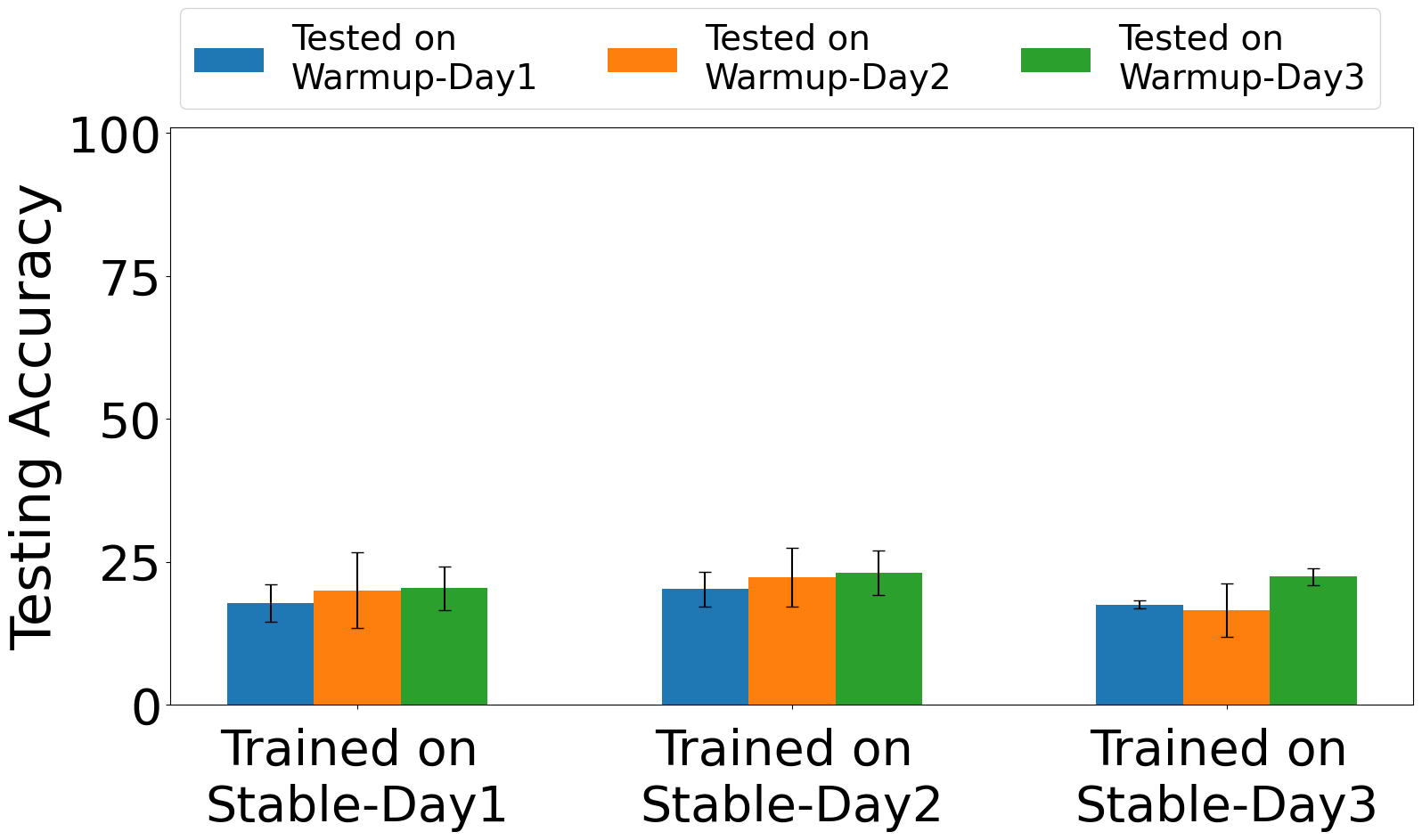}
   \label{TSTW_wired}}
   \hfill
   \subfloat[Wireless Setup]{
   \includegraphics[width=.47\columnwidth, height=.4\columnwidth]{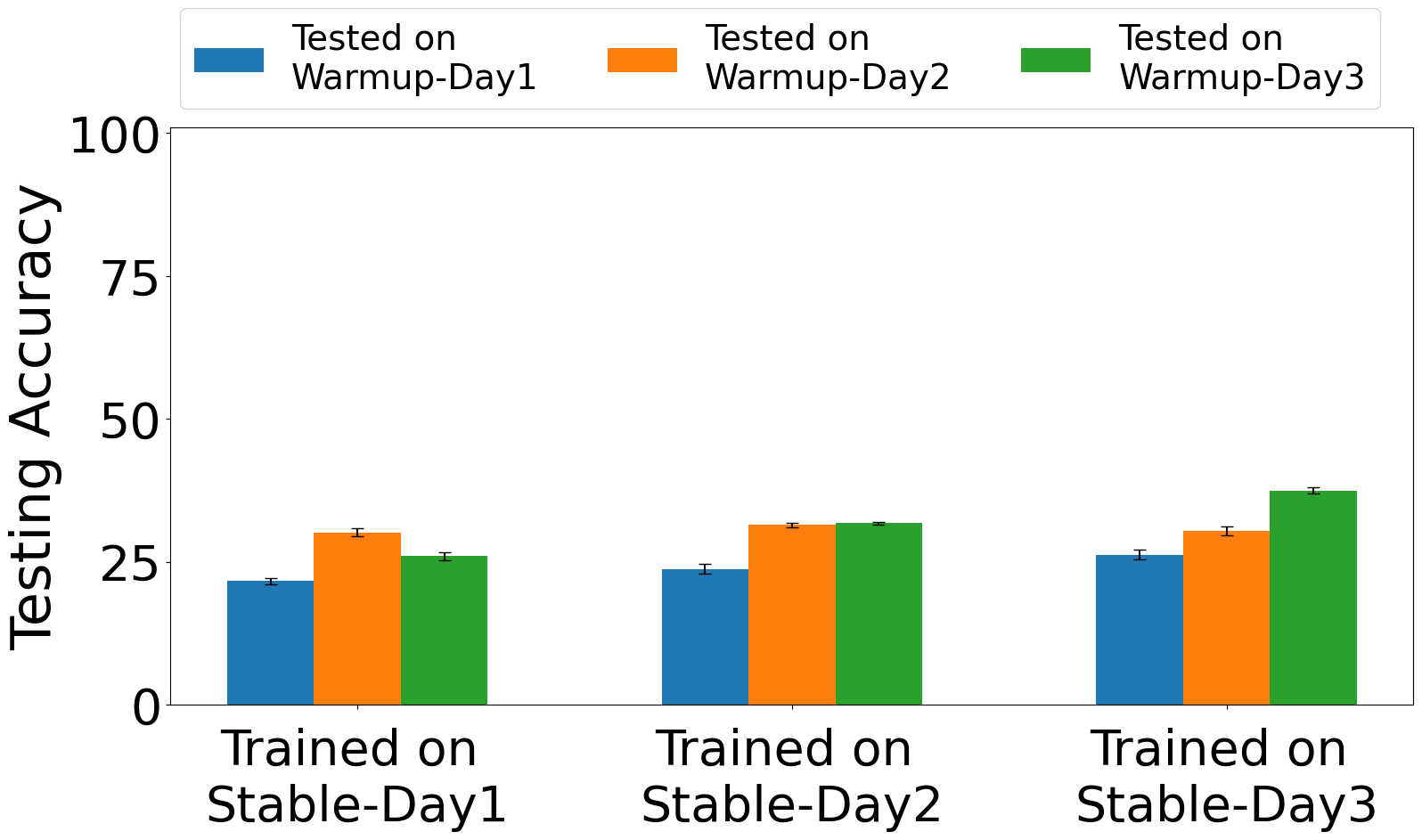}
   \label{TSTW_wireless}} 
   \\
   \caption{Exp. 3: training on Stable captures and testing on Warm-up captures}
\label{ex3}
\vspace{-0.2in}
\end{figure}
Now building on our findings that training and testing on stable data significantly mitigates the impact of hardware warm-up and that hardware warm-up is the main cause behind the observed fingerprinting time sensitivity, we assess in this experiment how systems trained on stable frames perform when tested on early warm-up frames. The experiment's results, shown in Fig.~\ref{ex3}, vividly illustrate the classifier's struggle when trained on stable captures but tested on early warm-up data, with accuracy declining to as low as 17.8\%. This stark decline, consistent across both wired and wireless settings and exacerbated in cross-day scenarios, underscores the critical impact of the warm-up phase on the fingerprinting accuracy consistency, and calls for an in-depth investigation into the behavior of hardware impairments, which are foundational to RF fingerprints, during this period, a largely unexplored area crucial for devising robust solutions to ensure seamless RF fingerprinting across all operational phases.

%% file: 4-anatomy.tex
RF fingerprints are distinctive patterns in RF signals arising from manufacturing variations, component tolerances, and hardware-induced impairments in RF devices. These imperfections, embedded during production, subtly shape the RF bursts and serve as raw data for fingerprinting methods to identify devices. Unlike human fingerprints, the origins of RF fingerprints are not well understood, casting doubt on their uniqueness and permanence \cite{tyler2023considerations}. However, many of these patterns are linked to specific hardware impairments. Studying these impairments individually provides critical insight into the time-sensitivity and variability of RF fingerprints, enabling the development of more reliable fingerprinting techniques. In this section, we introduce key hardware impairments that form the basis of our analysis and quantify their cumulative impact through a single, interpretable metric.

\subsection{Hardware Impairments Background}
\label{subsec:hw-bk}

\myitemizebegin
\item {\bf {Error Vector Magnitude (EVM)}:}
EVM measures the RMS error between received symbols and their ideal constellation points at symbol clock transitions, serving as a key indicator of modulation quality.
Mathematically, it can be expressed as $EVM[k] = \sqrt{I_{err}[k]^2 + Q_{err}[k]^2}$, where $k$ is the symbol index, $I_{err}=I_{ref} - I_{measure}$ is the In-Phase (I) error, and $Q_{err}= Q_{ref} - Q_{measure}$ is the Quadrature (Q) error. 
EVM serves as a unified error metric that captures the cumulative effect of transmitter impairments, such as IQ imbalance, phase noise, non-linearity, and frequency error, in a single measurement. While the error vector includes a phase component, this angle is typically random, as it depends on both the nature of the impairment and the (random) position of the symbol in the constellation.

\item {\bf {Magnitude Error}:}
As a standalone parameter, this error quantifies the deviation between the amplitude (magnitude) of the received signal and the ideal magnitude (or reference magnitude) of the signal constellation point. It is a measure of the signal's amplitude accuracy. Magnitude Error is typically represented as a percentage or in decibels (dB) and can be calculated as $IQ_{Mag\hspace{0.2em}error}[n] = |IQ_{ref}[n]| - |IQ_{measure}[n]|$.

\item {\bf {Phase Error}:}
As a standalone parameter, this error characterizes the phase deviation between the received signal and the ideal reference signal. It quantifies the angular error between the actual phase and the ideal phase of the signal constellation point and is usually expressed in degrees or radians. It can be calculated as $ IQ_{phase\hspace{0.2em} error}[n] = \angle{IQ_{ref}}[n] - \angle{IQ_{measure}}[n]$.

\item {\bf {IQ Origin Offset}:}
In communication systems, precise alignment of the in-phase (I) and quadrature (Q) components is essential for accurate signal demodulation. IQ offset, also known as origin IQ offset, refers to a displacement of the IQ origin from its ideal position, often manifested as a constant DC bias or carrier feedthrough, resulting from hardware imperfections such as analog or digital chain imbalances.
IQ Offset indicates the magnitude of the carrier feedthrough signal. When there is no carrier feedthrough, IQ offset is zero (-infinity dB) and is calculated by RMS averaging the measured IQ offset for each symbol in the measurement interval and is expressed relative to the average signal power. An IQ offset can be caused when the center (DC) carrier leaks into the signal. IQ offset can also be caused when the baseband signal has a DC offset which then shows up as (DC) carrier power when the baseband signal is upconverted.

\item {\bf {IQ Gain Imbalance}:}
IQ gain imbalance (IQ Gain Imb) compares the gain of the in-phase (I) signal with the gain of the quadrature phase (Q) signal. It is expressed as the ratio of the gain of the I component to the gain of the Q component of the signal. Mathematically, IQ gain imbalance can be represented as $\text{IQ Gain Imb} = 20 \log_{10} \left( \frac{G_I}{G_Q} \right)$,
%
where $G_I$ and $G_Q$ are the gains of the I and Q components, respectively. 
IQ gain imbalance, expressed in logarithmic terms, can take both positive and negative values. In single-carrier modulation, it appears as unequal scaling of the constellation points along the I and Q axes. In contrast, for OFDM systems like WiFi 802.11n, it introduces inter-carrier interference between each subcarrier and its mirror image relative to the center frequency.

\item {\bf {Quadrature Error}:}
IQ Quadrature Error (IQQuadErr), also known as Quadrature Skew, indicates the orthogonal error between the in-phase (I) and quadrature (Q) signals. Ideally, the I and Q signals should be orthogonal, i.e., 90 degrees apart. Quadrature skew error quantifies the deviation from this ideal orthogonality. For example, a quadrature skew error of 3 degrees means that the I and Q signals are 93 degrees apart. Mathematically, the quadrature error can be expressed as $\text{IQQuadErr} = \theta - 90^\circ$,
%
%
where $\theta$ represents the actual angle between the I and Q signals. 
A positive quadrature skew error rotates the Q axis counterclockwise relative to the I axis, creating an angle greater than 90 degrees in the upper-right quadrant. In OFDM systems like WiFi 802.11n, this error does not appear as a rotated constellation, since OFDM symbols reside in the frequency domain. Instead, quadrature skew causes each subcarrier to interfere with its mirror subcarrier, effectively smearing each constellation point into a scaled version of the full constellation.

\item {\bf {IQ Timing Skew}:}
Also known as IQ timing error, this impairment refers to the temporal misalignment between the in-phase (I) and quadrature (Q) components of a signal. Ideally, I and Q should be perfectly synchronized, but hardware imperfections, such as mismatched signal paths or component delays, can introduce skew, causing one component to lag behind the other.
Mathematically, IQ timing skew can be represented as $\text{IQ Timing Skew} = t_I - t_Q$,
where $t_I$ and $t_Q$ are the arrival times of the I and Q components, respectively. A positive value indicates that the I component leads the Q component, while a negative value indicates that the Q component leads the I component. IQ timing skew affects the synchronization of the I and Q components, leading to phase errors and distortions in the received signal. This can degrade the performance of modulation schemes, particularly those used in WiFi 802.11n, such as OFDM. In OFDM systems, timing skew can cause inter-symbol and inter-carrier interference, further complicating signal recovery.

\item {\bf {Carrier Frequency Offset (CFO)}:}
CFO, another important parameter in communication system \cite{smaini2012rf}, refers to the deviation of the carrier frequency from its intended or nominal value. This offset can result from various factors, including inaccuracies in the local oscillator or frequency synthesis components, Doppler shifts in wireless channels, and environmental conditions affecting the carrier frequency. Errors in RF frequency, LO frequency, or digitizer clock rate can all appear as carrier frequency errors.

\item {\bf {Symbol Clock Error (SCE)}:}
Symbol or chip clock error, also known as timing error, is an other critical parameter in digital modulation analysis \cite{smaini2012rf}. It refers to the deviation between the received signal's timing and the expected or nominal symbol or chip clock. The symbol clock represents the timing at which symbols are transmitted or received, and in the context of spread spectrum systems, ``chip" refers to the basic signaling unit. The error is calculated by averaging the symbol timing corrections during the measurement interval and presented in parts-per-million (ppm).

\item {\bf {Carrier Suppression Error (CSE)}:}
During the upconversion stage of transmission, some RF carrier leakage can pass through the mixer and appear at the output alongside the modulated signal. RF carrier suppression quantifies this leakage relative to the desired modulated output, typically expressed in decibels (dB). It provides a clear metric for assessing how effectively the carrier is suppressed.

\item {\bf {Average Burst Power}:}
Average burst power, the most commonly utilized power metric in WLAN measurement applications, represents the time-averaged power level of an 802.11 packet throughout the duration of a burst. This metric offers a comprehensive assessment of the signal strength during transmission intervals.
\myitemizeend

%% file: 5-behavior.tex
\subsubsection{\textbf{Measurement Setup}}
 \begin{figure}
 \setlength{\abovecaptionskip}{3pt}
     \subfloat[Wired Setup\label{subfig-1:wired_spec}]{%
       \includegraphics[width=0.23\textwidth, height = 0.18\textwidth]{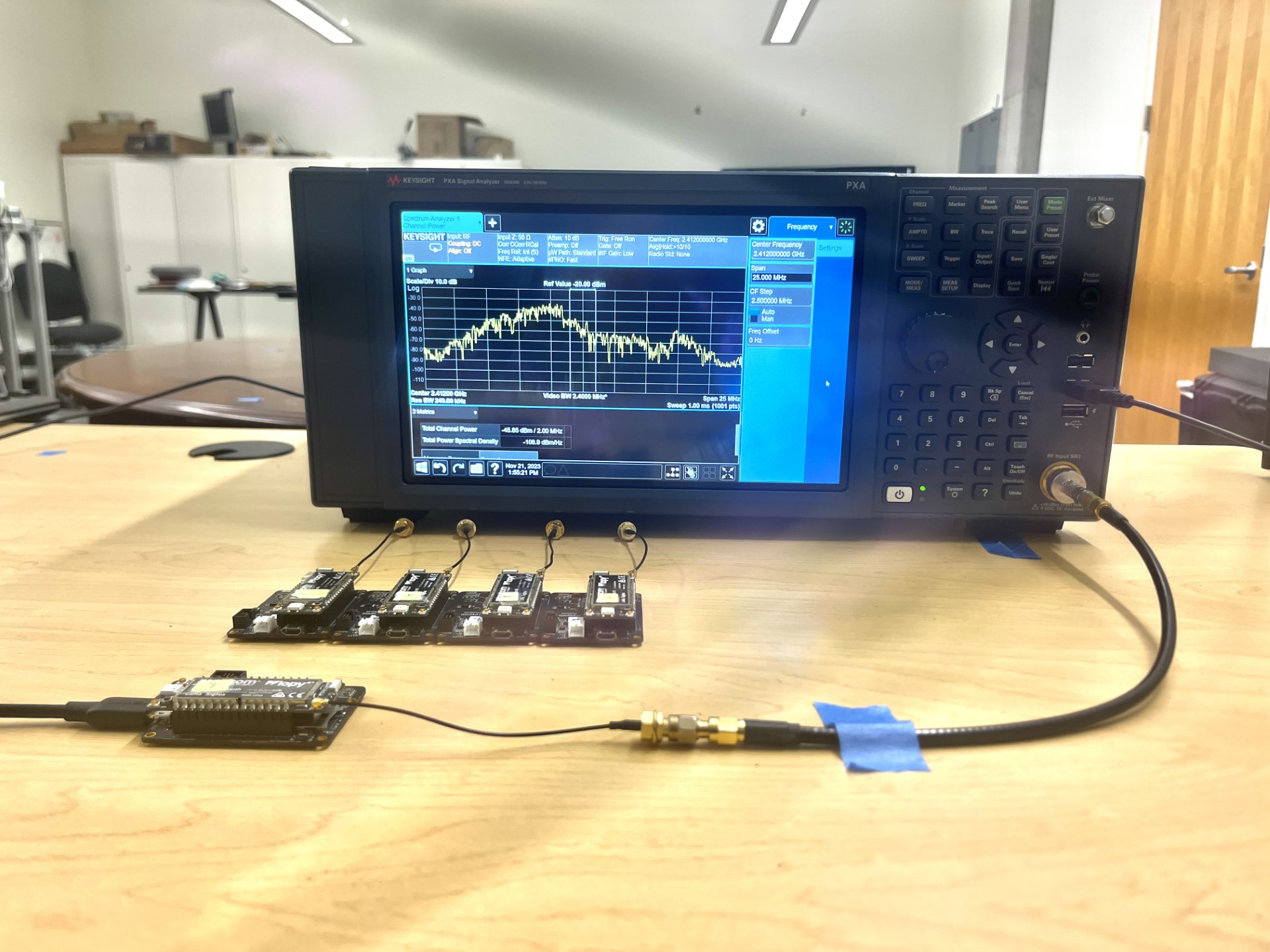}
     }
    \hspace{0.00001cm}
     \subfloat[Wireless Setup\label{subfig-2:wireless_spec}]{%
       \includegraphics[width=0.23\textwidth, height = 0.18\textwidth]{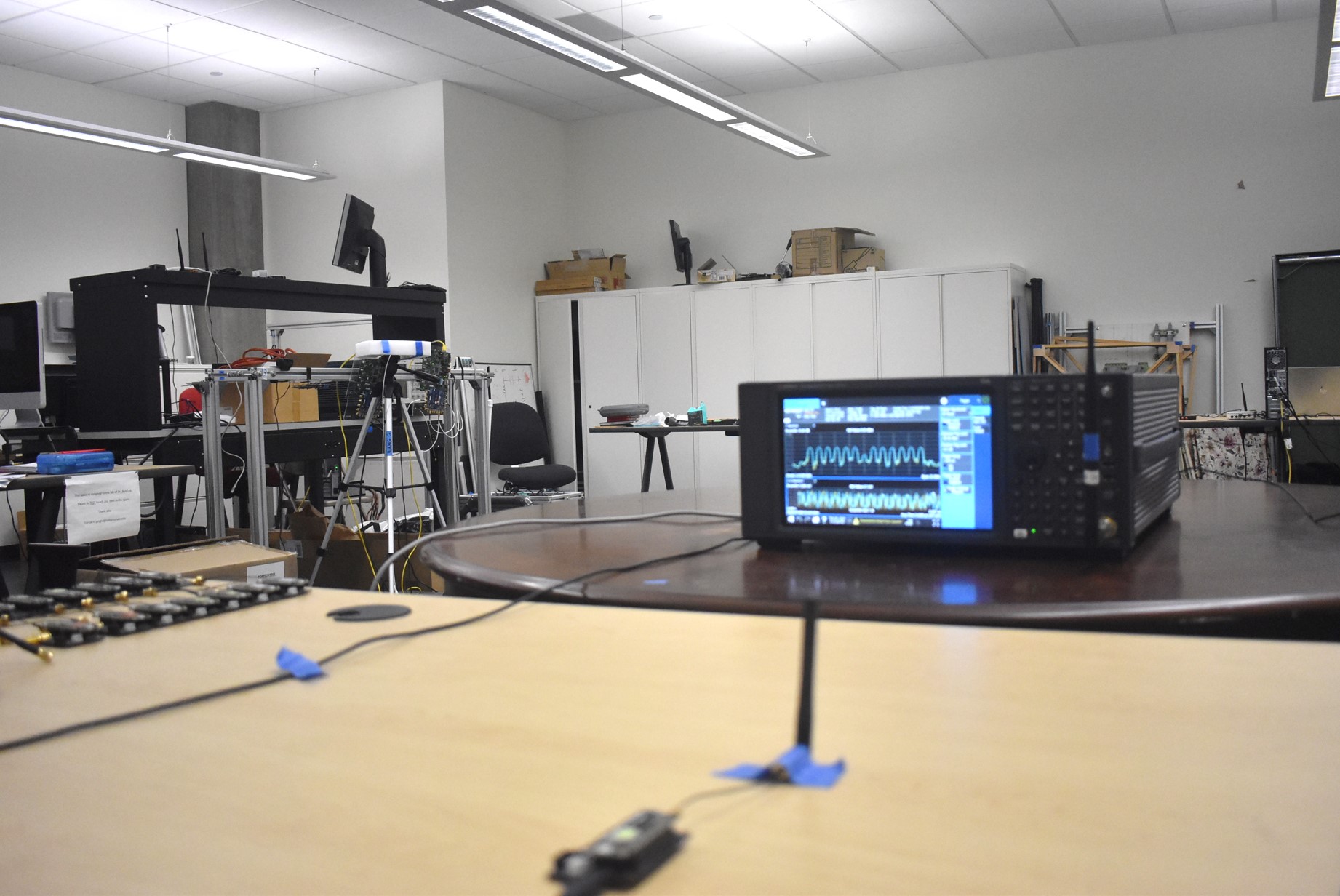}
     }
     \caption{The Hardware Impairment Measurement Setup}
     \label{fig:Imp_setup}
\end{figure}

\begin{figure*}  
    \setlength{\abovecaptionskip}{3pt}
    \includegraphics[width=\textwidth, height=0.35\textwidth]{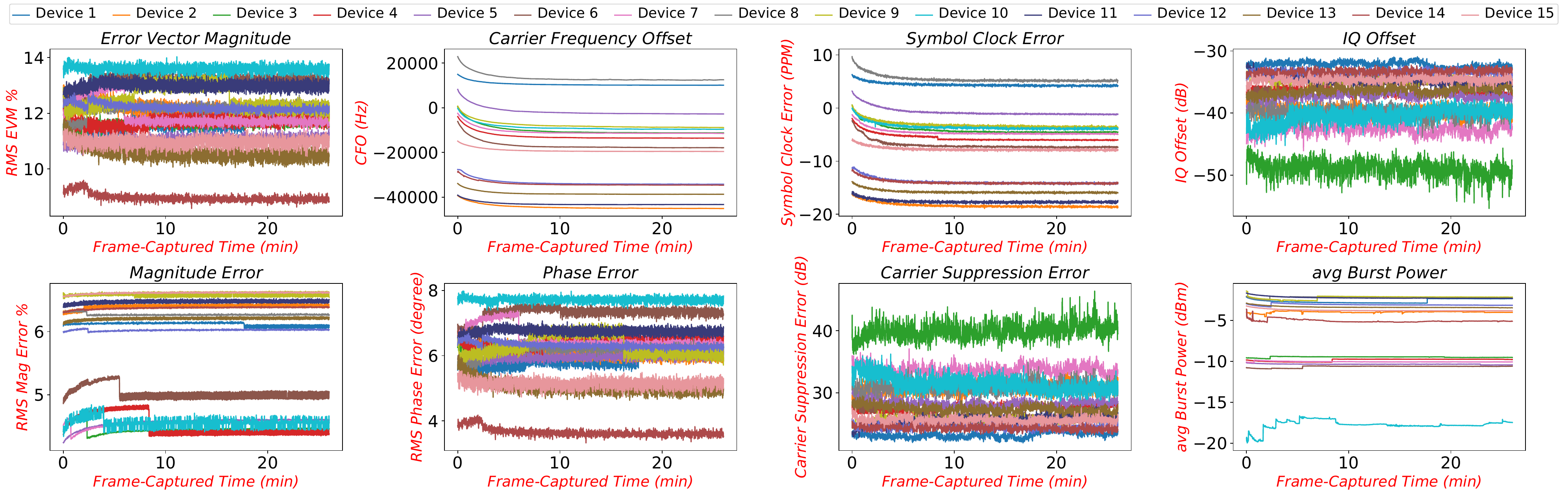}  
    \caption{The behavior of 8 hardware Impairments of 15 devices (Sending WiFi B packets) 
during the initial 30 minutes.}
    \label{fig:B_Imp}
\end{figure*}

\begin{figure*}  
    \centering
    \includegraphics[width=\textwidth, height=0.35\textwidth]{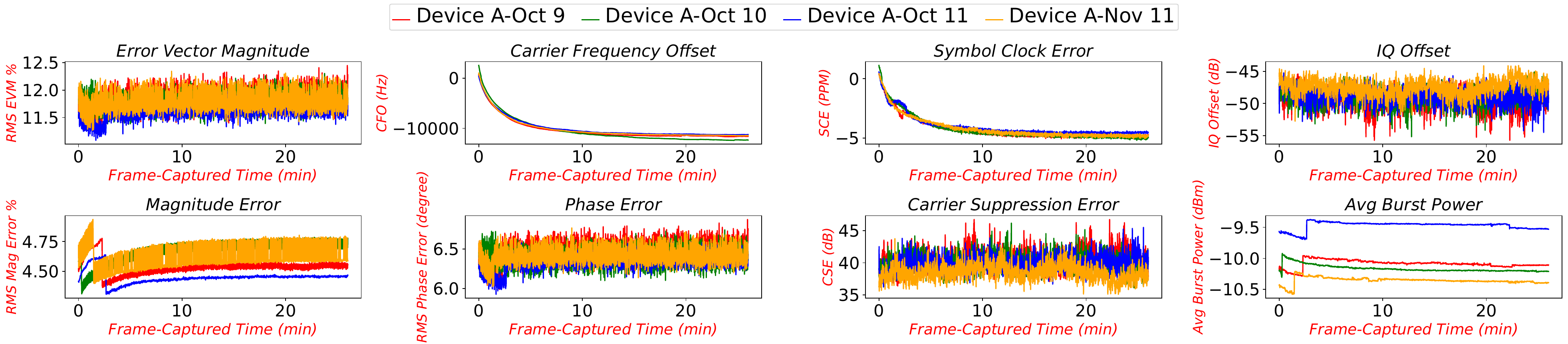}  
    \caption{The Behavior of Device A's Hardware Impairments Over a Month in a Wired Connection}
    \label{fig:dev6_Imp}
\end{figure*}

We established a dedicated measurement setup, as depicted in Fig.~\ref{subfig-1:wired_spec}, to closely monitor the behavior of various impairments of 15 Pycom devices both during and after the warm-up period over one month. Specifically, we established a wired connection for each Pycom device to interface with a Keysight PXA signal analyzer N9030B, running the WLAN 802.11x X-Series Measurement Application. Our Pycom devices were programmed to consistently transmit identical WiFi 802.11b packets at fixed intervals. Simultaneously, we configured the signal analyzer to sample incoming RF bursts at a rate of 35MSps and a bandwidth of 20MHz. To automate the measurement collection, we employed a Standard Commands for Programmable Instruments (SCPI) script. This script orchestrated the instrument to utilize the WLAN measurement application modulation analysis mode and calculate various hardware impairments for incoming packets triggered one at a time by the RF burst mode. Initiating the procedure involved powering On the transmitter and simultaneously running the script. Over the initial 30 minutes, encompassing both the warm-up and stabilization periods, the spectrum analyzer receives more than 2500 packets, extracts IQ samples, and calculates the corresponding impairments for each packet. We repeated the measurement collection process for all devices on four days over the course of one month (Oct 9, Oct 10, Oct 11, and Nov 11) to ensure behavior consistency and assess the aging impact on these impairments. This systematic approach ensured a comprehensive evaluation, enabling precise analysis and characterization of impairments.


\subsubsection{\textbf{Impairments' Behavior During \& After Warm-Up}}
The primary goal of this analysis is to examine the behavior and response of DL-based RF fingerprinting models when tested with frames transmitted during the warm-up period. We focus on eight key impairments discussed in Sec.~\ref{subsec:hw-bk}: EVM, CFO, SCE, IQ Offset, Magnitude Error, Phase Error, CSE, and Average Burst Power. Although DL models may not directly learn these impairment patterns, they provide valuable insights into the underlying hardware characteristics of the devices. 
Two key factors influenced the decision to select only these eight impairments:
(1) Measurement Consistency: Certain impairments, like phase noise floor and local oscillator leakage, could not be consistently estimated across all devices and frames with our measurement tools. To maintain the quality and comparability of training labels, we excluded impairments that exhibited low signal-to-noise ratios or unstable estimation performance.
(2) Relevance to Device-Level Identification: The eight selected impairments (e.g., CFO, EVM, IQ imbalance) are known to have distinct device-specific signatures and are frequently utilized in RF fingerprinting research. These impairments demonstrated greater variation across devices and a stronger correlation with classification accuracy.

Fig.~\ref{fig:B_Imp} shows the behavior of these hardware impairments for 15 (WiFi B) devices during the initial 30-minute operation period. Our observations reveal several noteworthy patterns:
\begin{myitemize}
    \item {\bf Pattern 1.} The impairments display dynamic behavior during the initial minutes of device operation, eventually stabilizing thereafter.

    \item {\bf Pattern 2.} CFO and SCE impairments undergo the most significant variation during warm-up, with notable changes from minute 1 to minute 12. For instance, the CFO of one device shifted from 2300 Hz at minute 1 to -12000 Hz by minute 12. Such drastic changes underscore the CFO's sensitivity to the warm-up effects, making it a pivotal factor in the transient behavior of RF signals during this period.

    \item {\bf Pattern 3.} The separability of impairments varies, with CFO and SCE standing out as the most distinguishable, showcasing the largest differences among devices.
\end{myitemize}
To verify the consistency of measurements and examine the long-term behavior of these impairments, we conducted four measurements over the span of one month. Notably, the impairments displayed consistent patterns across all four collection instances. Figure~\ref{fig:dev6_Imp} shows the measurements of the eight impairments for device A, captured between October and November 2023. This consistency aligns with the findings in Sec.~\ref{sec:exposing}, suggesting that once the devices stabilize, their behavior remains steady over time.

We want to mention that, in our experiments, we explicitly configured the tested Pycom devices to operate in IEEE 802.11b and 802.11n modes using standard-compliant settings and off-the-shelf chipsets, representative of those used in embedded and low-power Wi-Fi systems. Therefore, the warm-up behaviors observed and analyzed in this section are fundamental to all WiFi-capable hardware and the trends shown in Figures~\ref{fig:B_Imp}  and~\ref{fig:dev6_Imp} were consistently observed across different devices, days, and protocol modes, demonstrating stable and repeatable patterns that are neither a result of protocol switching nor unique to our specific test platform.

Based on these observations, we propose hypotheses to explain the performance drop observed when DL models are trained on stable datasets and tested with warm-up data. We hypothesize that DL models may overfit to stable impairment behaviors or rely heavily on a subset of impairments, likely those related to CFO and SCE, while neglecting other critical information within the RF fingerprint. Although some impairments show better distinguishability during the stable phase, their significant variability during the warm-up period makes them unreliable for classification in such scenarios.

This analysis provides researchers with the required insights to understand what might have gone wrong and how we can assist these models to latch into the correct features that would enable the development of more robust and reliable RF fingerprinting solutions across all operational phases.

%% file: 8-framework.tex
\begin{figure*}[t]  
    \centering
    \setlength{\abovecaptionskip}{3pt}
    \setlength{\abovecaptionskip}{3pt}
    \includegraphics[width=\textwidth, height=0.30\textwidth]{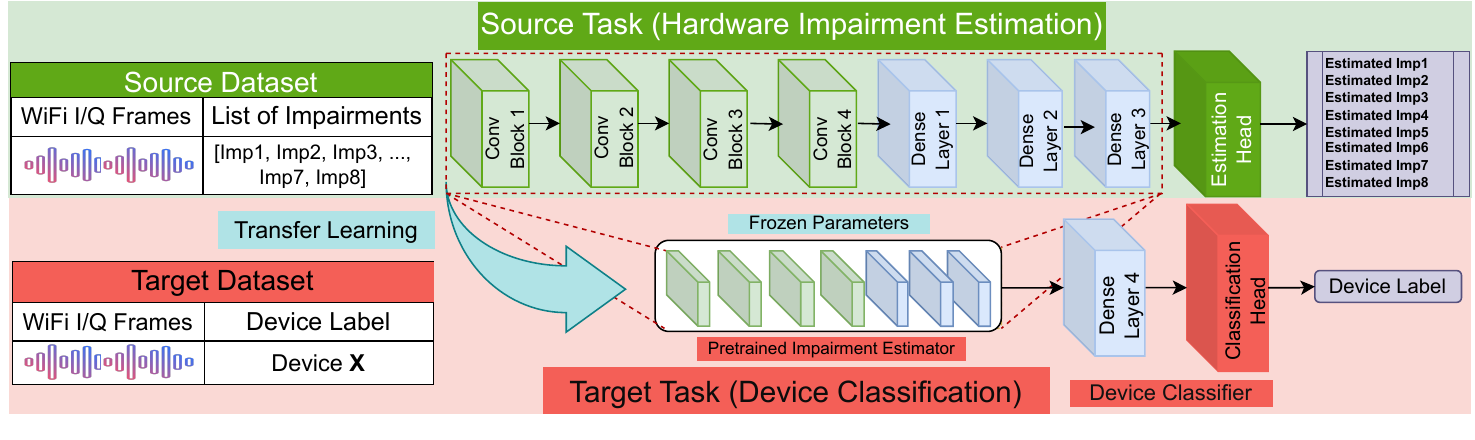}  
    \caption{Overview of \proposed: Sequential Transfer Learning on HW Impairment Estimation for Robust Device Classification}
    \label{fig:Imp}
\end{figure*}

\subsection{Intuition and Design Guidelines}
Drawing from insights learnt in previous sections, we identify key challenges and set a goal to develop a framework that performs reliably across all operational phases, particularly addressing the blind spot during the warm-up phase. We've distilled two critical observations and a set of hypotheses to guide our framework design: (1) hardware impairments exhibit stability during the post-warm-up phase, implying that training DL-based RF fingerprinting systems on stable captures effectively mitigates performance degradation in cross-day scenarios; and (2) systems trained on stable captures struggle when tested on frames transmitted during the warm-up phase. 

To elucidate the potential causes behind the second observation, we hypothesize that it may be attributed to: (1) system's overfitting on stable behavior, rendering it incapable of recognizing devices during the warm-up phase, (2) system's focusing on one or a subset of impairments that exhibit the most separability in the stable phase but experience significant variation in the warm-up phase, or/and (3) system's learning of features unrelated to the hardware characteristics of the device. 
Our goal is to leverage these insights to develop a framework that can be trained on stable captures yet performs admirably across all operational phases. 
To achieve this goal, the framework must (1) capture hardware impairment information effectively and (2) integrate it comprehensively into the final decision-making layers within the network. 
By focusing on hardware-impairment-specific information, which constitutes the unique fingerprint among devices, and understanding the interrelationships between impairments, we present this information comprehensively within the receptive field of the classification head. This contextual knowledge regarding the collective behavior of impairments enables the classification head to recognize devices even when some impairments deviate from their stable phase values. Hence, we propose \proposed, a framework that integrates sequential transfer learning \cite{ruder2019neural} for cross-phase adaptability and targeted impairment estimation to anchor the system in the immutable aspects of hardware identity, ensuring robust device recognition and identification across all operational phases. 

\subsection{High-Level Overview of \proposed}
At its core, \proposed~embodies two main components (Fig.~\ref{fig:Imp}): the pretrained impairment estimator and the device classification head. The foundation of our approach lies in the training of a deep single-input-multi-output CNN model, leveraging our source dataset. This dataset comprises time-domain I/Q representations of WiFi frames as input and encompasses eight hardware impairments as labels. Through rigorous training on this source task, we equip the model with the capability to concurrently and accurately estimate multiple impairments. Subsequently, we harness the expertise of this pretrained impairment estimator as a feature extractor for our device classification model. This fusion entails the integration of the pretrained impairment estimator with two additional dense layers that constitute the Device Classification Head. With this combination, we forge a cohesive framework poised for device classification. The crux of the matter lies in training the second model on the device classification task, utilizing the target dataset, which features time-domain I/Q data as input and device labels as the output. This strategic approach ensures that our model excels at device classification, harnessing the knowledge gleaned from the preceding impairment estimation phase to create a robust and effective RF fingerprinting system.

\subsection{Implementation Details of \proposed}
\subsubsection{Impairment Estimation Network Model} 
The core of our impairment estimation model, designed for the source task, operates on time-domain I/Q tensors with dimensions of 2x17550, encompassing the entire WiFi frame. The model's labels consist of eight key impairments (EVM, CFO, SCE, IQ offset, Mag Error, Phase Error, CSE, and Avg. Burst Power), normalized to ensure equitable attention from the network, thus averting suboptimal performance biases. The architecture of this network comprises four Convolutional blocks, each featuring a 2D convolutional layer, followed by Rectified Linear activation units (ReLU), and a 2D max-pooling layer (1x2 size with a stride of 1x2) to dimensionally reduce feature maps. The convolutional layers in these blocks employ filters of sizes 256, 128, 64, and 32, respectively. Two dense layers, outputting 256 and 128 neurons, are subsequently applied, each followed by ReLU activation layers. The output feature vector is then directed to the estimation head, which generates a list of estimated impairments for the received frame. 
Our estimation model jointly estimates the eight hardware impairments, allowing the model to exploit statistical correlations among impairments, share early feature representations, and avoid unnecessary over-parameterization. As shown later, our experiments show that this setup improves both estimation accuracy and downstream fingerprinting performance compared to training one model per impairment.

Our model is optimized using Mean Square Error (MSE) and utilizes the Adaptive Estimation Momentum (Adem) optimizer with a learning rate of 0.00025, determined through experimental fine-tuning within the range [$10^{-1}, 10^{-5}$]. We choose MSE becuase it is the standard choice for continuous regression tasks where all target dimensions are equally important. MSE penalizes larger errors more heavily than small ones, making it suitable for guiding the encoder to produce precise impairment estimates. Furthermore, MSE promotes smooth optimization and stable convergence during training, which we empirically observed in our experiments.

We formulated hardware impairment recognition as a regression problem rather than classification for two main reasons. First, impairments like CFO, EVM, and IQ imbalance are naturally continuous-valued, so regression avoids arbitrary discretization and yields more physically meaningful, interpretable outputs aligned with RF measurements (e.g., Hz, dB). Second, regression promotes a more generalizable latent space, which is particularly beneficial for sequential transfer learning, capturing gradual hardware variations more effectively than fixed classes.

\subsubsection{Device Classification Network Model} 
Following the completion of the impairment estimation model's training, we repurpose the pretrained CNN model (excluding the estimation head) as the feature extractor for our device classification model, tailored to the target task. Building upon the impairment estimator, we append a dense layer consisting of 256 neurons that transform the 128-size feature vector and relay it to the classification head. The classification head incorporates a dense layer with a softmax activation function, responsible for determining the radio classification probabilities. Both the device classification model and the impairment estimation network employ the same optimizer, although with the cross-entropy loss function for the former. 
For training the classifier, we opt for stable-state data (data collected after device stabilization), which offers a more consistent and reliable representation of the devices’ intrinsic hardware impairments, unaffected by transient startup anomalies. However, inference (e.g., device authentication) is not restricted to stable stage, but rather can take occur at any point in time, including during boot-up, reconnection, or other early operational states, where impairments are inherently less stable and more time-varying.
%



\subsection{Feasibility and Practical Considerations} 
By design, \proposed~mitigates the warm-up problem by separating the learning of hardware impairments from the device classification task. Specifically, the incorporated pre-training architecture, in which the encoder is optimized to directly estimate hardware impairments from I/Q samples, allows the model to learn a stable, warm-up-agnostic and transferable representation of device-specific characteristics, making it less sensitive to short-term thermal drift and hardware instability during the warm-up phase.

To achieve this, the pre-training phase uses samples from both the warm-up and stable phases of the device’s operation, enabling the encoder to learn a spectrum of impairment patterns and adapt to the temporal transitions in hardware behavior. Consequently, the downstream classification model benefits from representations that are robust to early startup variability, which is often the main cause of accuracy degradation in RF fingerprinting systems.

Regarding the feasibility of obtaining measurements of the hardware impairments, we emphasize that these measurements are typically obtained as a byproduct of the data collection process. For example, when using a spectrum analyzer, the estimated impairment values (e.g., CFO, EVM, IQ imbalance) are automatically extracted alongside the I/Q signals, without requiring any additional effort. Similarly, if alternative tools are used, the impairments can be derived through lightweight post-processing of the captured I/Q data. Therefore, the requirement to label 8 impairments per device does not introduce extra complexity or overhead during data collection. It is also equally important to note that pre-training is only performed once, offline, in a controlled setting. The final model can then be deployed in the field for efficient inference without requiring online estimation or labeling.

%% file: 9-evaluation.tex
\begin{figure*}  
    \centering
    \setlength{\abovecaptionskip}{4pt}
    \includegraphics[width=\textwidth, height=0.25\textwidth]{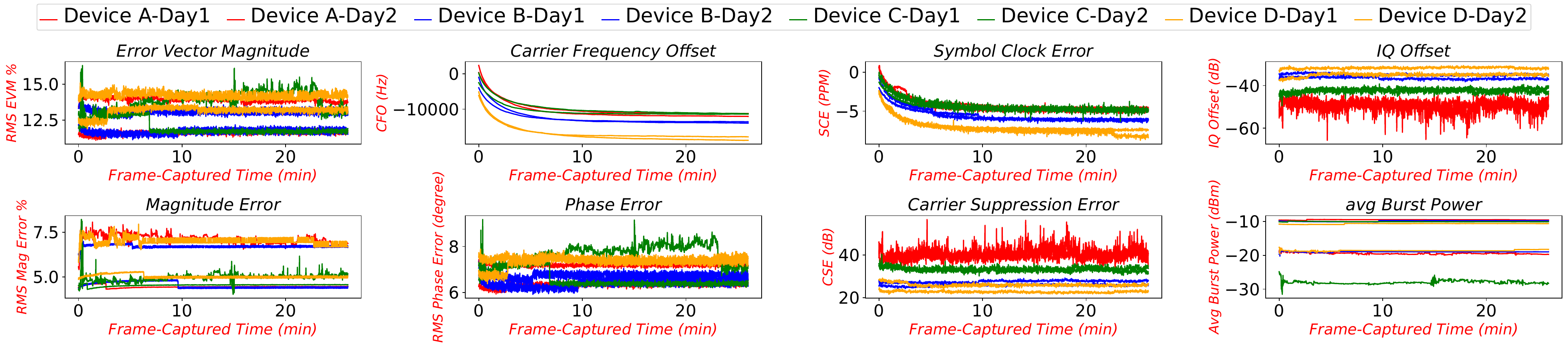}  
    \caption{Hardware impairment behaviors of 4 devices during the initial 30 minutes for the wireless setup: Day1 collection and Day2 collection are two weeks apart.}
    \label{fig:wireless_dataset}
\end{figure*}
We now present the performance of \proposed~using our comprehensive Indoor WiFi dataset. Our analysis focuses on its ability to bridge the detection gap observed during the warm-up period and its resilience to cross-day sensitivity. To provide context, we conduct comparative assessments by pitting our proposed framework against two baseline models: CNN-based~\cite{elmaghbub2021} and ResNet-based~\cite{jian2020deep} RF fingerprinting models. This comparative approach allows us to highlight the unique strengths and advantages of our framework.  

\subsection{WiFi Dataset Description} \label{subsec:wifi-collection}
In our evaluation of the impact of the warm-up phase in Section~\ref{sec:setup}, we described and used a dataset consisting of WiFi 802.11b frames only, which were specifically designed to isolate and analyze the warm-up effect. In this section, which focuses on evaluating the proposed framework's ability to overcome the warm-up effect, we also consider datasets collected from WiFi 802.11n frames to assess robustess across different protocols. In addition, each recorded WiFi (b and n) frame is annotated with hardware impairments, such as CFO, IQ imbalance, and EVM, which serve as training labels for the impairment estimation component of \proposed.

\subsubsection{WiFi 802.11b}
We captured both wired and wireless datasets of 15 Pycom devices transmitting WiFi 802.11b using the HR/DSSS physical-layer mode. The initial data collection involved gathering a wireless dataset (Day 1 dataset), followed by a deliberate two-week gap before resuming data collection to obtain the second wireless dataset (Day 2 dataset). Wireless data offers the advantage of capturing real-world environmental interactions, providing a comprehensive view of signal behavior in typical usage scenarios. Figure~\ref{fig:wireless_dataset} illustrates the impairment behavior of four devices across the two datasets. Within each dataset, we captured 3000 frames per device, spanning the initial 30 minutes of each device's operation, ensuring the inclusion of both the warm-up and stable phases across various days. Additionally, we collected wired data, which eliminates environmental variables, thus offering a controlled setting to focus on hardware-specific impairments.   
Employing the same measurement setup outlined in Sec. IV.~\ref{subsec:Imp_behavior}, all devices used the same antenna which is positioned 1 meter away from the spectrum analyzer, as visually depicted in Fig. \ref{fig:Imp_setup}. The resulting datasets encompass over 180k frames, where each frame is characterized by its time-domain I/Q samples, represented as (2x17550) dimensions, and, notably, includes the corresponding 8 key hardware impairments (EVM, CFO, Symbol Clock Error, IQ Offset, Magnitude Error, Phase Error, Carrier Suppression Error, and average burst power). These datasets not only serve as a pivotal component in the RF fingerprinting research, offering unprecedented insights into device characteristics through the real measurements of hardware impairments but also hold the potential for broader applications, such as impairment estimation and beyond.   

\subsubsection{WiFi 802.11n}
To assess the applicability of our framework on other protocol standards and to evaluate cross-protocol performance, we used the same transmitters to send the same message using the WiFi 802.11n protocol, which employs different modulation and data rates compared to WiFi 802.11b. Our WiFi 802.11n transmitters utilize OFDM modulation with QAM-16 for subcarrier modulation and a 20MHz bandwidth. Similar to the previous datasets, we captured data in both wired and wireless setups, using the same measurement setup. The resulting datasets encompass over 96k frames, where each frame is characterized by its time-domain I/Q samples, represented as (2x1012) dimensions, and include the corresponding 8 key hardware impairments (EVM, CFO, Symbol Clock Error, IQ Offset, IQ Gain Imbalance, Quadrature Error, IQ Timing Skew, and Pilot EVM). 

It is important to iterate that the WiFi 802.11b and 802.11n datasets described in this section differ from those presented in Section~\ref{sec:setup}. Specifically, here, each captured and recorded frame is annotated with hardware impairments that serve as training labels for \proposed's impairment estimation component. As shown later, these datasets enable a broader evaluation of the framework's ability to predict impairments and generalize across variations in protocol and time.
Both WiFi 802.11 b and n datasets can be accessed and downloaded from NetSTAR Lab at {\color{blue} {\it https://research.engr.oregonstate.edu/hamdaoui/datasets}}.

\subsection{Evaluation Methods and Metrics}
We begin by assessing the effectiveness and ability of our proposed hardware impairment estimator in simultaneously predicting 8 key impairments from time-domain raw I/Q frames, regardless of the device's operational phase. 
For this, we utilize the Mean Absolute Error (MAE) metric to measure the accuracy of our estimator when applied to unseen frames. 
MAE quantifies the average magnitude of the absolute differences between estimated values and actual/measured values. Unlike other metrics that may square the error (such as MSE), MAE treats all errors equally, making it more interpretable and less sensitive to outliers.
Mathematically, MAE is defined as:
\begin{equation}
\text{MAE} = \frac{1}{n} \sum_{i=1}^{n} \left| y_i - \hat{y}_i \right|
\end{equation}
\noindent
where \( n \) is the total number of measurements, \( y_i \) is the measured value, and \( \hat{y}_i \) is the estimated value.

Furthermore, we explore the influence of the number and selection of impairments to be simultaneously estimated on the overall estimation accuracy. To ensure a robust evaluation, we partitioned our dataset into training (60\%), validation (20\%), and testing (20\%) subsets, encompassing both the warm-up and stable phases of all tested devices' operation.

We then evaluate \proposed's performance across different operational phases of the devices. In particular, we examine how the framework performs during the unstable, warm-up period. To achieve this, we train the framework using post-stabilization data and test it using frames captured at different intervals within the warm-up period, spanning the initial 2 to 6 minutes. Our primary evaluation metric is per-frame accuracy, offering a robust measure of the classification performance. To facilitate meaningful comparisons, we benchmark the performance of \proposed~against two baseline models, CNN and ResNet, thus enabling a thorough assessment of its effectiveness and superiority.
All models are trained using the Keras library on an NVIDIA GeForce RTX 2080 Ti GPU system, with training extending over 20 epochs.

\subsection{Performance Analysis of the Impairments Estimator}
\begin{figure*}
 \setlength{\abovecaptionskip}{3pt}
    \subfloat[EVM \label{subfig:evm-wired}]{%
       \includegraphics[width=0.25\textwidth, height = 0.21\textwidth]{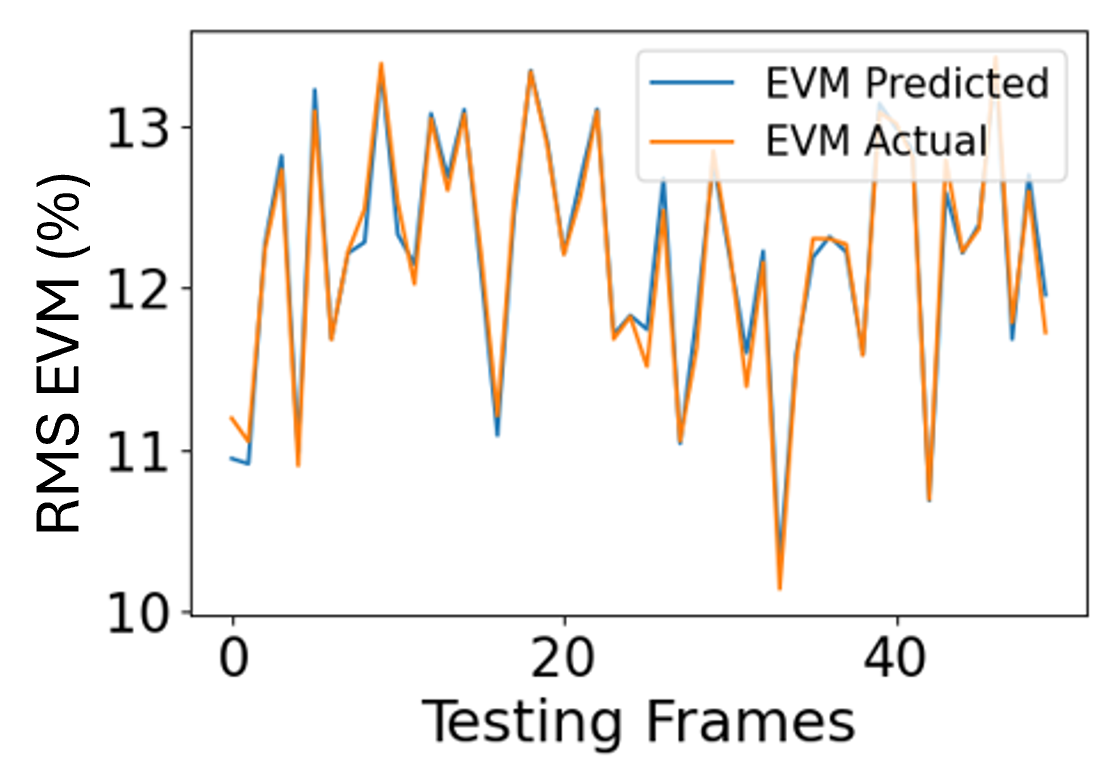}}
    \hspace{-0.2cm}
        \subfloat[CFO\label{subfig:cfo-wired}]{%
       \includegraphics[width=0.25\textwidth, height = 0.225\textwidth]{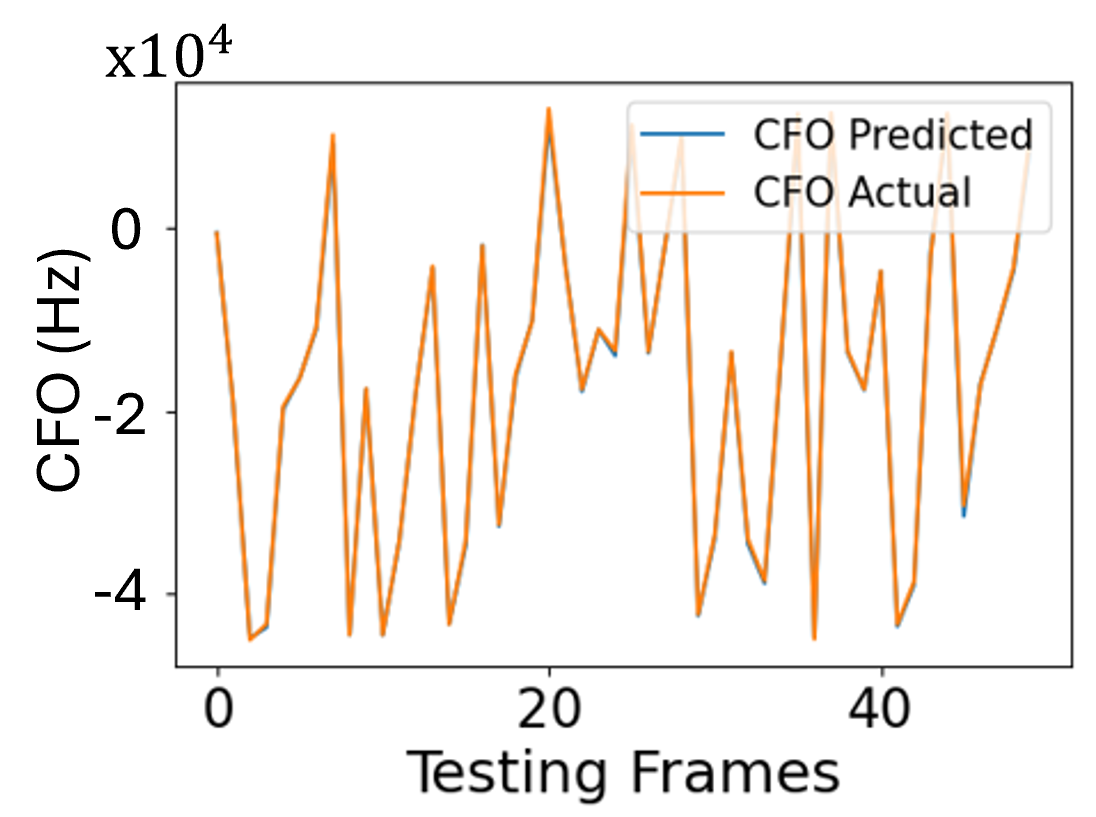}}
    \hspace{-0.2cm} 
        \subfloat[SCE\label{subfig:sce-wired}]{%
       \includegraphics[width=0.25\textwidth, height = 0.21\textwidth]{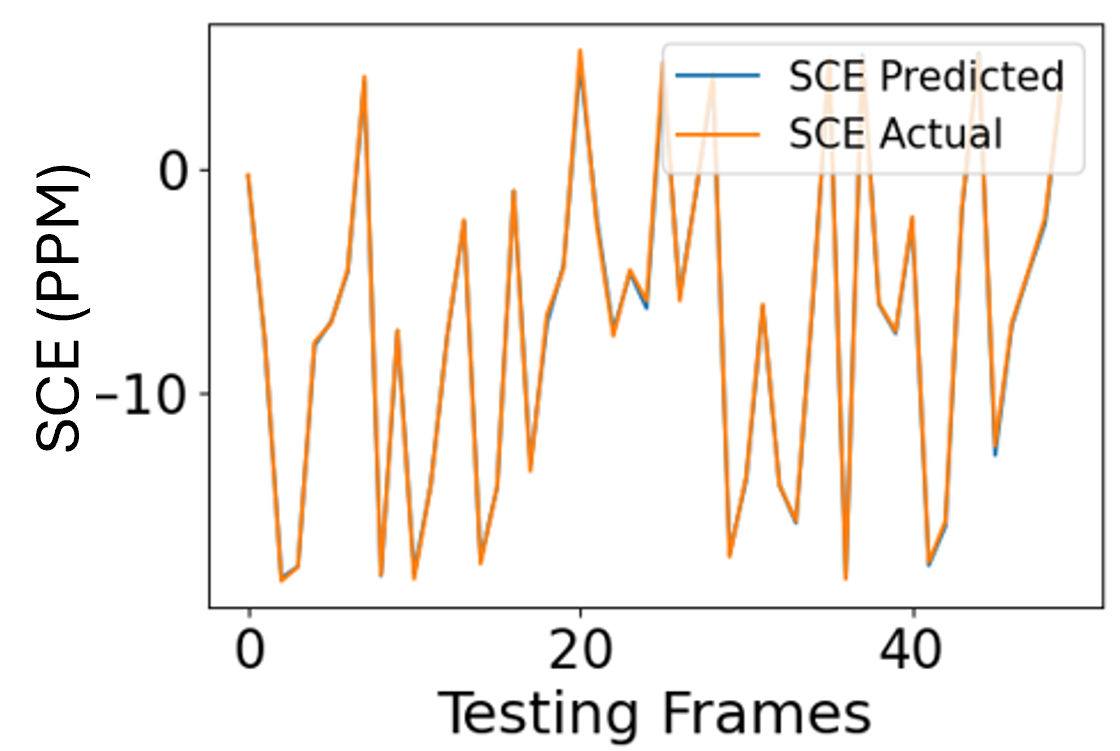}}
    \hspace{-0.1cm}
        \subfloat[IQ Offset\label{subfig:iq-offset-wired}]{%
       \includegraphics[width=0.25\textwidth, height = 0.215\textwidth]{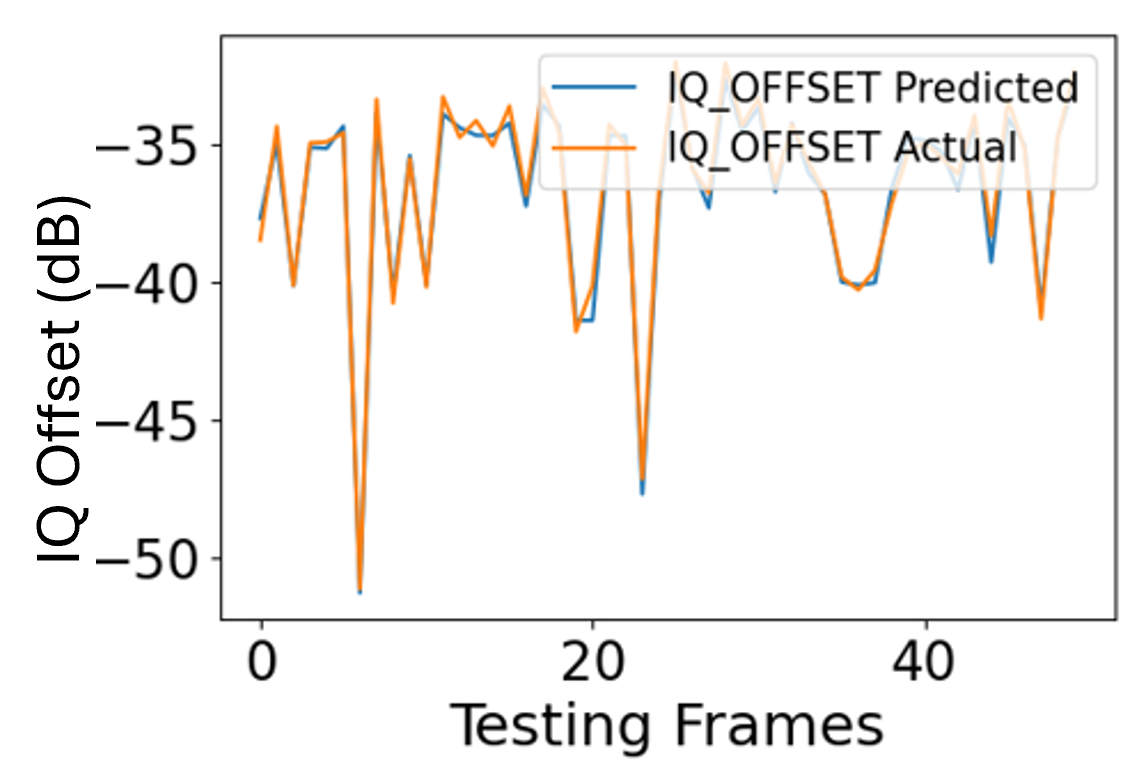}} 
\vspace{-0.1cm}
    \subfloat[Mag Error \label{subfig:mag-wired}]{%
       \includegraphics[width=0.25\textwidth, height = 0.21\textwidth]{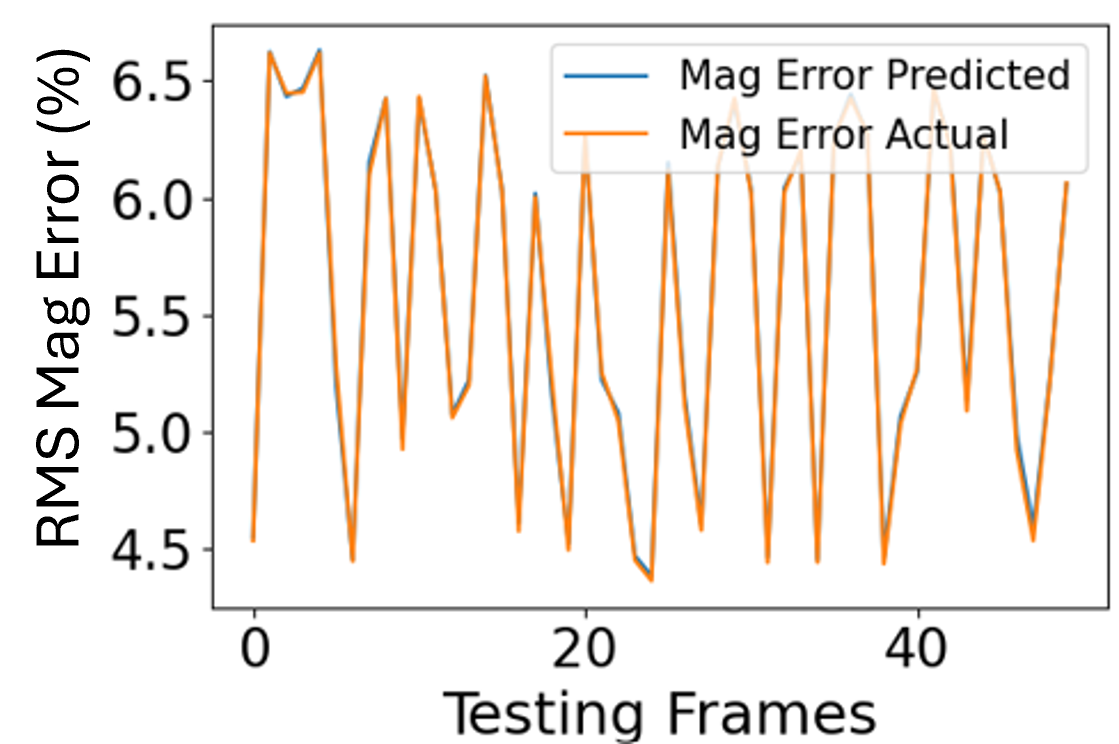}}
    \hspace{-0.2cm}
        \subfloat[Phase Error\label{subfig:phase-wired}]{%
       \includegraphics[width=0.25\textwidth, height = 0.21\textwidth]{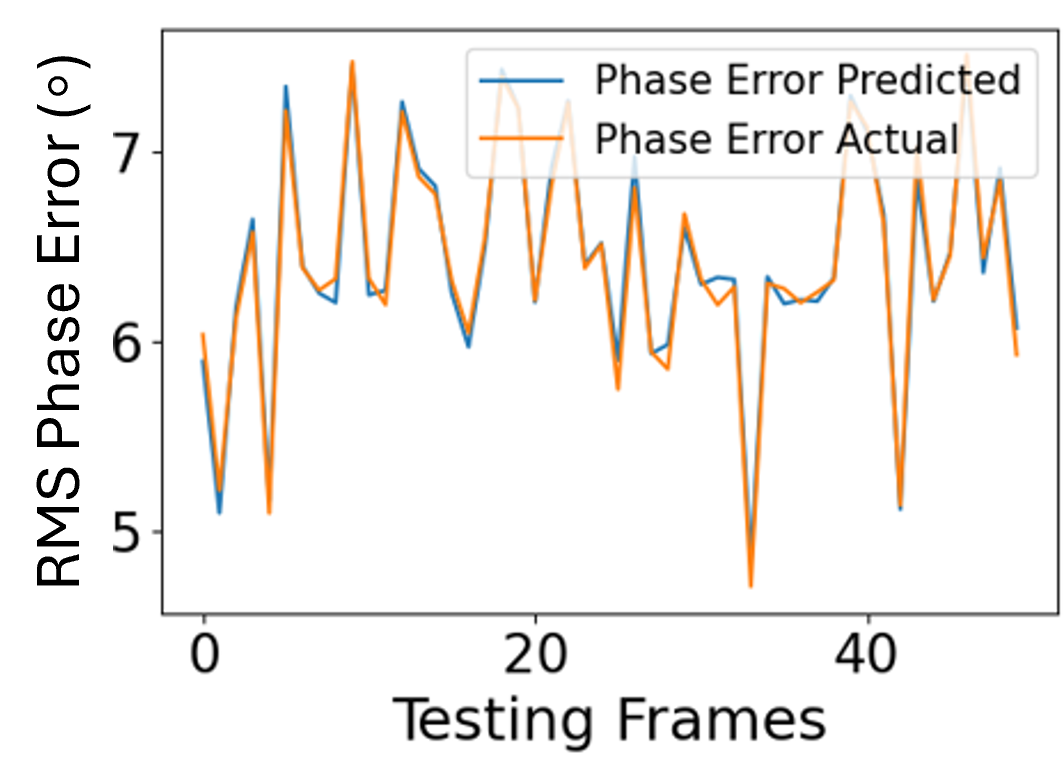}}
    \hspace{-0.2cm} 
        \subfloat[CSE\label{subfig:cse-wired}]{%
       \includegraphics[width=0.25\textwidth, height = 0.21\textwidth]{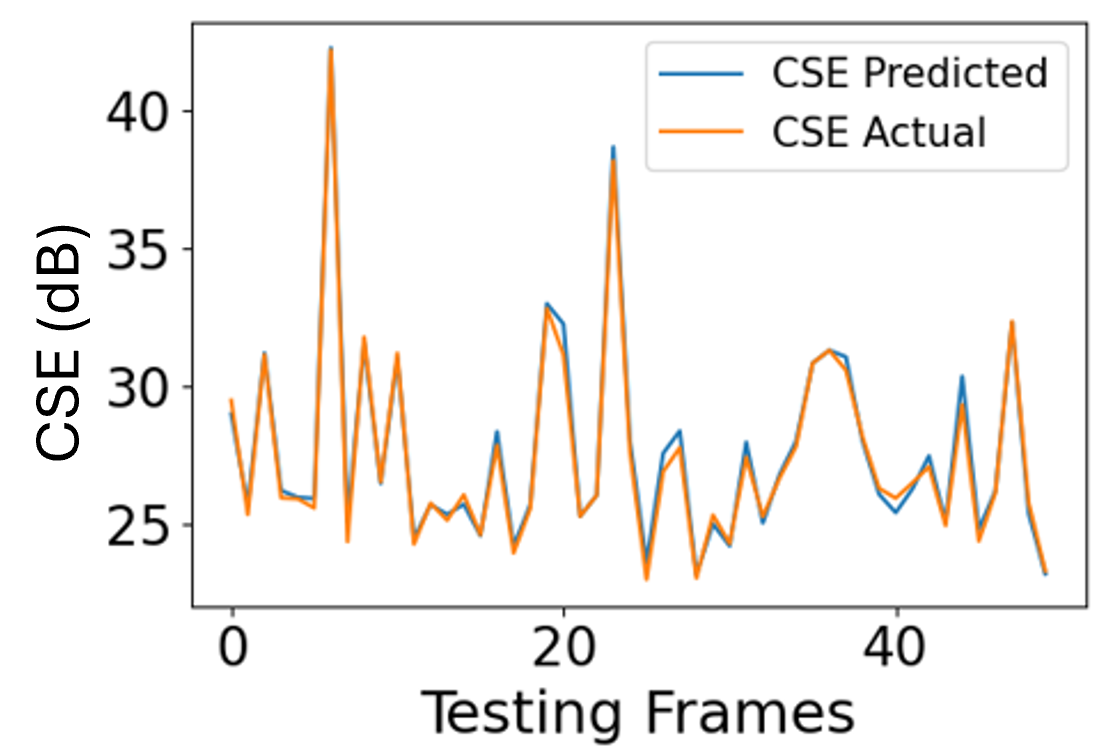}}
    \hspace{-0.1cm}
        \subfloat[Avg Burst Power\label{subfig:power-wired}]{%
       \includegraphics[width=0.25\textwidth, height = 0.22\textwidth]{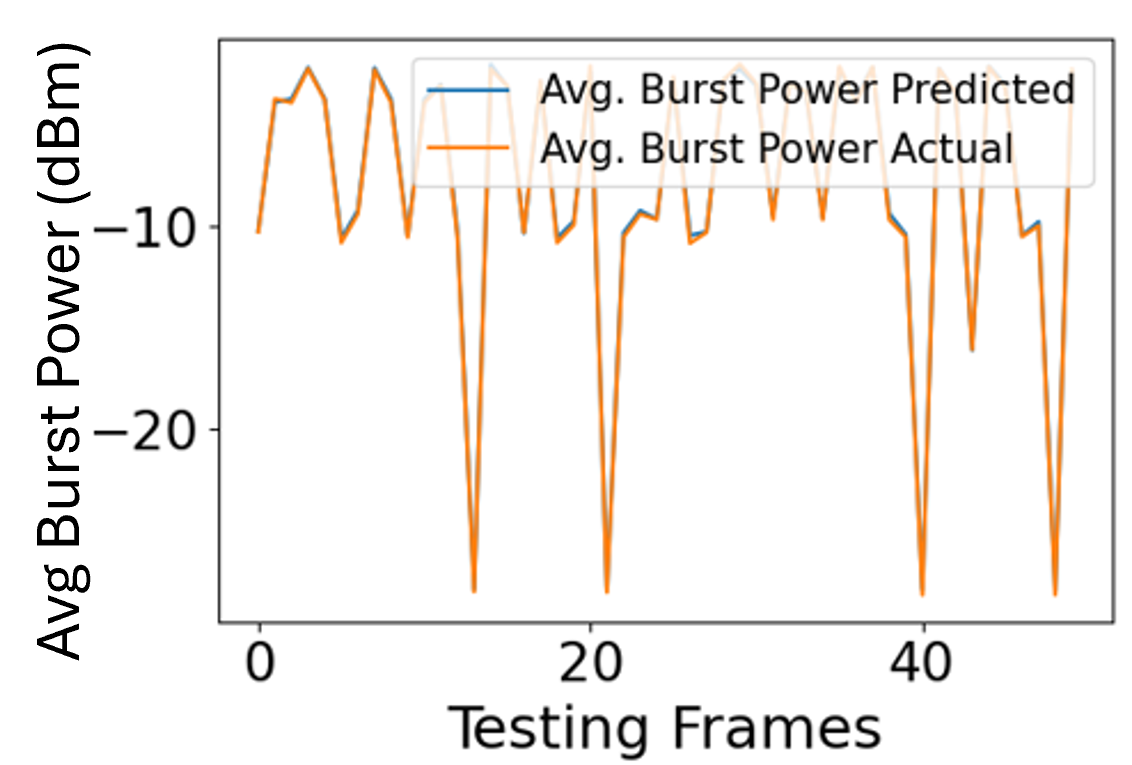}} 
\caption{Estimation of 8 impairments using \proposed's Impairment Estimator: Wired setup with 15 devices}
\label{fig:Imp_est_wired}
\end{figure*}



\begin{figure*}
 \setlength{\abovecaptionskip}{3pt}
    \subfloat[EVM \label{subfig:evm-wireless}]{%
       \includegraphics[width=0.25\textwidth, height = 0.21\textwidth]{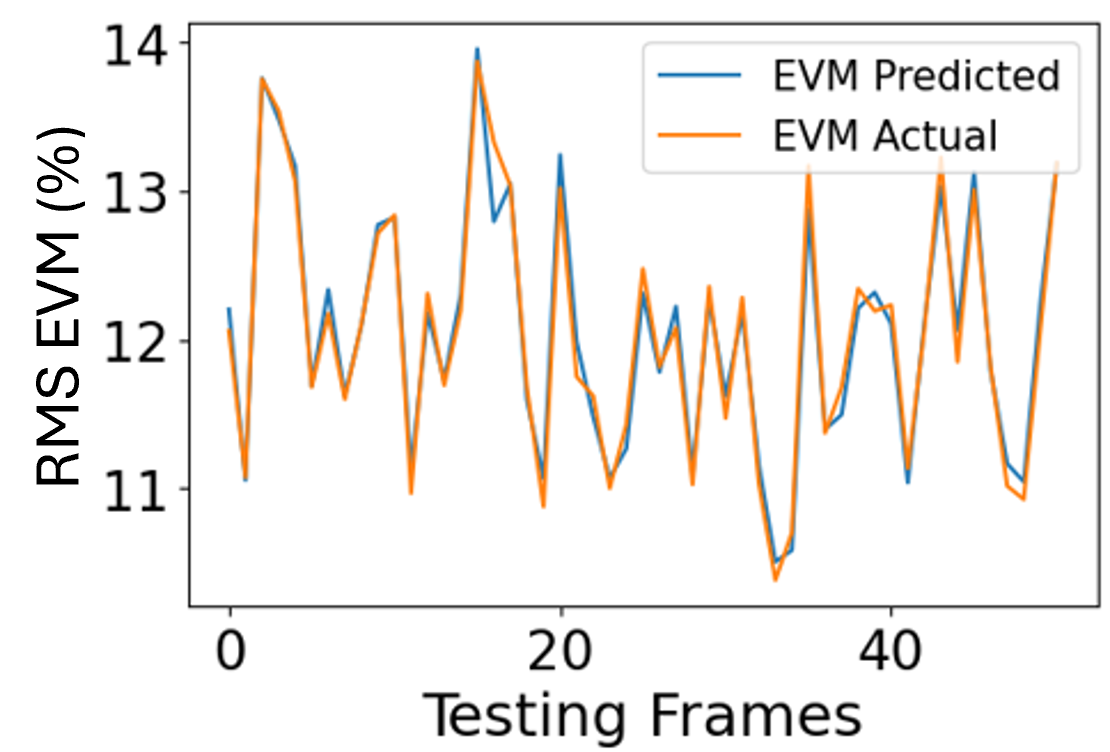}}
    \hspace{-0.2cm}
        \subfloat[CFO\label{subfig:cfo-wireless}]{%
       \includegraphics[width=0.25\textwidth, height = 0.23\textwidth]{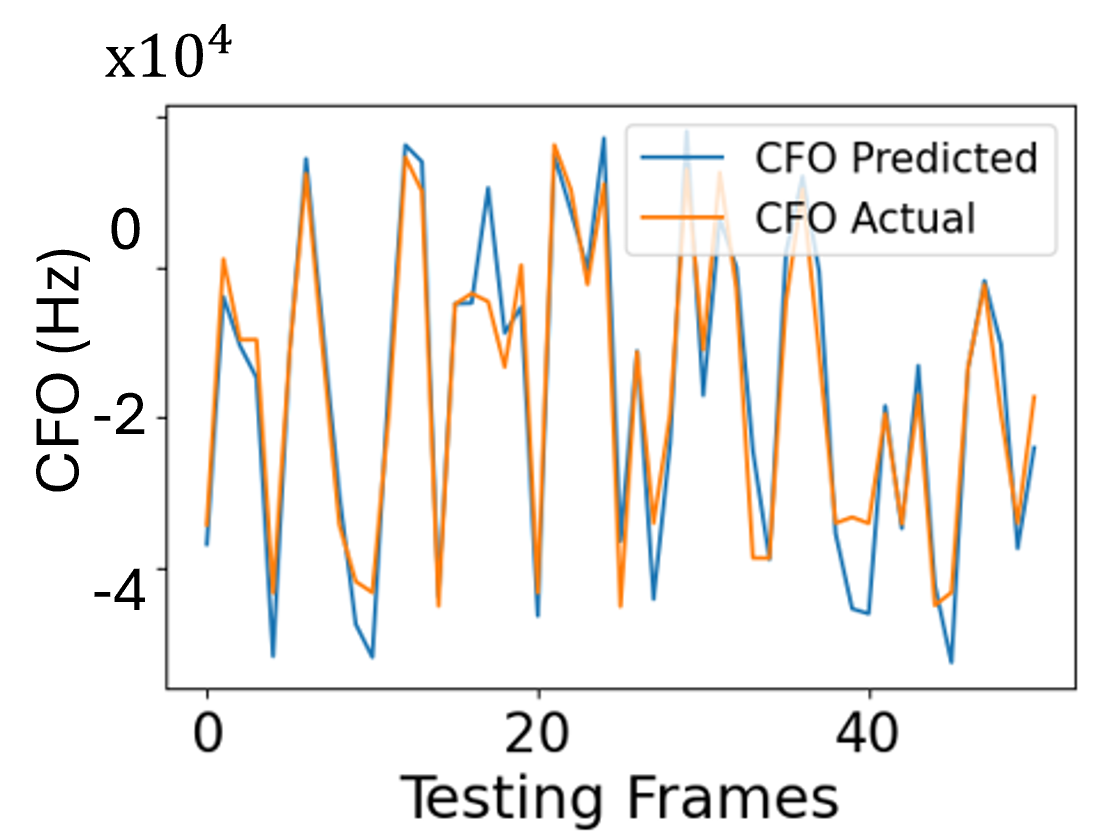}}
    \hspace{-0.2cm} 
        \subfloat[SCE\label{subfig:sce-wireless}]{%
       \includegraphics[width=0.25\textwidth, height = 0.21\textwidth]{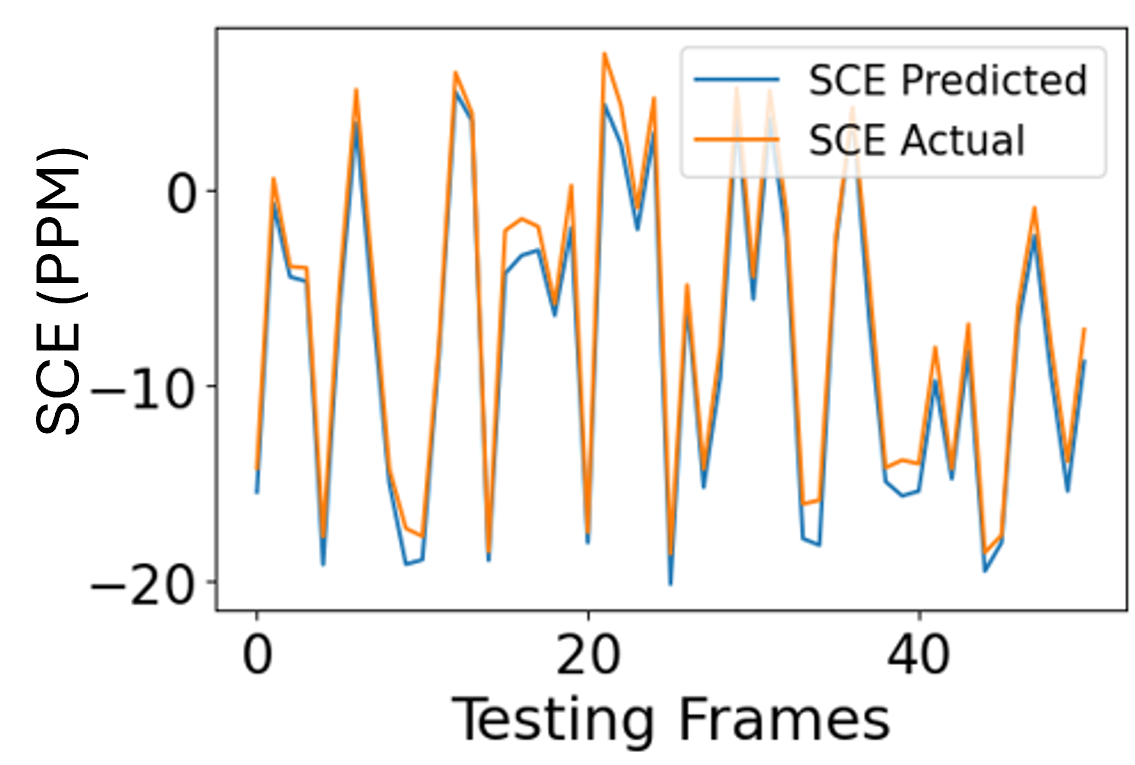}}
    \hspace{-0.1cm}
        \subfloat[IQ Offset\label{subfig:iq-offset-wireless}]{%
       \includegraphics[width=0.25\textwidth, height = 0.21\textwidth]{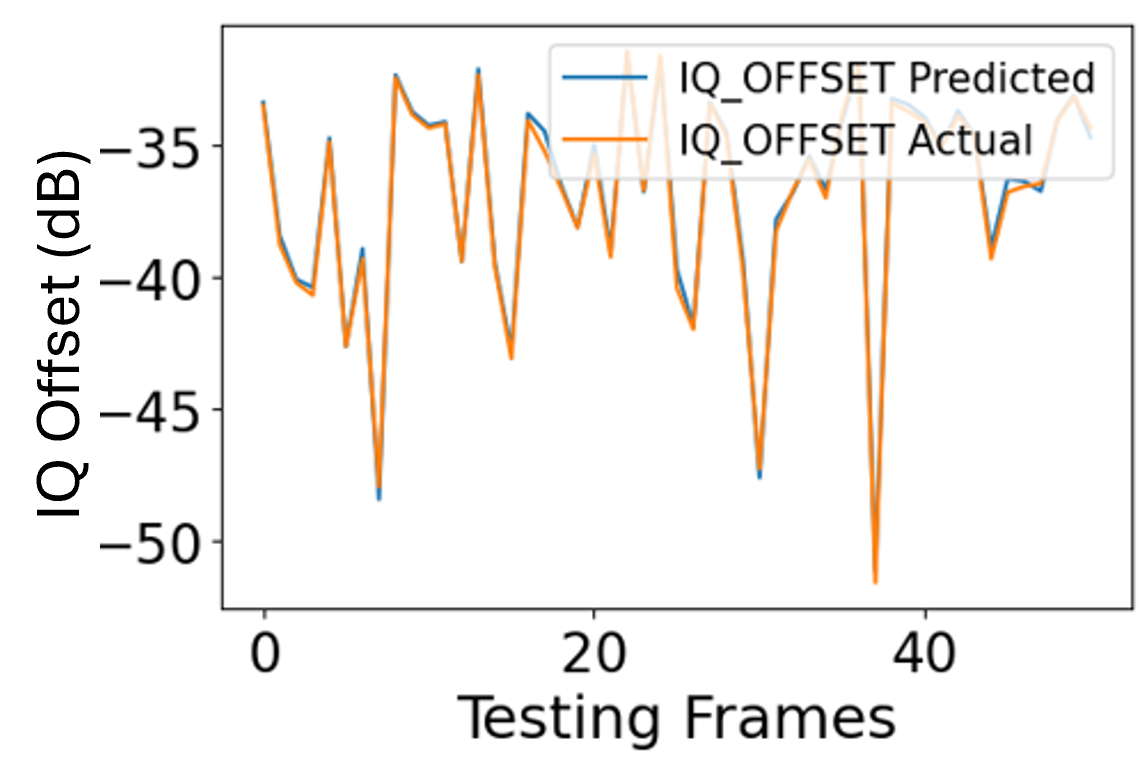}} 
\vspace{-0.1cm}
    \subfloat[Mag Error \label{subfig:mag-wireless}]{%
       \includegraphics[width=0.25\textwidth, height = 0.21\textwidth]{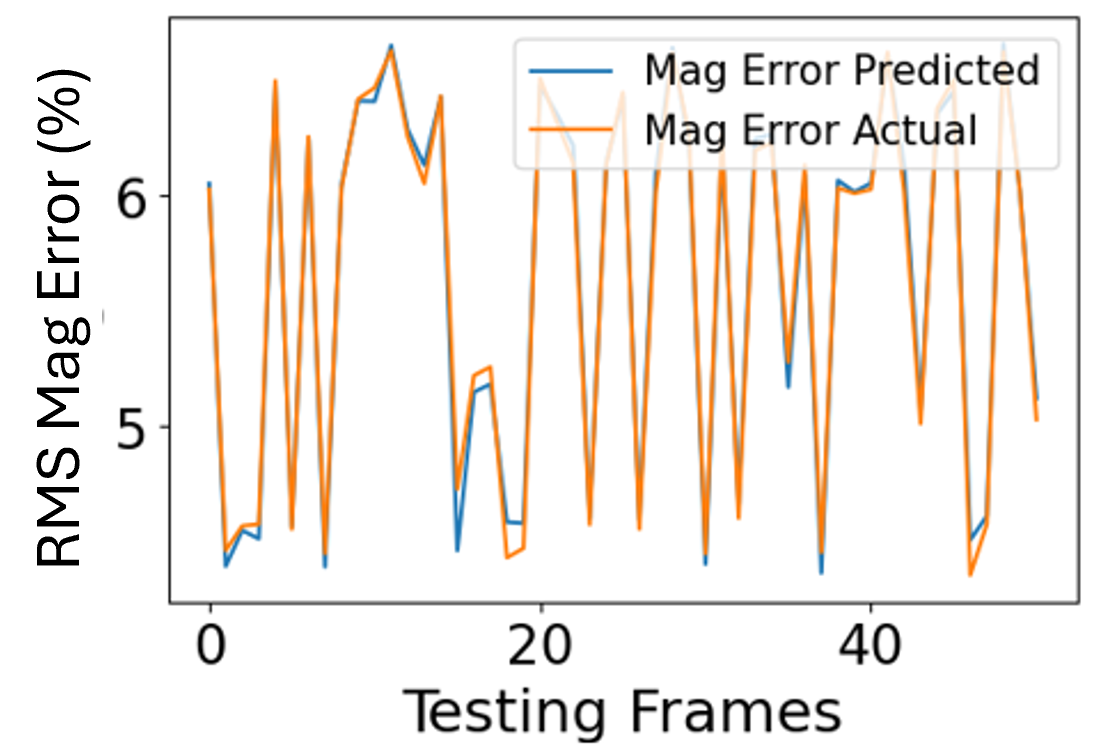}}
    \hspace{-0.2cm}
        \subfloat[Phase Error\label{subfig:phase-wireless}]{%
       \includegraphics[width=0.25\textwidth, height = 0.215\textwidth]{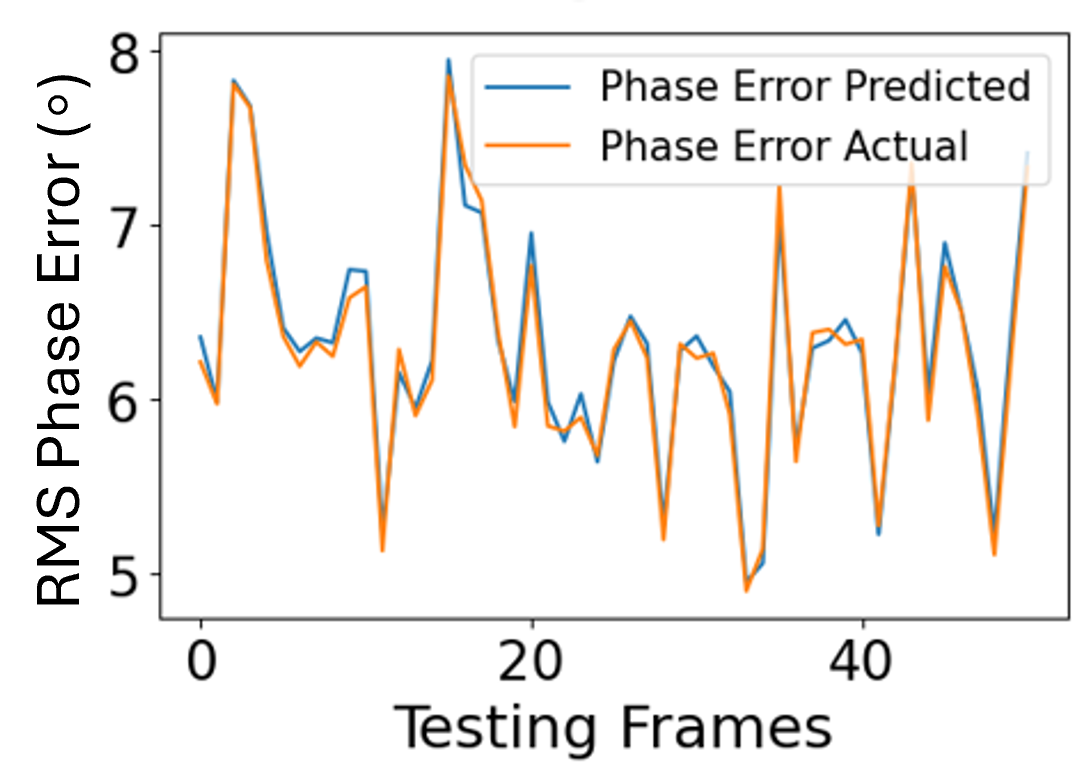}}
    \hspace{-0.2cm} 
        \subfloat[CSE\label{subfig:cse-wireless}]{%
       \includegraphics[width=0.25\textwidth, height = 0.21\textwidth]{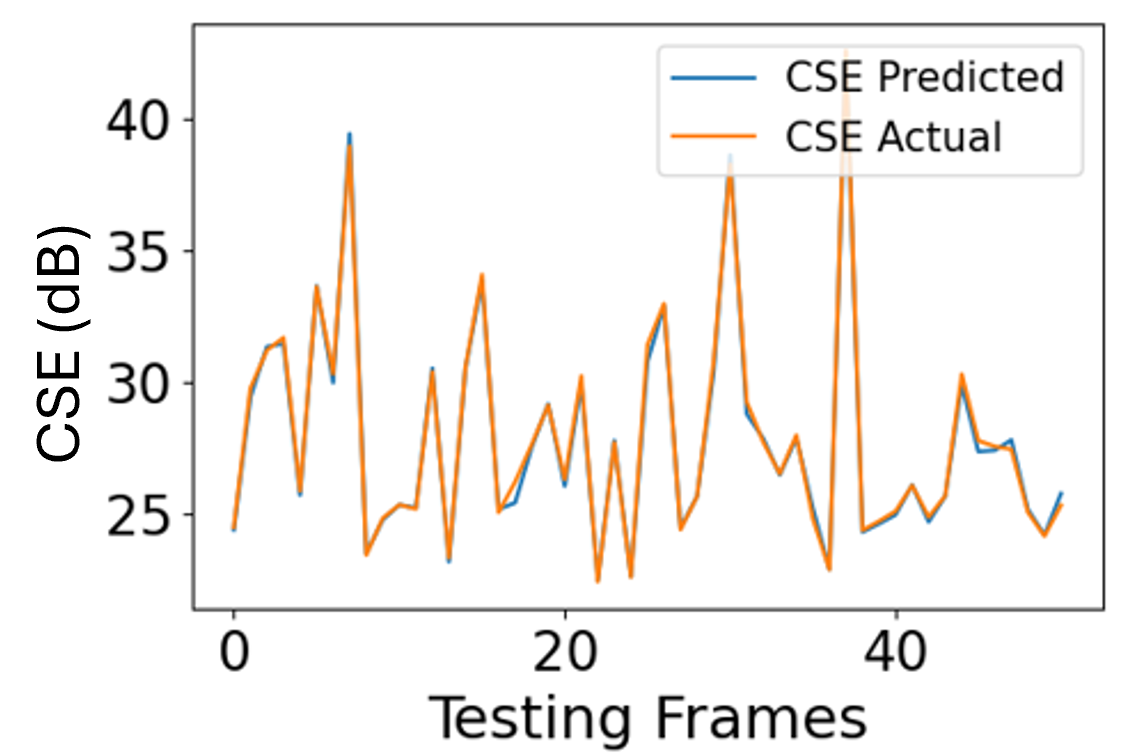}}
    \hspace{-0.1cm}
        \subfloat[Avg Burst Power\label{subfig:power-wireless}]{%
       \includegraphics[width=0.25\textwidth, height = 0.215\textwidth]{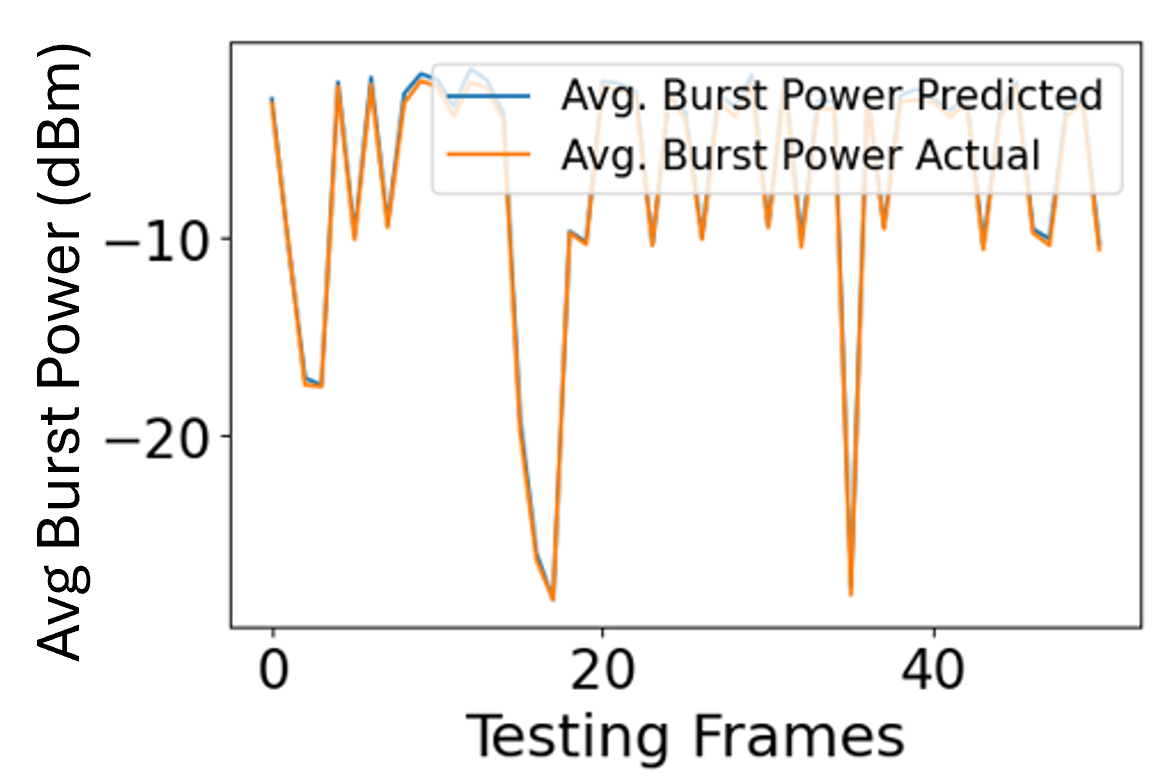}} 
\caption{Estimation of 8 impairments using \proposed's Impairment Estimator: Wireless setup with 15 devices}
\label{fig:Imp_est_wireless}
\end{figure*}


Our evaluation involved a comprehensive analysis of our impairment estimator model on WiFi 802.11b packets from 15 Pycom devices, collected during the initial 30 minutes of device operation, covering both the warm-up and stable phases. We assessed both wired and wireless scenarios to evaluate the model's performance with and without channel effects. In both scenarios, we estimated the following impairments: EVM, CFO, SCE, Mag Error, Phase Error, Carrier Suppression Error, and Avg Burst Power. To avoid biases, we normalized these impairments to fall between 0 and 1, given their significantly varying ranges. The input to our impairment estimator was the entirety of the WiFi packet with a dimension of [2x17550]. In the wired scenario, the average MAE for the concurrent estimation of the normalized eight distinct hardware impairments was 0.004. This accuracy is visually affirmed in Fig.~\ref{fig:Imp_est_wired}, where a perfect overlap between predicted and actual impairment measurements for 50 randomly selected testing frames highlights our network's capability to accurately map raw I/Q frames to their corresponding impairments. Furthermore, the model maintained its performance in the wireless setting, despite the additional challenge posed by channel impact. Fig.~\ref{fig:Imp_est_wireless} demonstrates this capability, showing predicted impairments closely matching the measured values for 50 random packets. This robust performance across both settings underscores the effectiveness of our impairment estimator model in handling diverse operational conditions, ensuring reliable and accurate impairment detection essential for robust RF fingerprinting.

\begin{figure*}
 \setlength{\abovecaptionskip}{3pt}
        \subfloat[After 2 Mins\label{subfig:2w}]{%
       \includegraphics[width=0.25\textwidth, height = 0.21\textwidth]{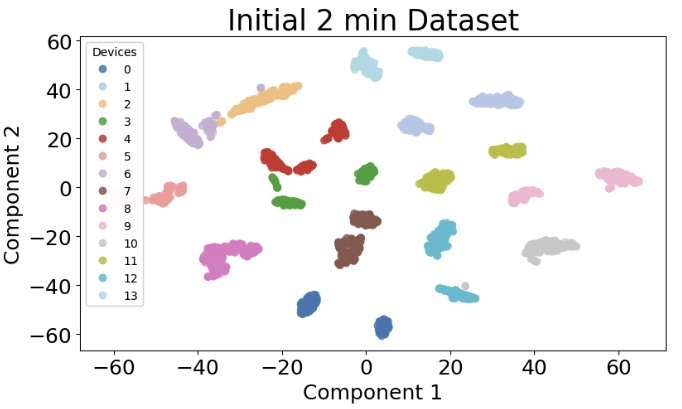}}
    \hspace{-0.2cm}
        \subfloat[After 4 Mins\label{subfig:4w}]{%
       \includegraphics[width=0.25\textwidth, height = 0.21\textwidth]{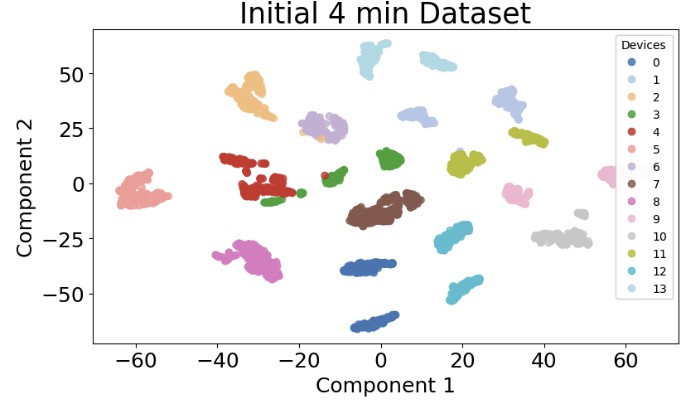}}
    \hspace{-0.1cm} 
        \subfloat[After 6 Mins\label{subfig:6w}]{%
       \includegraphics[width=0.25\textwidth, height = 0.215\textwidth]{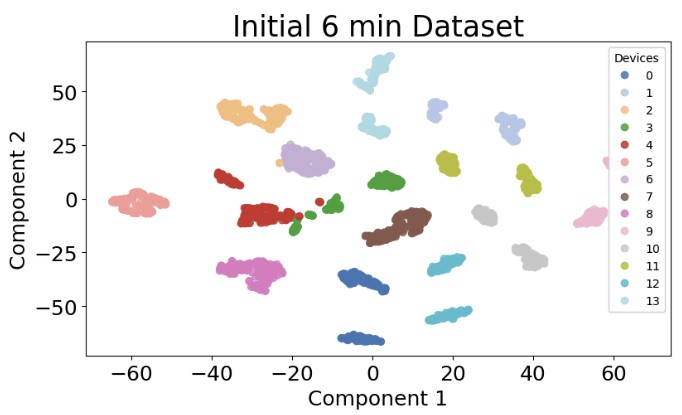}}
    \hspace{-0.1cm}
        \subfloat[After 12 Mins (stable)\label{subfig:sw}]{%
       \includegraphics[width=0.25\textwidth, height = 0.21\textwidth]{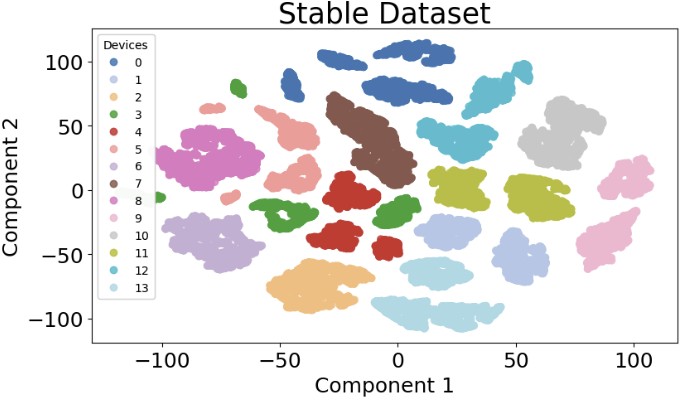}} 
\caption{t-SNE plot of the second-to-last layer of the impairment estimator: Wired scenario.}
\label{fig:tsne_wired}
\end{figure*}

\begin{figure*}
 \setlength{\abovecaptionskip}{3pt}
        \subfloat[After 2 Mins\label{subfig:2w}]{%
       \includegraphics[width=0.25\textwidth, height = 0.21\textwidth]{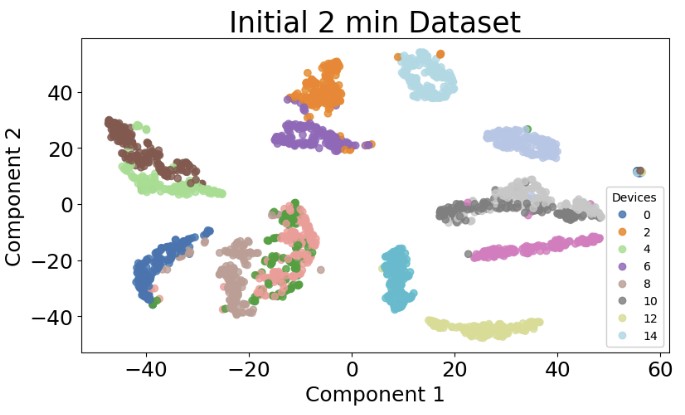}}
    \hspace{-0.2cm}
        \subfloat[After 4 Mins\label{subfig:4w}]{%
       \includegraphics[width=0.25\textwidth, height = 0.21\textwidth]{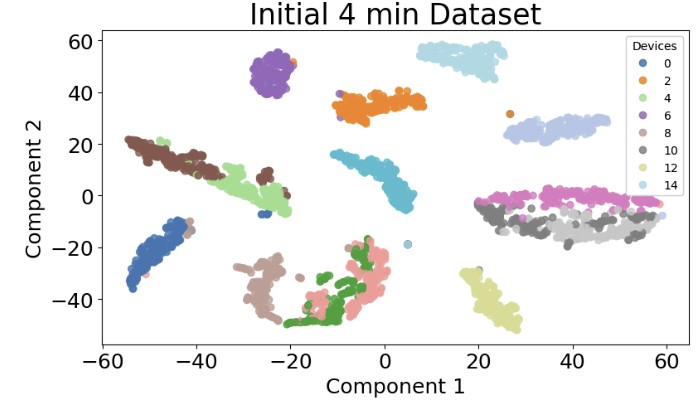}}
    \hspace{-0.1cm} 
        \subfloat[After 6 Mins\label{subfig:6w}]{%
       \includegraphics[width=0.25\textwidth, height = 0.215\textwidth]{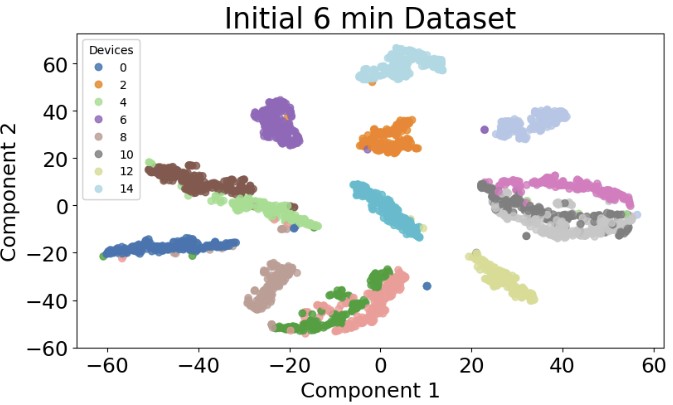}}
    \hspace{-0.1cm}
        \subfloat[After 12 Mins (stable)\label{subfig:sw}]{%
       \includegraphics[width=0.25\textwidth, height = 0.21\textwidth]{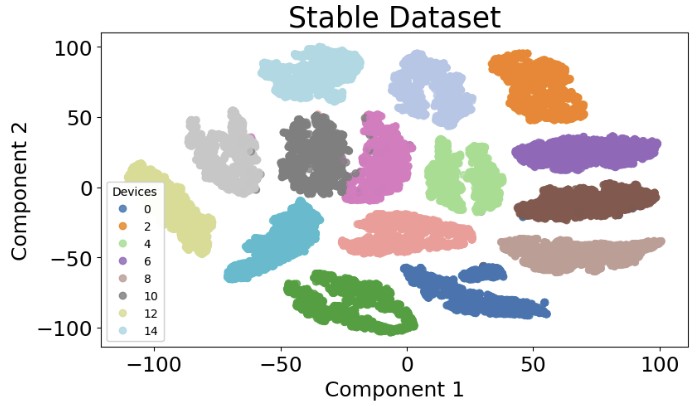}} 
\caption{t-SNE plot of the second-to-last layer of the impairment estimator: Wireless scenario.}
\label{fig:tsne_wireless}
\end{figure*}

\comment{
\begin{figure*}[t]  
    \centering
    \setlength{\abovecaptionskip}{3pt}
    \setlength{\abovecaptionskip}{3pt}
    \includegraphics[width=\textwidth]{Figures/Impairments/tsne_wired.png}  
    \caption{t-SNE plot of the second-to-last layer of the impairment estimator: Wired scenario.}
    \label{fig:tsne_wired}
\end{figure*}

\begin{figure*}[t]  
    \centering
    \setlength{\abovecaptionskip}{3pt}
    \setlength{\abovecaptionskip}{3pt}
    \includegraphics[width=\textwidth]{Figures/Impairments/tsne_wireless.png}  
    \caption{t-SNE plot of the second-to-last layer of the impairment estimator: Wireless scenario.}
    \label{fig:tsne_wireless}
\end{figure*}
}

To assess the capabilities of \proposed~ in leveraging convolutional blocks from the impairment estimator for robust device classification, we employed the t-distribution stochastic neighbor embedding (t-SNE) method \cite{van2008visualizing} to visualize the output of the feature extractor. This visualization focuses on the embeddings of WiFi packets to explore how the latent representations vary across devices and operational phases, particularly during the warm-up periods compared to stable conditions. The t-SNE plots, as illustrated in Fig.~\ref{fig:tsne_wired} for the wired scenario and Fig.~\ref{fig:tsne_wireless} for the wireless scenario, demonstrate distinct clusters for each device. These clusters are color-coded and each represents a different device, exhibiting a well-defined separation in the stable dataset. This clear differentiation, evidenced by the compact and non-overlapping nature of the clusters, underscores the feature extractor's accuracy and reliability in device identification during stable phases. During the initial minutes of operation, as shown in the subsequent plots for the 2-minute, 4-minute, and 6-minute datasets, the clusters maintain distinguishability but with slight overlaps in certain cases, such as between devices 2 and 6 or 3 and 4. Interestingly, the t-SNE plots showcase a consistent cluster map across the studied warm-up intervals but it looks like a reflection across a central axis when compared with the cluster map of the stable phase. Despite this transformation, the relative distances and relationships among the clusters are still maintained, indicating that the impairment estimator can effectively serve as a reliable base model for device classification across different operational phases.

In the wireless scenario, Fig.~\ref{fig:tsne_wireless}, the separability of the clusters during the warm-up phases is less pronounced, with more instances of overlapping. This suggests potential challenges in achieving the same high classification accuracy in wireless environments as in wired scenarios during device warm-up periods, due to the inherent variability and noise in wireless signal transmission.

Overall, these observations validate that the impairment estimation task is well-suited as a source task in sequential transfer learning applications aimed at enhancing device fingerprinting capabilities. This adaptation not only preserves device separability across phases but also accentuates the potential for \proposed~ to consistently deliver high-performance device classification, even in dynamically changing operational conditions.

\subsection{Performance Analysis of \proposed}

\subsubsection{Resiliency to Hardware Warm-Up}
In evaluating the robustness of \proposed~during device warm-up, we consider both wired and wireless scenarios, offering a comprehensive view of the system's performance. This dual-scenario analysis is crucial, as wired setups eliminate the channel effects present in wireless communications, allowing us to isolate hardware variations from environmental interference while those effects are incorporated in the wireless case. 

\begin{figure}[t!]
\setlength{\abovecaptionskip}{2pt}
\subfloat[Wireless Scenario]{
   \includegraphics[width=\columnwidth]{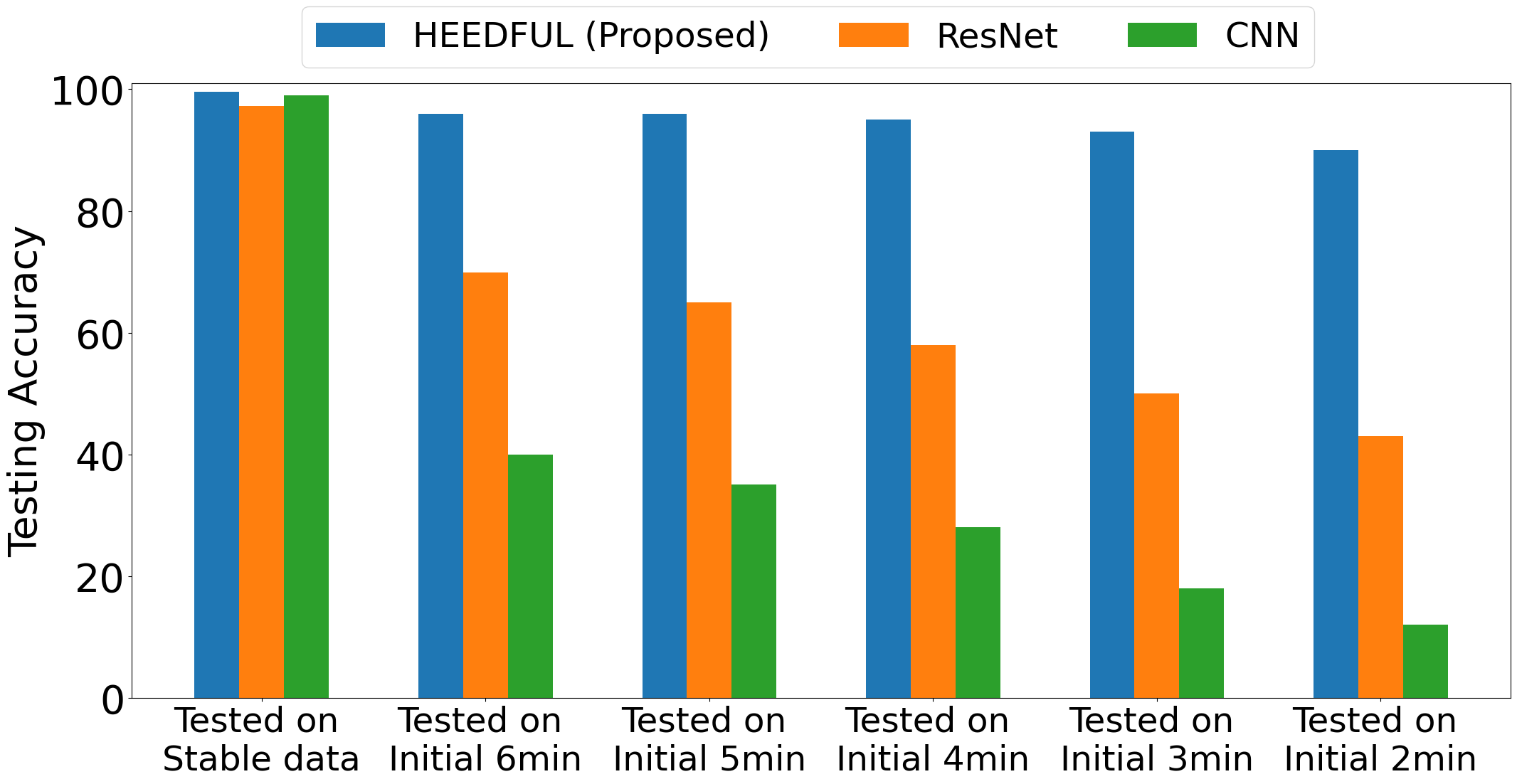}
   \label{RFImp_wireless}}
    \vfill
   \subfloat[Wired Scenario]{
   \includegraphics[width=\columnwidth]{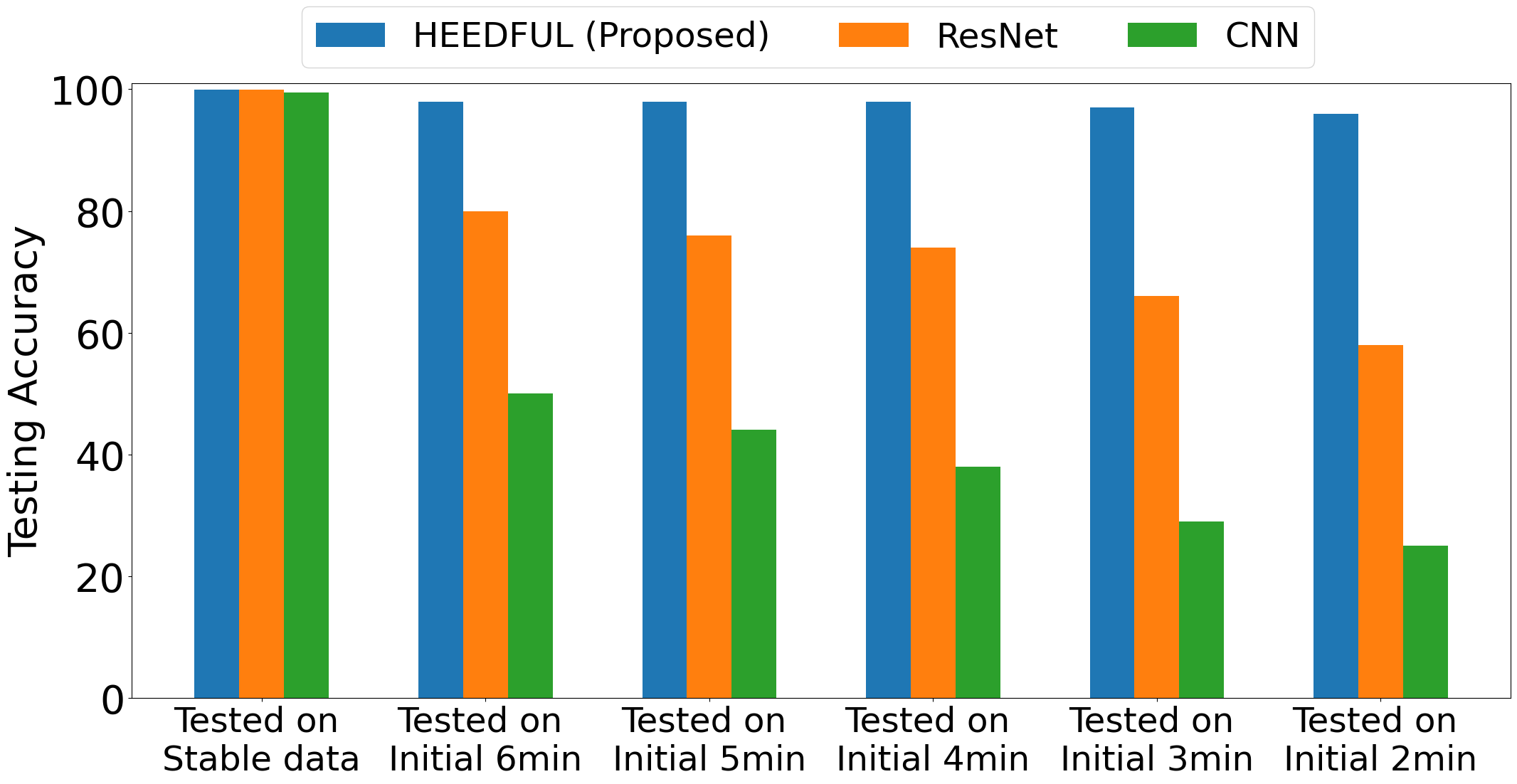}
   \label{RFImp_wired}}
    \\
   \caption{Accuracy results during warm up: All models are trained on Stable Day 1}
\label{results}
\end{figure}

For the wireless scenario, our evaluation reveals \proposed's superior resilience in handling hardware variations that occur during the initial warm-up phase. Despite the channel effects, \proposed~maintains high accuracy levels, significantly outperforming baseline models across all intervals.  As depicted in Fig.~\ref{RFImp_wireless}, all three models achieve near-perfect classification when tested with stable data. More interestingly, when tested with packets transmitted within the initial 6 minutes of device operation, \proposed~achieves an accuracy of 96\%, compared to 70\% and 40\% for ResNet and CNN, respectively. This trend continues even in the most challenging intervals, with \proposed~attaining 90\% accuracy at the 2-minute mark, where CNN and ResNet struggle with accuracies of 12\% and 43\%, respectively.

On the other hand, in the wired scenario (Fig.~\ref{RFImp_wired}), where channel effects are absent, all models show improved performance, consistent with the previously discussed t-SNE plots, with \proposed~still leading significantly. In the stable phase, the three frameworks achieve nearly perfect classification accuracy here as well. Within the initial 6 minutes of device operation, \proposed~achieves an accuracy of 97\%, compared to 65\% and 79\% for ResNet and CNN, respectively, as shown in Fig.~\ref{RFImp_wired}. \proposed~maintains excellent classification at the outset, with a slight decrease to 91\% accuracy in the most unstable 2-minute interval, demonstrating an ability to handle rapid hardware stabilization variations. In contrast, CNN and ResNet exhibit a steeper decline in performance, with accuracy dropping to 52\% and 46\% respectively in the same interval.

%

\subsubsection{Robustness to Cross-Day Scenario} 
\begin{figure}
    \centering
    \setlength{\abovecaptionskip}{2pt}
    
{%
        \includegraphics[width=\columnwidth]{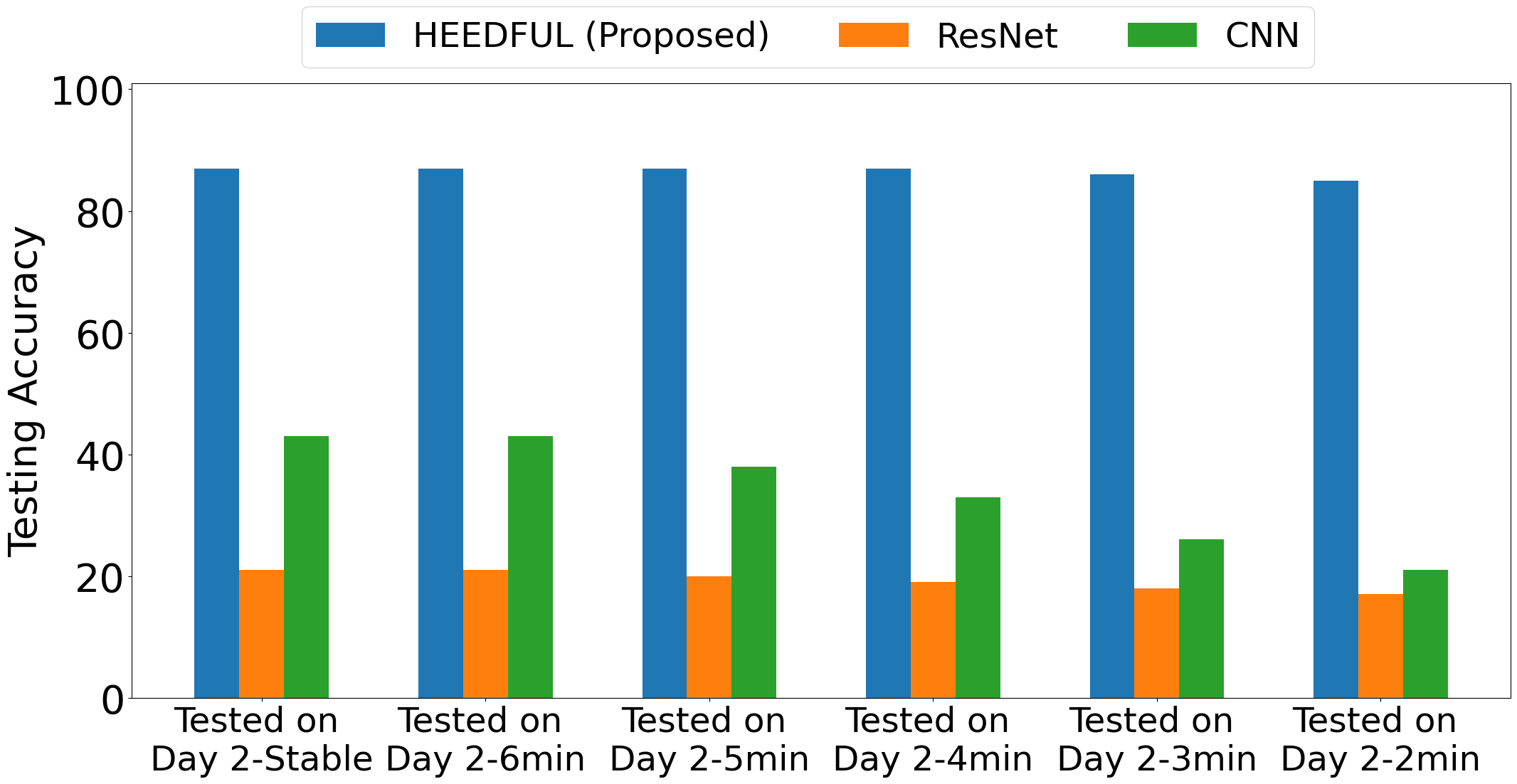}%
    }
    \caption{Accuracy results of wireless cross-day scenario: All models are trained on the Stable Day 1 data and tested on Day 2 dataset, which was collected two weeks later.}
    \label{fig:cross-day}
\end{figure}

We now extend our evaluation and assess how well the models perform when tested on frames captured at Day 2, two weeks later from Day 1. Our results, depicted in Fig.~\ref{fig:cross-day}, show that \proposed~consistently maintains a high classification accuracy of 87\% across the warm-up intervals even when tested on Day 2 dataset. In contrast, ResNet struggles to attain a classification accuracy above 20\%, while CNN's highest classification accuracy is 43\% within the initial 6-minute interval, dropping to as low as 21\% in 2-minute intervals. The underwhelming performance of the CNN and ResNet benchmarks underscores the challenges of learning RF fingerprinting in the complex RF environment without the guidance of domain knowledge. And thus, the consistent high performance of \proposed~represents then a significant advancement toward achieving robust device identification solutions.
\subsubsection{Sensitivity to Selected Impairments}

\begin{figure}
\subfloat[Multiple Impairments Evaluation]{
   \includegraphics[width=\columnwidth]{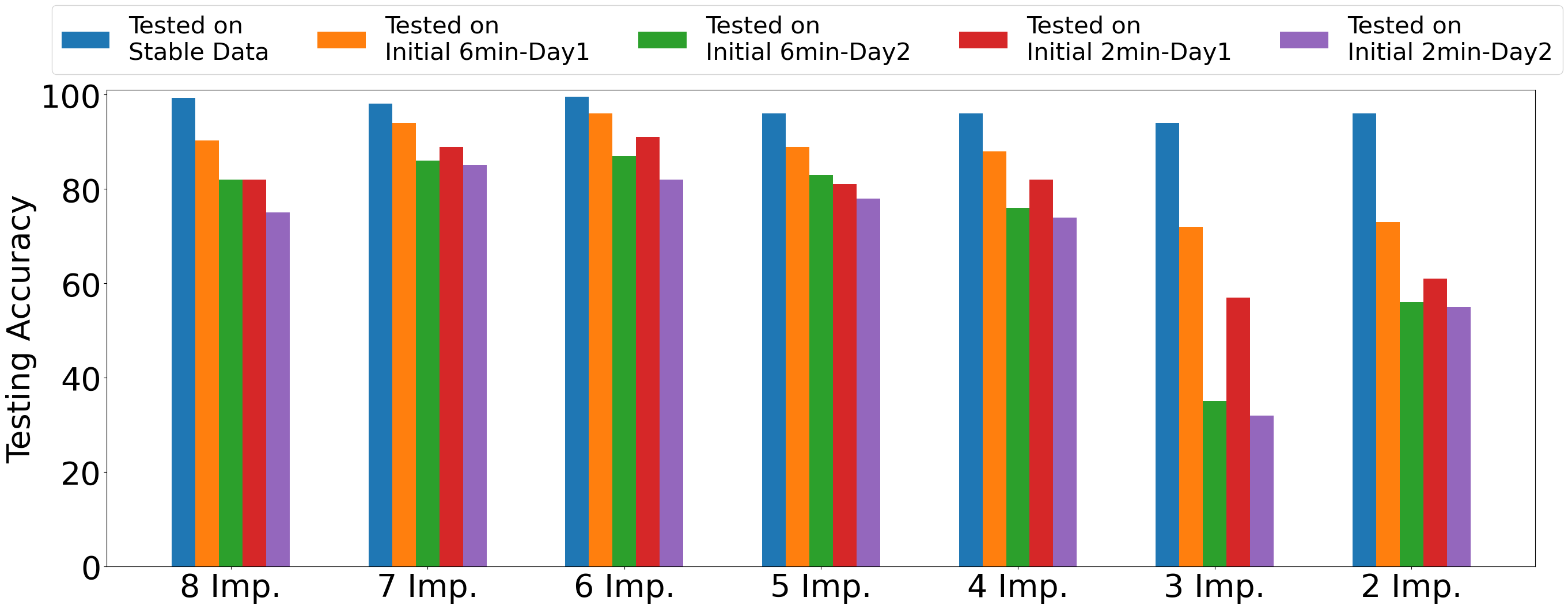}
   \label{imp_impact_multi}}
    \hfill
\subfloat[Single Impairment Evaluation]{
   \includegraphics[width=\columnwidth]{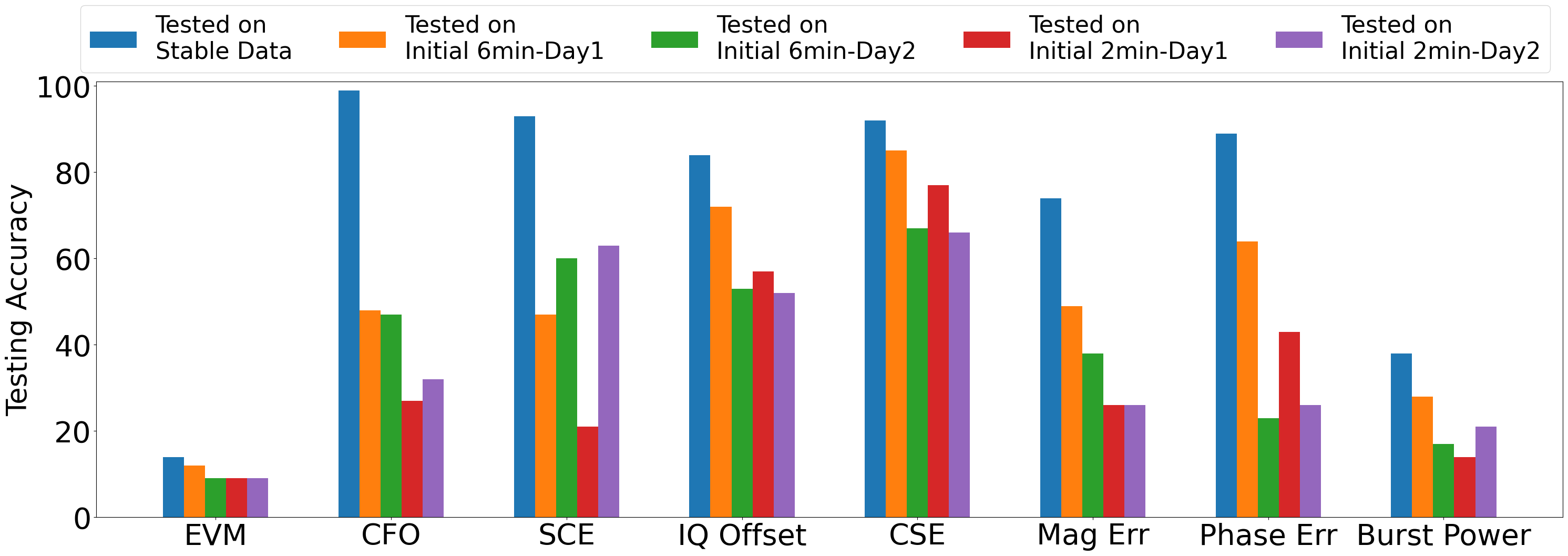}
   \label{imp_impact_ind}}
    \\
   \caption{Impact of impairment selection: all models are trained on wireless Stable Day 1 data}
\label{impact_results}
\end{figure}

One notable feature of \proposed~lies in its comprehensive grasp of hardware impairments, seamlessly integrated into the device classification head. Consequently, it becomes pivotal to examine the impact of the number and specific choice of impairments used to train the impairment estimator model in the source task. Our experiments involved training the impairment estimator on various sets of impairments, detailed in Table \ref{tab:est}, and assessing their performance in stable and warm-up intervals across the two datasets. Illustrated in Fig. \ref{impact_results}, \proposed~exhibits consistent accuracy with different-impairments-based estimators in post-stabilization captures, while variations emerged in the warm-up intervals. The multi-impairments evaluation, as depicted in Fig. \ref{imp_impact_multi}, highlights superior performance with the 8, 7, and 6 impairments-based systems for \proposed, while the 3-imp and 2-imp-based systems struggled the most in the warm-up intervals. This underscores the advantage of providing a comprehensive knowledge of impairments to the classifier head. However, the impairments' selection is crucial as well. Systems based on 7 and 6 impairments outperformed the 8-imp-based system, attributed to the inclusion of avg. burst power in the impairment list, which is different across datasets due to varied power sources. Additionally, the 3-imp-based system fared worse than the 2-imp-based system within warm-up intervals, attributed to two of the three impairments, CFO and SCE, displaying significant variance between stable and warm-up intervals. Further delving into individual impairment behavior, we evaluated \proposed~with the impairment estimator trained on each impairment separately. The reported testing accuracies, plotted in Fig.~\ref{imp_impact_ind}, reaffirmed the observed impairment behaviors across the two datasets, shown in Fig.~\ref{fig:wireless_dataset}. CFO-based and CSE-based systems displayed the highest accuracy in the stable phase but performed poorly in warm-up intervals, both on the same and different days. Reflecting their studied behavior, IQ-offset-based and CSE-based systems showed less accurate classification in the stable interval but exhibited more stable performance in warm-up intervals. EVM and Burst power-based systems demonstrated the poorest performance among impairments.

\begin{table}
\setlength{\abovecaptionskip}{4pt}
  \caption{Mean Absolute Error (MAE) when varying the number of estimated impairments: Wired setting.} 
  \label{tab:est}
 \scriptsize
  \begin{tabular}{p{0.2cm}p{6.9cm}p{.4cm}}
    \toprule
     \# Imp. & \centering List of the Estimated Impairments & MAE \\
     
    \midrule
    8 &  \centering EVM, CFO, SCE, IQ Offset, Mag Err,  CSE, Phase Err, Burst Power & 0.004 \\
    7 &  \centering EVM, CFO, SCE, IQ Offset, Mag Err, CSE, Phase Err & 0.004 \\
    6 & \centering EVM, CFO, SCE, IQ Offset, Mag Err, CSE & 0.004  \\
    5 & \centering EVM, CFO, SCE, IQ Offset, CSE & 0.004  \\
     4 & \centering EVM, CFO, IQ Offset, CSE & 0.004  \\
     3 & \centering CFO, IQ Offset, CSE & 0.007  \\
     2 & \centering CFO, IQ Offset & 0.007  \\
     1 & \centering CFO & 0.0008 \\
    \bottomrule
  \end{tabular}
\end{table}

\subsubsection{{Sensitivity of Fingerprinting Accuracy to Impairment Estimation}}
While the design of \proposed~is centered around the idea that robust impairment estimation serves as a foundational step toward resilient RF fingerprinting, it is important to explicitly analyze how variations in estimation quality impact downstream classification accuracy. Our observations consistently indicate a positive correlation between the fidelity of impairment estimation and the accuracy of device identification---particularly during the warm-up phase. We do not present additional confusion matrices or regression-conditioned plots in this context, as this relationship is already implicitly captured across multiple experiments. For instance, the baseline models, which can be interpreted as operating with no impairment awareness (i.e., estimation accuracy $\approx$ 0), consistently underperform across warm-up intervals. In contrast, \proposed, which leverages detailed impairment estimation as an intermediate task, exhibits significantly higher classification accuracy in these same intervals---up to 96\% vs. 70\% and 40\% for ResNet and CNN, respectively, in the wireless scenario (Fig.~\ref{RFImp_wireless}). This trend is reinforced when comparing the wired and wireless scenarios. As discussed earlier, the wired setup eliminates channel-related noise, allowing for more precise impairment estimation. Accordingly, \proposed~achieves slightly higher classification accuracies in the wired case (e.g., 97\% vs. 96\% in the stable phase), with even more pronounced improvements during early warm-up intervals (91\% vs. 90\% at 2 minutes). These differences are echoed in the t-SNE embeddings, where better impairment estimation (as in the wired case) results in more well-separated device clusters, indicating that the fingerprinting model has learned more discriminative representations. Together, these findings validate the underlying hypothesis of \proposed: the more accurately a model can infer the latent hardware impairments, the better it can learn stable, device-specific features for classification. This insight is not only theoretically intuitive but also empirically supported across our evaluations. Rather than adding more figures, we make this relationship explicit in our narrative to highlight that impairment estimation quality is not just a component of our pipeline—it is the key enabler of its robustness.


\subsection{Cross-Protocol Adaptation Capability}
\begin{figure}[t!]
\setlength{\abovecaptionskip}{2pt}
\subfloat[Wired Scenario]{
   \includegraphics[width=\columnwidth]{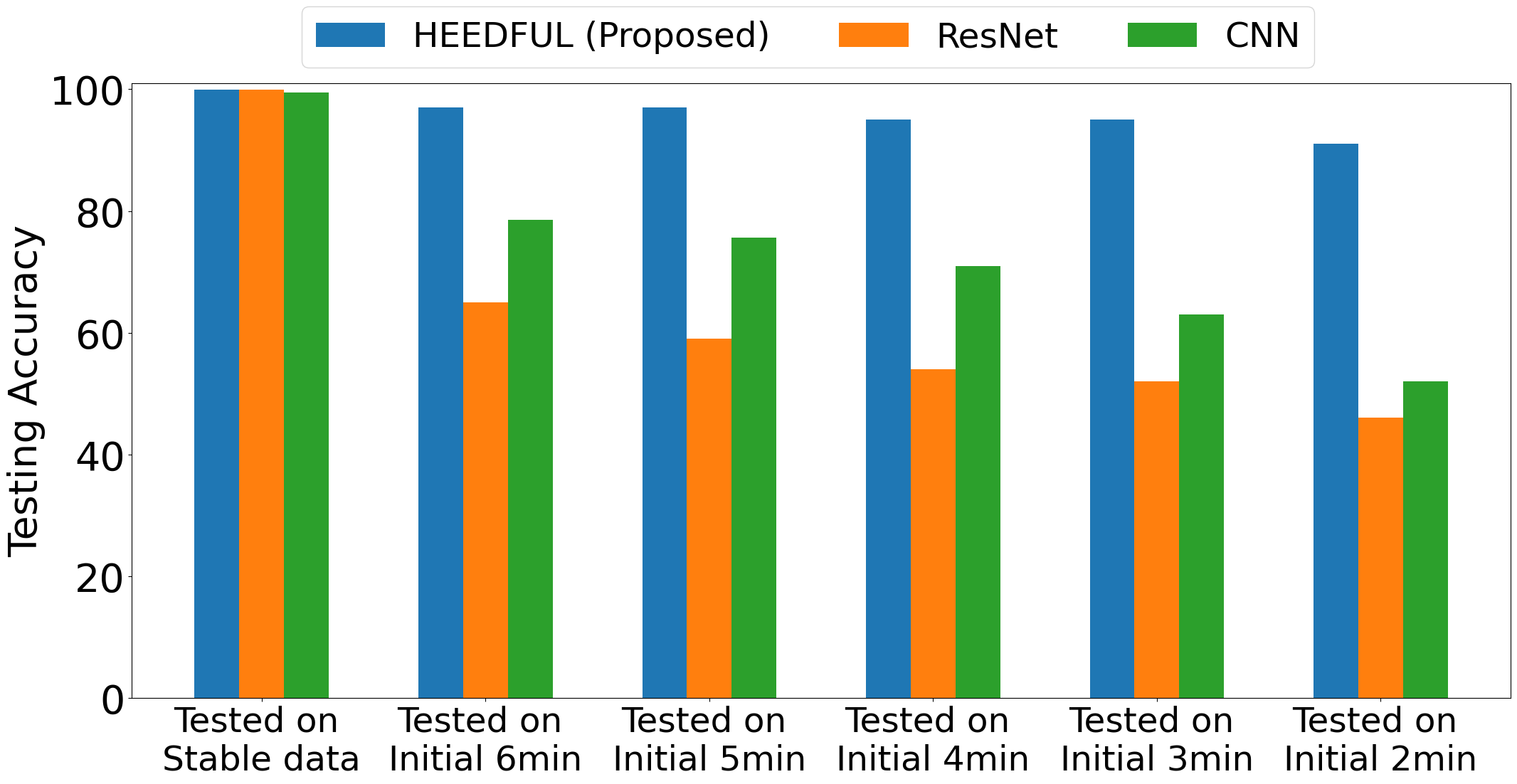}
   \label{wired_N}}
    \vfill
   \subfloat[Wireless Scenario]{
   \includegraphics[width=\columnwidth]{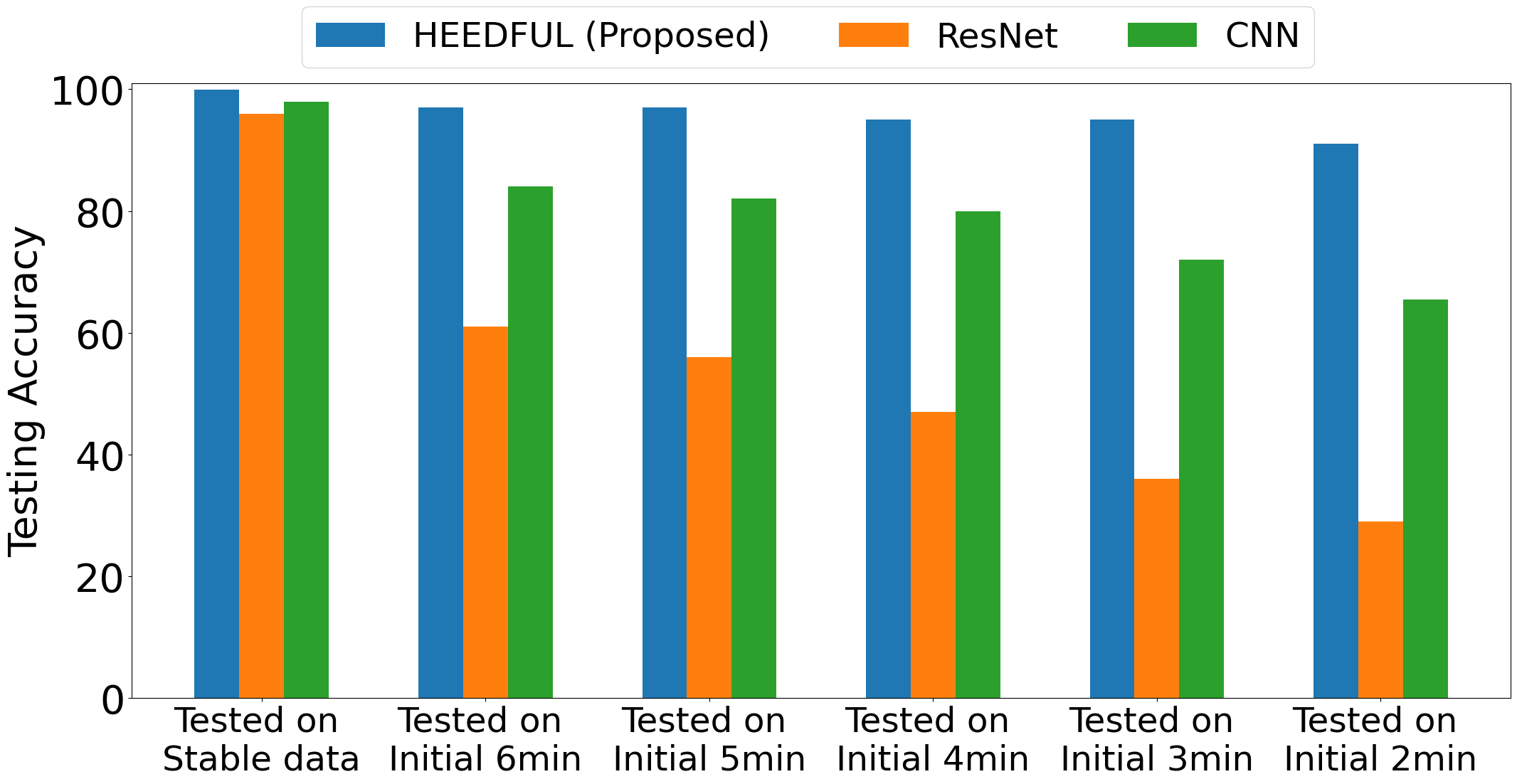}
   \label{wireless_N}}
    \\
   \caption{Accuracy results: All models are trained on Stable data of WiFi type N standard.}
\label{N_results}
\end{figure}
Devices often support multiple IEEE 802.11 standards, such as 802.11b (referring to as WiFi B), 802.11g (WiFi G), and 802.11n (WiFi N), and have the capability to switch between these protocols based on network configurations and environmental factors. This backward compatibility enhances network flexibility but introduces complexities for RF fingerprinting systems. The ability of devices to seamlessly transition among these standards means they may present varying signal characteristics influenced by the specific properties of each protocol, including differences in modulation schemes, data rates, and channel utilization. For instance, devices like the Pycom boards used in our testbed can adjust their communication protocol to optimize transmission speed or adapt to network conditions. Such adaptability, while advantageous for network efficiency and performance, poses significant challenges for RF fingerprinting techniques. A robust RF fingerprinting system must therefore be versatile enough to accurately identify devices regardless of the active communication standard. It must efficiently parse and interpret the unique RF signatures associated with each standard, which can range from 11 Mbps in WiFi B up to 600 Mbps in WiFi N. This capability is crucial for maintaining reliable device authentication and ensuring security across a dynamically changing RF environment.

We begin by evaluating \proposed's performance on the same testbed, this time using the WiFi N standard for both wired and wireless settings. WiFi N operates at a higher transmission rate compared to WiFi B, which means that the same message will be transmitted more quickly. Consequently, when captured at the same sampling rate, the packet size for WiFi N, in terms of I/Q samples, is [2x1012], compared to the larger packet size for WiFi B. Therefore, we modified the input layer of the impairment estimator to accommodate the different input shape of WiFi N packets. The impairment estimator is trained to estimate the following set of impairments: EVM, CFO, SCE, I/Q imbalance, Quadrature error, Pilot EVM, and I/Q timing skew. Then, the whole framework is trained on stable WiFi N data and tested with packets coming from the different operational phases. As depicted in Fig.~\ref{N_results}, \proposed~consistently outperforms both CNN and ResNet models across all testing intervals, demonstrating its robust adaptability to different WiFi standards. For example, in the wireless setting, \proposed~achieves a testing accuracy of 97\% during the initial 6 minutes and 91\% during the most challenging 2-minute interval. Similarly, in the wired setting, \proposed~maintains a high accuracy of 97\% at the 6-minute mark and 91\% at the 2-minute mark. Note that in this wireless scenario, we had to downsize our testbed from 15 to 9 devices. This adjustment became necessary due to the unavailability of the required instrumentation both during and after the data collection period. These results indicate \proposed's capability to handle the various operational modes of these devices effectively, ensuring accurate device identification regardless of the communication standard in use.

\begin{figure}
    \centering
    \includegraphics[width=\columnwidth]{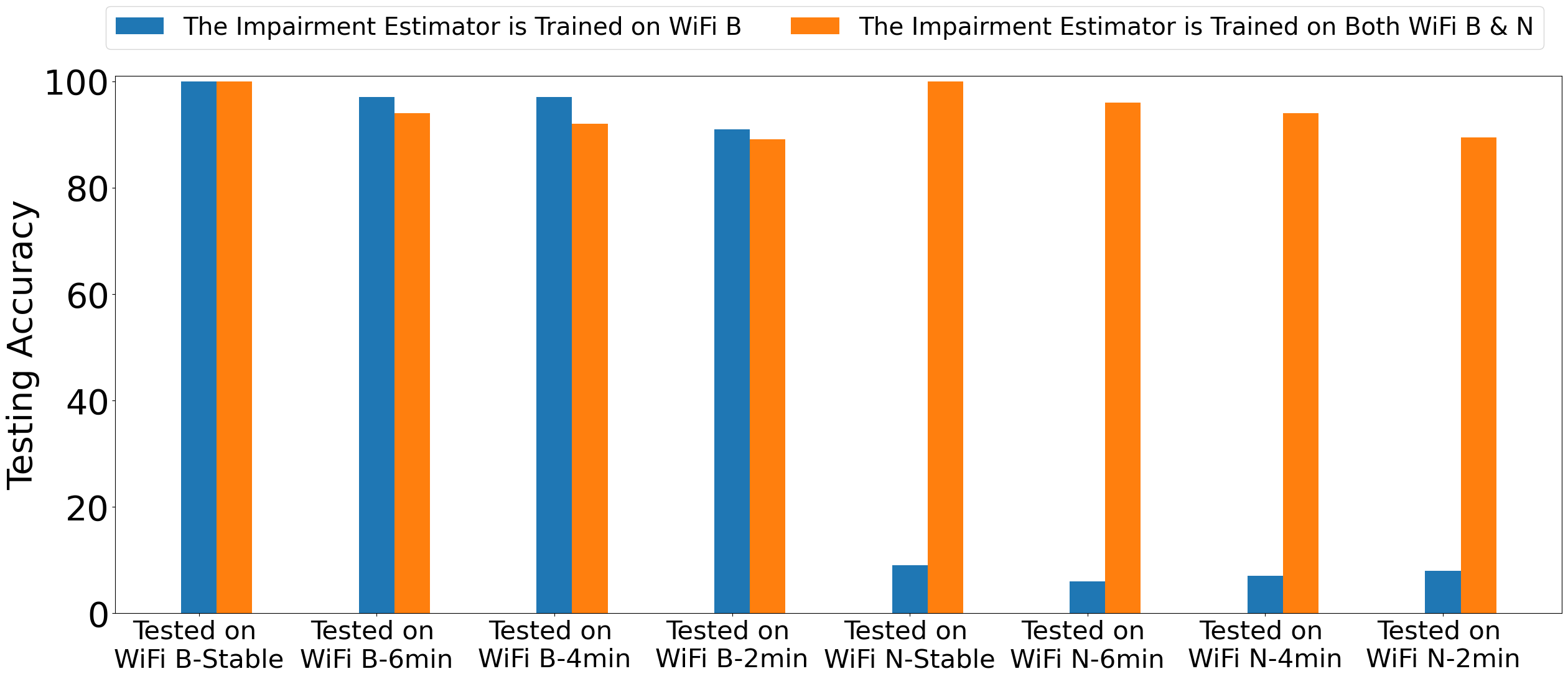} 
    \caption{Accuracy results of the wired cross-protocol scenario: the two models trained on WiFi-B Stable data and tested on both WiFi-B \& N datasets.}
    \label{fig:cross_protocol_eval}
\end{figure}

\begin{figure}
\setlength{\abovecaptionskip}{2pt}
\subfloat[Time-Domain I/Q signal from the same device]{
   \includegraphics[width=\columnwidth]{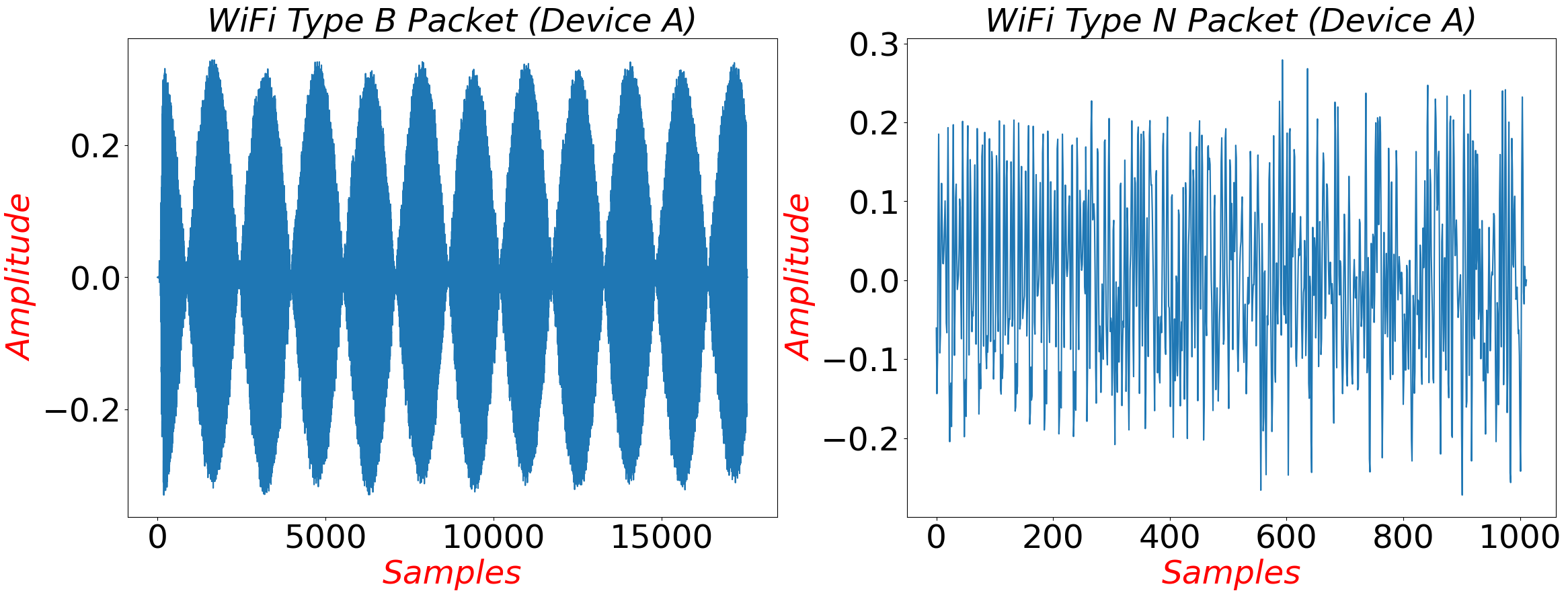}
   \label{IQ_B_N}}
    \vfill
   \subfloat[Hardware impairment behavior of the same device]{
   \includegraphics[width=\columnwidth]{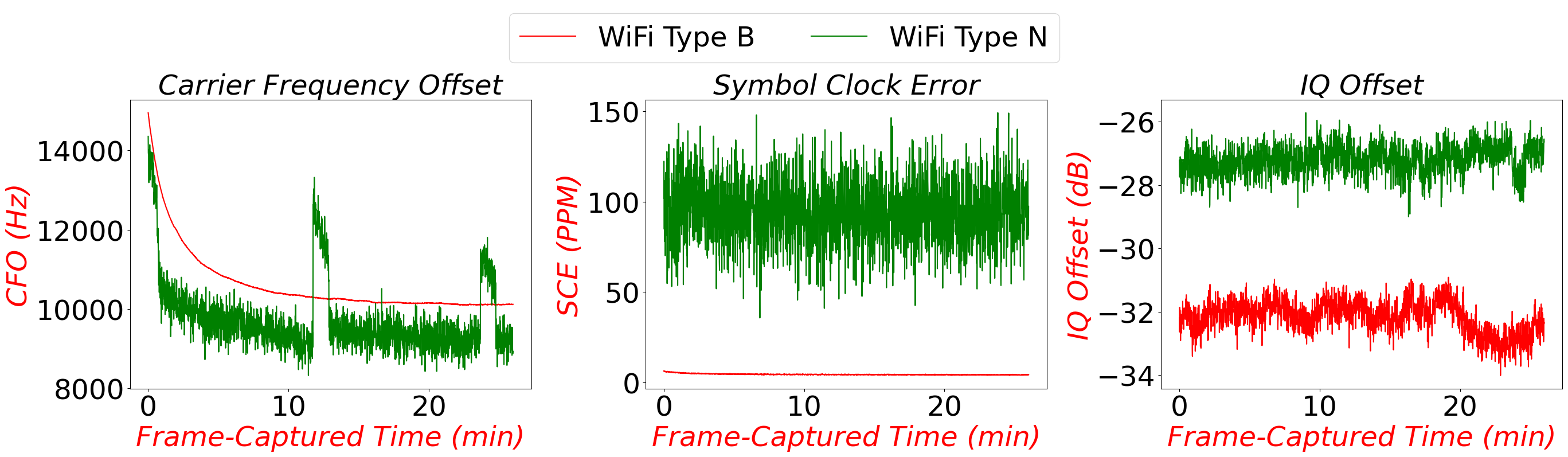}
   \label{Imp_B_N}}
    \\
   \caption{Comparison between WiFi B \& WiFi N packets of the same Pycom device: wired setup.}
\label{protocol_comp}
\end{figure}

We now evaluate \proposed's robustness when the impairment estimator is trained exclusively on WiFi type B data but tested on WiFi type N data. To accommodate the different packet sizes of the two WiFi standards, we made the input layer of the impairment estimator length versatile and replaced the Flatten layer before the fully-connected layer with a global average pooling layer. This modification ensures that \proposed~can handle the varying lengths of packets from different WiFi standards. As depicted in Fig. \ref{fig:cross_protocol_eval}, \proposed~trained on WiFi B achieves high testing accuracies of 97\% and 91\% during the stable phase and initial 6-minute interval, respectively, of the WiFi type B datasets. However, the accuracy significantly drops to 9\% - 6\% when tested on the WiFi N datasets. This performance decline highlights the challenge posed by the different signal characteristics between WiFi B and WiFi N. For instance, as illustrated in Fig.~\ref{IQ_B_N}, the I/Q representation of the same message from the same device varies drastically between the two standards. This variance makes it difficult for a system trained exclusively on one standard to recognize the same device transmitting using another standard. Interestingly, the impairments of the same device also differ between the two standards, as shown in Fig.~\ref{Imp_B_N}. This figure presents the behavior of three key impairments—--CFO, Symbol Carrier Error, and IQ offset—--of the same device transmitting WiFi type B and N packets over the first 30 minutes. These differences in impairment characteristics further complicate the task for the system trained solely on WiFi type B data. In contrast, when the impairment estimator is trained in both WiFi B and N, \proposed~maintains excellent performance on the WiFi type N datasets, achieving 100\% and 96\% accuracy during the stable phase and initial 6-minute interval, respectively, and 94\% and 89\% during challenging 4-minute and 2-minute intervals. This suggests that training on a diverse set of protocols enables the system to generalize better across different WiFi standards, achieving high testing accuracies even when devices switch between them. The observed performance improvements can be attributed to the comprehensive learning of signal characteristics and impairments across multiple standards. By exposing the model to the variations inherent in both WiFi B and N during training, \proposed~becomes adept at identifying the unique RF fingerprints of devices, regardless of the protocol in use. This ability to handle cross-protocol transitions is crucial for ensuring reliable device identification and authentication in dynamic network environments, where devices frequently switch between different WiFi standards.

We want to mention that the selection of WiFi B and N for evaluating \proposed's cross-protocol adaptability was primarily driven by the hardware constraints of the Pycom devices used in our experiments, which natively support only WiFi B (DSSS) and WiFi N (OFDM) transmission modes without requiring hardware or firmware modifications. 
Nevertheless, we emphasize that the objective of our study was to evaluate the general behavior of hardware impairments across different types of physical layers. Notably, WiFi B and N represent two distinct PHY technologies —-- Direct-Sequence Spread Spectrum (DSSS) and Orthogonal Frequency Division Multiplexing (OFDM) --- which introduce significantly different modulation, channelization, and transmission characteristics. By including both DSSS-based (WiFi B) and OFDM-based (WiFi N) protocols, our experimental evaluation already captures fundamental diversity in signal structures and hardware impairment manifestations.
Furthermore, the proposed \proposed~framework is inherently protocol-agnostic, as it operates on raw I/Q samples and relies on low-level signal distortions that are present across all WiFi generations, irrespective of the specific modulation or standard. Therefore, while our experiments were conducted using WiFi B and WiFi N due to hardware availability, the methodology and findings are expected to generalize to other WiFi standards, including 802.11a, g, ac, and ad.

\subsection{Beyond Pycom Devices}
Our evaluations demonstrate that \proposed~generalizes well across diverse domains, achieving strong RF fingerprinting performances in both cross-day and cross-protocol scenarios. It consistently maintains high accuracy during both stable operation and the initial warm-up phase when tested on WiFi-enabled Pycom signals. Importantly, the \proposed~framework is not limited to Pycom devices; it is applicable to any embedded wireless device equipped with an RF communication front end, including drones, phones, and IoT gadgets.
For instance, Drones, like other wireless systems, are equipped with RF front-ends that suffer from hardware-induced impairments such as CFO, IQ imbalance, and EVM — all of which stem from manufacturing imperfections and component aging. These impairments are further complicated in drones by the presence of thermal instability during power-up (warm-up effects) and by frequent changes in the wireless channel due to motion, elevation shifts, and varying propagation conditions. Both the transient behavior of impairments during startup and the non-stationarity of the channel can degrade traditional RF fingerprinting models. \proposed~directly addresses these challenges through a sequential transfer learning design that first learns to estimate and abstract the underlying hardware impairments, even in the presence of temperature-related drift, and then transfers that knowledge to a downstream classifier. Moreover, its evaluation across wired and wireless conditions demonstrates robustness to environmental variability, making it more suitable than conventional approaches for mobile and thermally sensitive platforms like drones.

\section{Toward Explainable AI for RF Fingerprinting}
While the surge in deep learning-based approaches has advanced the performance of RF fingerprinting systems, most of these models often operate as black boxes, making them difficult to interpret or validate—particularly in mission-critical applications such as wireless security, spectrum enforcement, and device authentication. As the reliability and accountability of AI systems come under increasing scrutiny, there is a growing consensus in both academia and industry on the importance of explainable AI (XAI) techniques in wireless communication contexts.

Our proposed \proposed~framework contributes to this emerging paradigm by embedding interpretability directly into the fingerprinting process. Rather than mapping raw I/Q samples directly to device identities, \proposed~first learns to estimate a set of physical-layer hardware impairments—factors that are known to be device-specific and inherently tied to the analog front-end. This intermediate estimation acts as a form of structured reasoning, grounding the classification decision in domain-relevant features. From a system design perspective, this two-step pipeline serves as a bridge between domain knowledge and AI, making the overall architecture more explainable, trustworthy, and aligned with the underlying physics of signal generation.

Recent efforts in explainable AI have emphasized saliency maps \cite{simonyan2013deep}, SHAP values \cite{lundberg2017unified}, and attention-based methods \cite{vaswani2017attention} to interpret decisions made by neural networks in image and natural language processing. In wireless, however, the nature of the signal data—being complex-valued, high-rate, and protocol-dependent—poses new challenges. Nevertheless, works such as \cite{zhao2024cross} have begun exploring interpretable RF fingerprinting by identifying the most salient spectral features across domains. Similarly, \cite{yin2025noise} uses denoising diffusion models to isolate device-relevant components, thereby enhancing robustness and interpretability.

These advancements signal a broader trend toward domain-guided interpretability in wireless AI. For instance, \cite{gizzini2023towards} proposes XAI‑CHEST, a framework that quantifies the contribution of different input features in deep channel estimation tasks, providing transparency at the PHY layer. Likewise, \cite{brik2024explainable} emphasize the role of XAI within Open RAN environments, where explainability is essential for the trustworthiness of closed-loop control decisions in the RIC. Furthermore, \cite{munir2023neuro} develops a neuro-symbolic XAI Twin for zero-touch IoE network management, combining black-box predictions with symbolic reasoning for better human interpretability. These efforts collectively demonstrate that explainability is increasingly being treated as a first-class design objective across wireless applications from beamforming to interference management to security analytics.

Building on these insights, we advocate for the broader adoption of structured and physics-informed XAI frameworks in RF fingerprinting and beyond. Approaches that combine domain knowledge, such as modulation structure, protocol timing, or analog impairments, with learned representations offer a promising path toward both generalization and transparency. In this regard, \proposed~is a first step toward that vision. By teaching the model to recognize interpretable, physics-grounded characteristics before performing device identification, our system facilitates deeper understanding and accountability of its decisions.

Future work could expand on this by incorporating attention mechanisms focused on impairment regions, applying post-hoc tools such as SHAP or concept-based explanations, or integrating explainability constraints into the loss function to balance transparency with accuracy. Additionally, real-time deployment scenarios such as ORAN, 6G network slicing, or cognitive spectrum enforcement could benefit from such interpretable designs—ensuring AI not only enhances performance but also earns stakeholder trust.

As AI continues to permeate RF applications—from 6G waveform classification to real-time channel prediction—the need for interpretable and reliable models will become increasingly urgent. Explainable RF fingerprinting, supported by domain-aware architectures like \proposed, will play a critical role in ensuring that the adoption of AI in wireless systems is not only effective but also trustworthy.

%% file: 10-conclusion.tex
This work unveils the significant impact of hardware warm-up on DL-based RF fingerprinting, highlighting its role in cross-domain performance degradation. We introduced \proposed, equipped with sequential transfer learning and targeted impairment estimation, showcasing exceptional accuracy on warm-up captures in both the same-day and cross-domain evaluation for both WiFi type B \& N standards. 
Our evaluations demonstrate \proposed's ability in achieving remarkable classification accuracies during the initial device operation intervals, as well as in cross-day and cross-protocol scenarios, maintaining high accuracy during both the stable and initial warm-up phases when tested on WiFi signals. 
Furthermore, we released comprehensive RF fingerprinting datasets, featuring time-domain I/Q samples labeled with their corresponding hardware impairments. 

While \proposed~introduces an effective approach for improving the robustness of RF fingerprinting, it has some limitations, which open up opportunities for future work.

\begin{itemize}
    \item {Dependency on Impairment-Labeled Data:} The most significant limitation lies in  \proposed's reliance on access to impairment-labeled training data, typically obtained offline via spectrum analyzers. This limits scalability and applicability in real-time or unlabeled field environments. Future directions to overcome this include: 
    (i) Self-supervised learning of impairments, where models could be trained to learn proxy impairment features using contrastive or clustering-based methods that promote embedding consistency across captures from the same device. 
    (ii) Pre-training on synthetic RF data with simulated impairments, followed by fine-tuning on real data. 
    (iii) Teacher-student learning or weak supervision trained with lab-labeled impairments and then used to generate soft labels or embeddings for new unlabeled data in deployment. 
    (iv) Online adaptation using pseudo-labeling or domain alignment to adjust representations when exposed to new environments or protocols, minimizing reliance on handcrafted labels.

\item {Limited Protocol Evaluation:} While  \proposed~operates on raw I/Q samples and is theoretically protocol-agnostic, it is only evaluated on 802.11b and 802.11n. Newer standards (e.g., 802.11ac/ax/ad) introduce challenges like higher-order modulations and mmWave frequencies. Enhancements such as protocol-aware fine-tuning, modulation-aware models, or multi-protocol training could extend generalization and help \proposed~adapt to protocol-specific signal characteristics while preserving its physical-layer interpretability.

\item {Controlled Environment Assumptions:} Our reported experiments were conducted primarily in controlled indoor environments with limited mobility and low channel variability. However, real-world wireless deployments --- particularly in urban, vehicular, or drone-based scenarios --- exhibit frequent channel impairments due to multipath fading, Doppler shifts, and interference. While \proposed's impairment-guided learning helps mitigate the entanglement between channel effects and hardware signatures (especially when training on both wired and wireless data), further robustness that account for real-world settings could be achieved by (i) incorporating mobility-aware data augmentation strategies, (ii) using domain randomization to simulate realistic channel dynamics, or (iii) explicitly modeling the channel via auxiliary inputs or multi-task learning. These additions could enable the framework to maintain high fingerprinting accuracy even under rapidly changing channel conditions and non-stationary environments.

\item {Open-Set Detection:} While the current implementation assumes a fixed set of devices, extending \proposed~to support open-set recognition---where the model can reject unseen devices---and continual learning---where the system can learn new devices on-the-fly without catastrophic forgetting---is a promising direction for future work. These capabilities would allow \proposed~to transition from a lab-scale proof-of-concept into a fully deployable RF fingerprinting solution in dynamic environments such as smart homes, industrial networks, and drone swarms.

 \end{itemize}